\documentclass[traditabstract]{aa} 
\usepackage{graphicx,natbib,longtable,txfonts,lscape}

\newcommand{\kms}{$\rm{\,km \,s}^{-1}$}

\newcommand{\Ha} {H$\alpha$}  
\newcommand{\Hb} {H$\beta$}

\newcommand{\WHz}{\>{\rm W}\,{\rm Hz}^{-1}}

\usepackage{hyperref}
\usepackage{upgreek}

\hypersetup{
  colorlinks   = true, 
  urlcolor     = blue, 
  linkcolor    = red, 
  citecolor    = blue 
}

\begin{document}

\titlerunning{The LOFAR view of massive early-type galaxies}
\authorrunning{A. Capetti and M. Brienza }

\title{The LOFAR view of massive early-type galaxies:\\ Transition
  from radio AGN to host emission.}  \author{A. Capetti\inst{1} \and
  M. Brienza\inst{2,3}}

\institute{INAF - Osservatorio Astrofisico di Torino, Strada
  Osservatorio 20, I-10025 Pino Torinese, Italy \and
  INAF - Osservatorio di Astrofisica e Scienza dello Spazio di
  Bologna, via Gobetti 93/3, 40129 Bologna, Italy \and Dipartimento di
  Fisica e Astronomia, Universit\`a di Bologna, Via P. Gobetti 93/2} \date{}

\abstract{We extend the study of the radio emission in early-type
  galaxies (ETGs) in the nearby Universe (recession velocity $<7,500$
  \kms) as seen by the 150 MHz Low-Frequency ARray (LOFAR)
  observations and extend the sample from giant ETGs to massive
  ($\sim 6 \times10^{10} - 3 \times10^{11} {\rm M}_\odot$) ETGs
  (mETGS) with $-25 < M_K < -23.5$. Images from the second data
  release of the LOFAR Two-metre Sky Survey were available for 432
  mETGs, 48\% of which are detected above a typical luminosity of
  $\sim 3 \times 10^{20} \WHz$. Most (85\%) of the detected sources
  are compact, with sizes $\lesssim 4$ kpc. The
  radio emission of 31 mETGs is extended on scales ranging from 2 to 180 kpc (median
  12 kpc). In several cases, it is aligned with the host galaxy. We set a
  limit of $\lesssim$1\% to the fraction of remnant or restarted objects, which is
  $\lesssim$16\% of the extended sources.

  We found that the properties of the radio sources are connected with
  the stellar mass of the ETGs (the median radio power, the fraction
  of extended radio sources, and the link with the large-scale
  environment). However, these results only describe statistical
  trends because the radio properties of sources of similar stellar
  mass and environment show a large spread of radio properties. These
  trends break at the lowest host luminosities ($M_K>-24.5$). This
  effect is strengthened by the analysis of even less massive ETGs,
  with $-23.5 < M_K < -21.5$. This suggests that at a mass of $\sim 2
  \times10^{11} {\rm M}_\odot$ , a general transition occurs from
  radio emission produced from radio-loud active galactic nuclei (AGN)
  to processes related to the host galaxy and (or) radio quiet AGN. At
  this luminosity, a transition in the stellar surface brightness
  profile also occurs from S\'ersic galaxies to those with a depleted
  stellar core, the so-called core galaxies. This finding is in line
  with previous results that indicated that only core galaxies host
  radio-loud AGN.}
\keywords{galaxies: active --  galaxies: jets} 
\maketitle

\section{Introduction.}
\label{intro}

Radio-mode feedback, that is, the interplay between the transfer of
energy and matter from relativistic jets to the external medium and
the accretion powering active galactic nuclei (AGNs), is an essential
ingredient in the evolution of galaxies (e.g.,
\citealt{fabian12}). The energy injected by relativistic jets is
thought to quench star formation and produce the exponential cutoff
at the bright end of the galaxy luminosity function
\citep{croton06}. Powerful radio sources are almost invariably
associated with early-type galaxies (ETGs) with high black hole
masses, $M_{\rm SMBH} \gtrsim 10^8 M_\odot$
\citep{baldi10,chiaberge11}. In order to understand how radio-mode
feedback operates, it is necessary to explore the properties and the
origin of the radio emission in massive ETGs to study their
morphology and to explore their duty cycle.

In recent years, the advent of new-generation radio facilities has
allowed major progress in this field, especially because much lower flux limits than previously
accessible could be explored. For example, it has become clear that the extended and
powerful radio galaxies (hundreds of kiloparsec, kpc, with L$\gtrsim10^{24}$ W
Hz$^{-1}$), which dominated the historical 3C and B2 radio
samples, clearly represent only a small part of the entire
radio AGN population. The bulk of radio AGN instead has low radio
powers and sizes smaller than a few kpc (e.g.,
\citealt{best05a,baldi15,hardcastle19}). In analogy to the historical
radiogalaxy classification into FRI or FRII ( \citet{fanaroff74}), this
population of compact radio sources was named FR0s. It is still debated why most radio sources do not develop large-scale jets, but it is likely due to a combination of different
intrinsic properties of the black hole (e.g., mass and spin) and the mass
of the host, its environment, and the jet duty cycle.

Even the classical FRI/FRII dichotomy (low power/edge-darkened versus
high-power/edge-brightened radio galaxies) has recently started to be
questioned. Large samples of extended sources at high sensitivity have
indeed shown that there is a very large overlap in luminosity for the
two morphologies. In particular, a population of
low-power FR IIs was revealed (e.g., \citealt{capetti17,mingo19,mingo22}). They
appear to be associated with lower-mass hosts and reach radio
luminosities lower by three orders of magnitude of the traditional
FR break ($\sim 10^{25}$ W Hz$^{-1}$ at 1.4 GHz
\citealt{ledlow96}). Two main scenarios have long been discussed in
the literature to explain the difference between FRI and FRII
morphology. The first scenario suggests that FRIs live in more massive
galaxies and denser environments, which are able to decelerate their
jets with respect to FRIIs. The second scenario suggests instead that there
is a more intrinsic difference in the jet formation of these two
classes of sources, or that it might be related to a different accretion
process (radiatively efficient versus radiatively inefficient). However,
while the large predominance of FRI radio galaxies seem to be
associated with radiatively inefficient accretion, FRII sources can
clearly be both: Up to 65\% of the luminous FRIIs (L$_{150}
\gtrsim 10^{26}$ W Hz$^{-1}$) are radiatively inefficient (e.g.,
\citealt{miraghaei17,mingo22}.)

To explore all this further, in \citet{capetti22}, hereafter Paper I,
we studied the properties and the origin of the low-frequency radio
emission in the most luminous giant ETGs (gETGs) in the local Universe
(recession velocity $\rm \leq 7,500 \ km \ s^{-1})$. Low-Frequency
ARray (LOFAR, \citealt{shimwell17}) observations at 150 MHz that were
obtained as part of the LOFAR Two-metre Sky Survey (LoTSS) were
available for 188 out of the 489 selected galaxies at the time of this
analysis.  We confirmed the positive correlation between the stellar
luminosity of gETGs and their median radio power that has been
reported by previous studies
\citep{colla75,fanti78,sadler89,wrobel91a,wrobel91b,best05b,brown11,sabater19}
and also the very wide spread of the radio luminosity (several orders
of magnitude) at a given host mass.  Two-thirds of the detected gETGs
are compact, with sizes $\lesssim 4$ kpc. This confirms the prevalence
of compact radio sources in nearby ETGs \citep{capetti17}. The radio
emission of 46 (24\%) gETGs is extended, with sizes up to $\sim$300
kpc, and at least 80\% of these have a \citet{fanaroff74} type I
morphology. Based on the morphology and spectral index of the extended
sources, we classified $\sim$30\% of them as candidate remnant or
restarted sources. Optical spectroscopy (available for 44 gETGs)
indicates that the nuclear gas in 7 gETGs is ionized by young stars,
which suggests a contribution to their radio emission from
star-forming regions. Although all the 25 most luminous ($M_K <-25.8$)
gETGs are detected at 150 MHz, they are not all currently turned on:
At least 4 of them are remnant sources, and at least one source is
likely powered by star formation.

We here extend our analysis to less massive ETGs (mETGs) with $-23.5 <
M_K <-25$ in the same redshift range (see below for the
details of the sample selection). The selected galaxies lie
  above the exponential cutoff of the near-infrared luminosity
  function of local ETGs, which occurs at $M_K \sim $ -22.3 (see,
  e.g., \citealt{bell03b}). ETGs can be separated into two classes
based on their surface brightness profile (SBP): The SBPs of S\'ersic galaxies
(SGs) are well represented by a single S\'ersic
function \citep{sersic68}, while it deviates in core-S\'ersic galaxies (CSGs) from a S\'ersic law in that the core is partially depleted
\citep{graham03,trujillo04,scott13}. SGs are less luminous than CSGs.
The transition occurs at $M_K \sim -24.5$ \citep{capetti06},
although the two classes coexist at luminosities $-23.5 \gtrsim M_K \gtrsim -25$ . \citeauthor{capetti06} found that the radio-loudness
of AGN that are hosted by ETGs is univocally related to the brightness
profile of the host: CSGs invariably host radio-loud nuclei, while SGs are
associated with radio-quiet nuclei. The gETGs are expected to be CSGs
based on their luminosity, while the mETGs straddle the luminosity
separation between the two classes. Therefore, the sample of ETGs we
study in this work covers the crucial mass range in which the
transition between SGs and CSGs (and radio-loud/radio-quiet AGN)
occurs.

We also consider a sample of even less massive ETGs
(lmETGs) with $-21.5 < M_K <-23.5$ in our study. The sample corresponds to a mass range of
$\sim 0.8\times10^{10} - 6 \times10^{10} {\rm M}_\odot$. The lmETGs are thought to be all SGs
based on their near-infrared luminosity.

\begin{figure}
\includegraphics[scale=0.5]{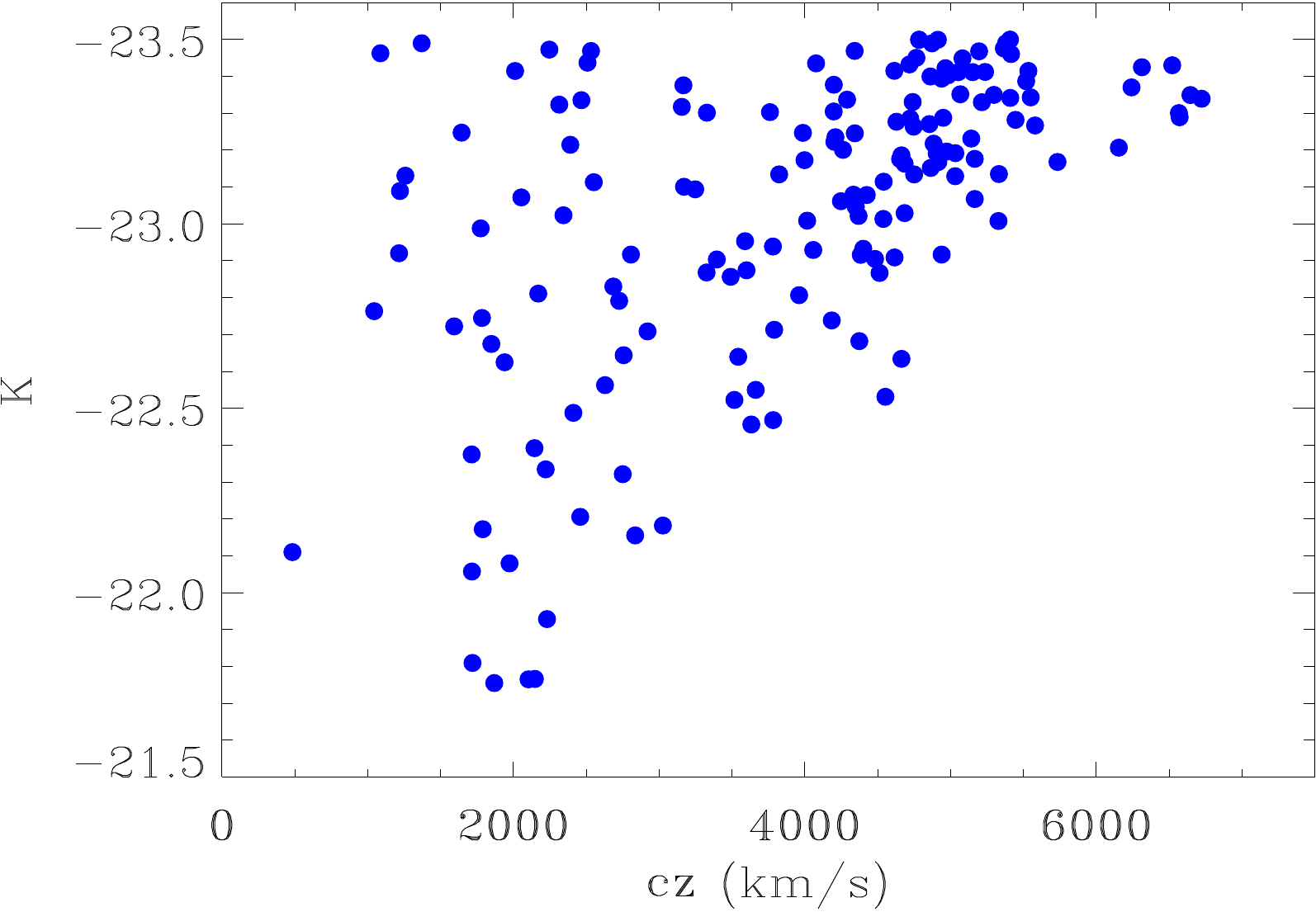}
\caption{Distribution in the $cz$ vs. absolute magnitude plane of
  the lmETG sample. The empty region in the bottom right corner is due
  to the threshold of the 2MASS sample we used for the near-infrared
  selection of these galaxies. }
\label{kz}
\end{figure}

\begin{figure*}[h]
\includegraphics[scale=0.5]{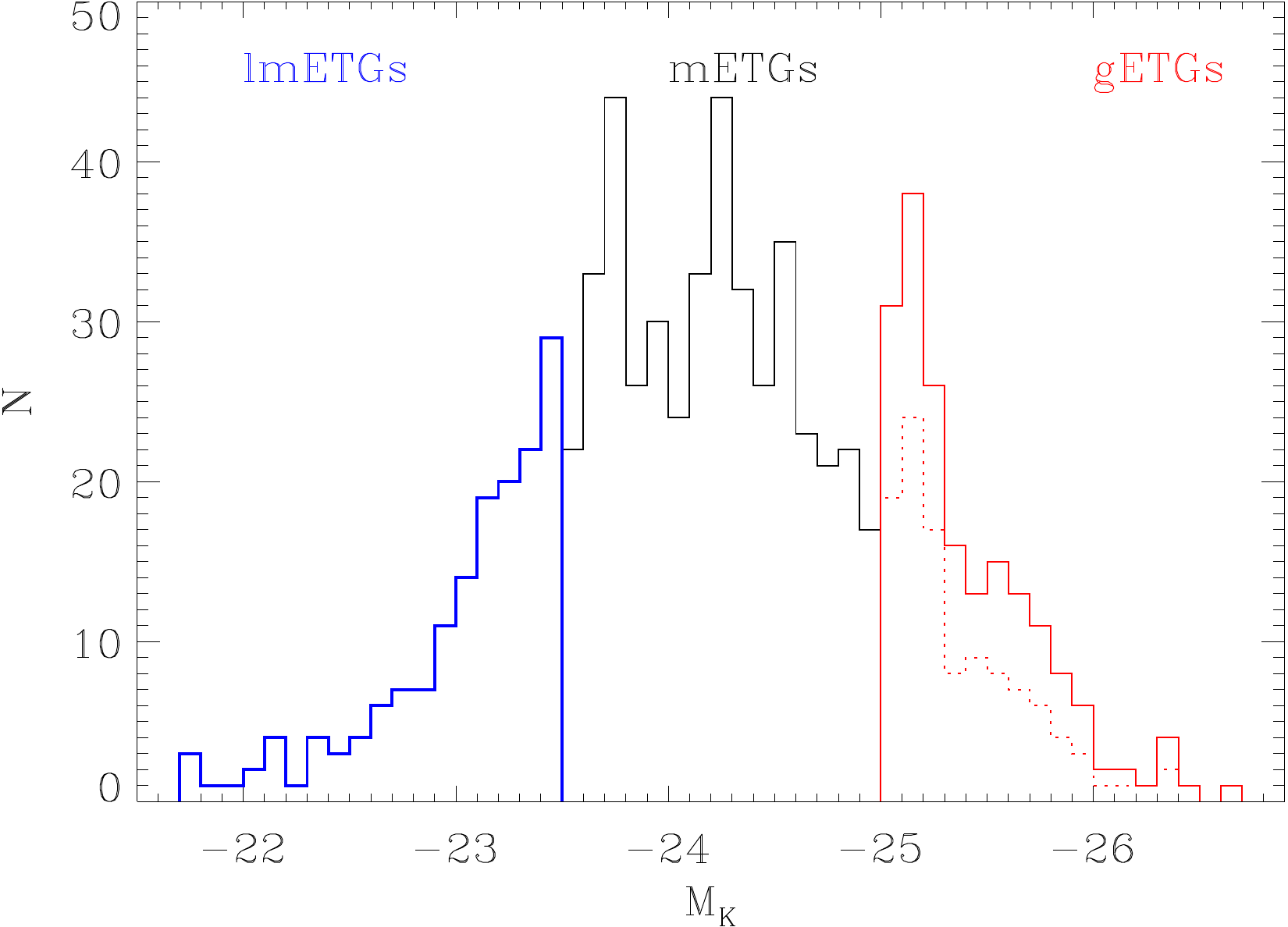}
\includegraphics[scale=0.5]{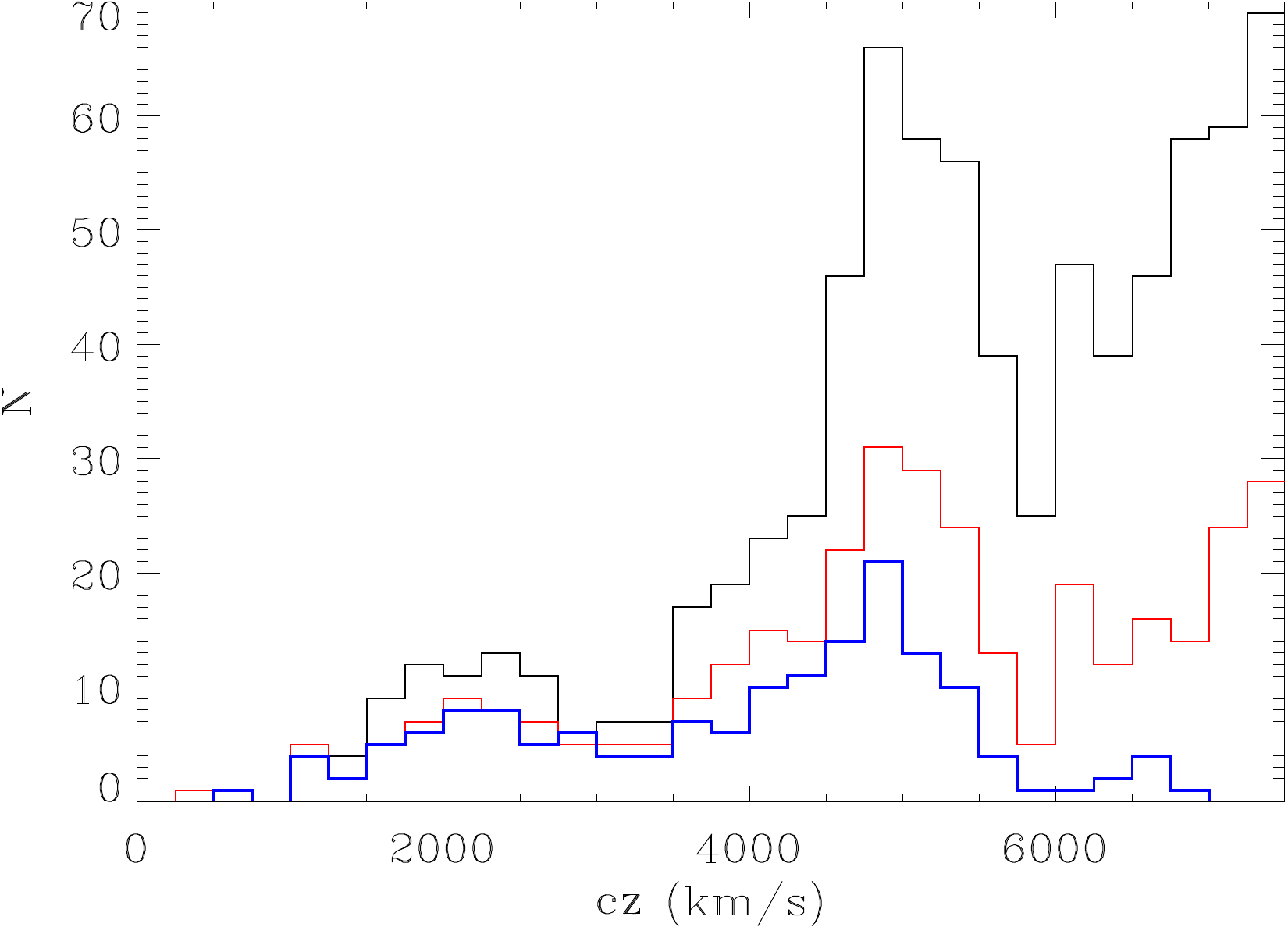}
\caption{Distributions of absolute K-band magnitude (left) and
  recession velocity (right) for the three ETG samples. The red
  histograms show the gETGs studied in Paper I (the dashed histogram
  in the left panel represents the gETGs included in the LoTSS DR2
  area, the same region in which the mETGs and lmETGs were selected), the
  black histogram shows the mETGs sample, and the blue histogram represents the
  lmETGs. The lmETG sample is not complete, and this is clearly seen in
  the reduced number of galaxies below $M_K>-23.5$ and for $v \gtrsim
  6000$ \kms.}
\label{kzhist}
\end{figure*}

The properties of radio-AGN are also likely to be affected by the
large-scale environment. \citet{best07} studied the connection between
environment and radio activity in nearby groups ($z<0.1$) identified
in observations of the Sloan Digital Sky Survey (SDSS,
\citealt{york00b}) by \citet{vonderlinden07}. They found that brightest
group and cluster galaxies (BCGs) are more likely to host a radio-loud
AGN than galaxies of the same stellar mass. Within $\sim$20\% of the
virial radius, cluster galaxies exhibit an enhanced likelihood of
radio-loud AGN activity. \citet{croston19} found that only 10\% of the
radio AGN are located in halos with $M>10^{14} M_{\odot}$. Within this
population, they found that the radio luminosity of the brightest AGN
depends on the richness, and that the radio luminosity negatively
depends on the distance from the cluster center. These results
indicate that the nuclear activity is related to the large-scale
environment and not to the host stellar mass alone.

\citet{capetti20b} explored the properties of the large-scale
environment of the compact FR~0 radio galaxies associated with nearby
($z<0.05$) ETGs. They reported that FR~0s are located in regions with an
average galaxy density that was lower by a factor of two than for
FR~Is. This difference is driven by the large fraction (63\%) of FR~0s
that are located in groups formed by fewer than 15 galaxies. FR~Is
rarely (17\%) inhabit an environment like this. The authors interpreted the
differences in environment between FR~0s and FR~Is as due to an
evolutionary link between local galaxies density, black hole spin, jet
power, and extended radio emission. The analysis of the environment of
local ETGs can be used to test these conclusions and strengthen their
statistical significance.

The paper is organized as follows: In Sect. 2 we describe the sample
of the selected sources and the available radio observations with
LOFAR and from surveys at 1.4 GHz. In Sect. 3 we present the main
results for the mETGs, including a description of the radio morphology
and spectral shape. The connection between radio properties and
environment is the subject of Sect. 4. The information available for
the mETGs from optical spectroscopy are described in Sect. 5. In
Sect. 6 we draw a comparison of the properties of gETGs and mETGs. The
analysis is then extended to even lower-mass ETG in Sect. 7, which  are
used to set our results on a firmer basis. In Sect. 8 we discuss the
results, which are summarized in Sect. 9, where we also draw our
conclusions.

\section{Sample selection and the LOFAR observations.}
\label{sample}

\begin{figure}
\includegraphics[scale=0.5]{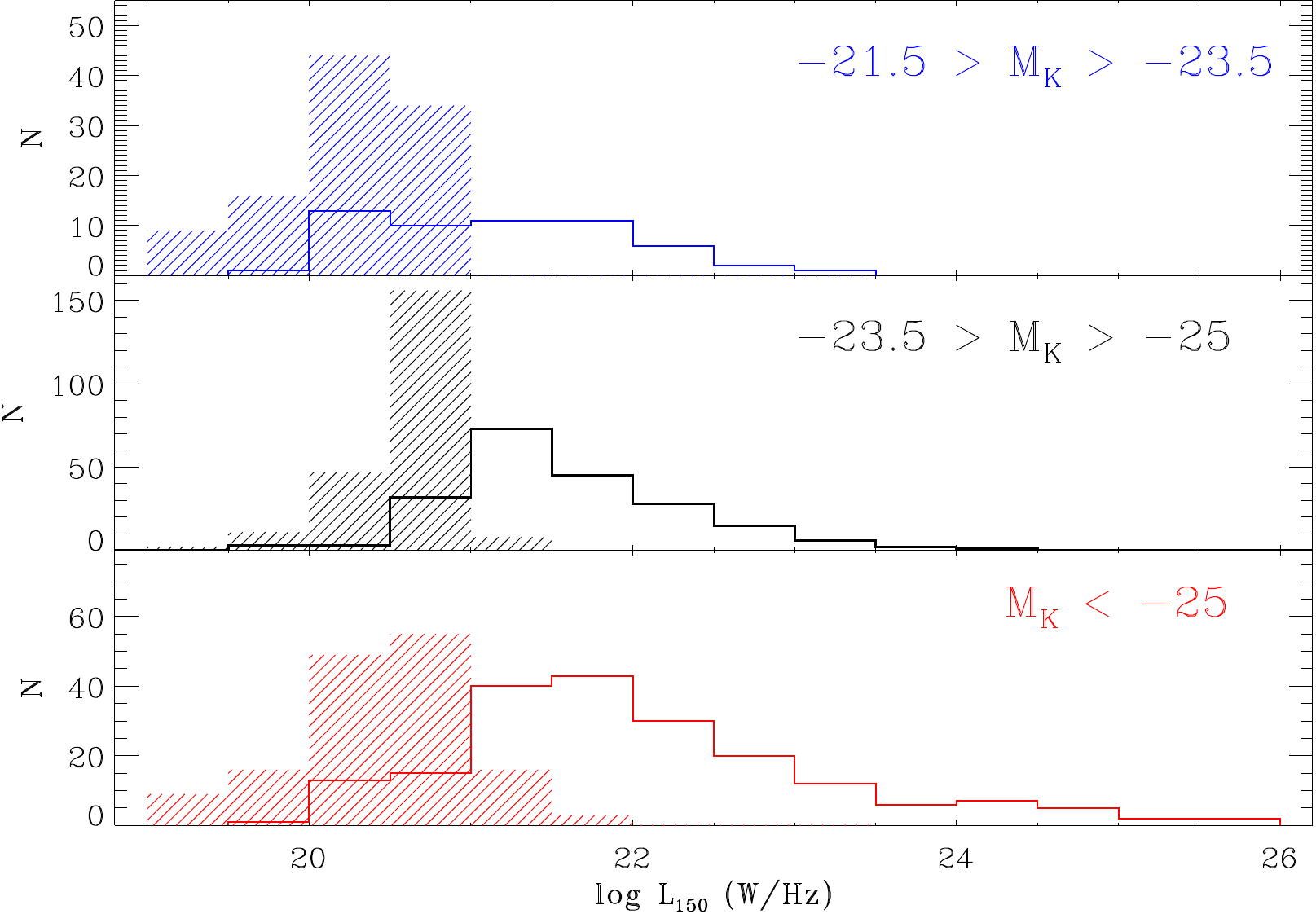}
\caption{Distribution of the 150 MHz luminosity of the three samples
  of ETGs: (top) lmETGs, (middle) mETGs, and (bottom) the gETGs
  studied in Paper I. The dashed histograms correspond to the upper
  limits of the undetected sources.}
\label{lum}
\end{figure}

\begin{figure*}
\includegraphics[scale=0.27]{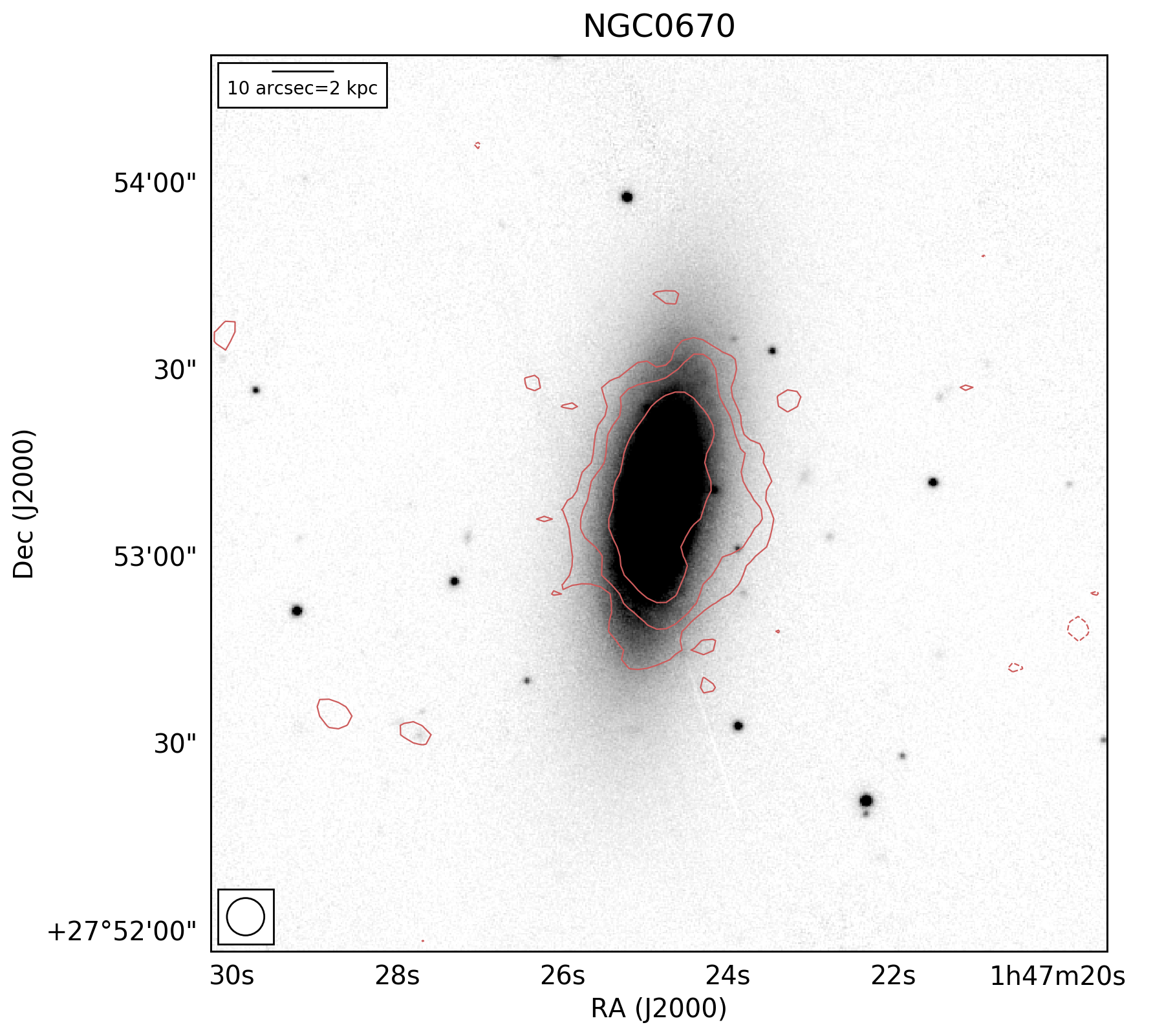}
\includegraphics[scale=0.27]{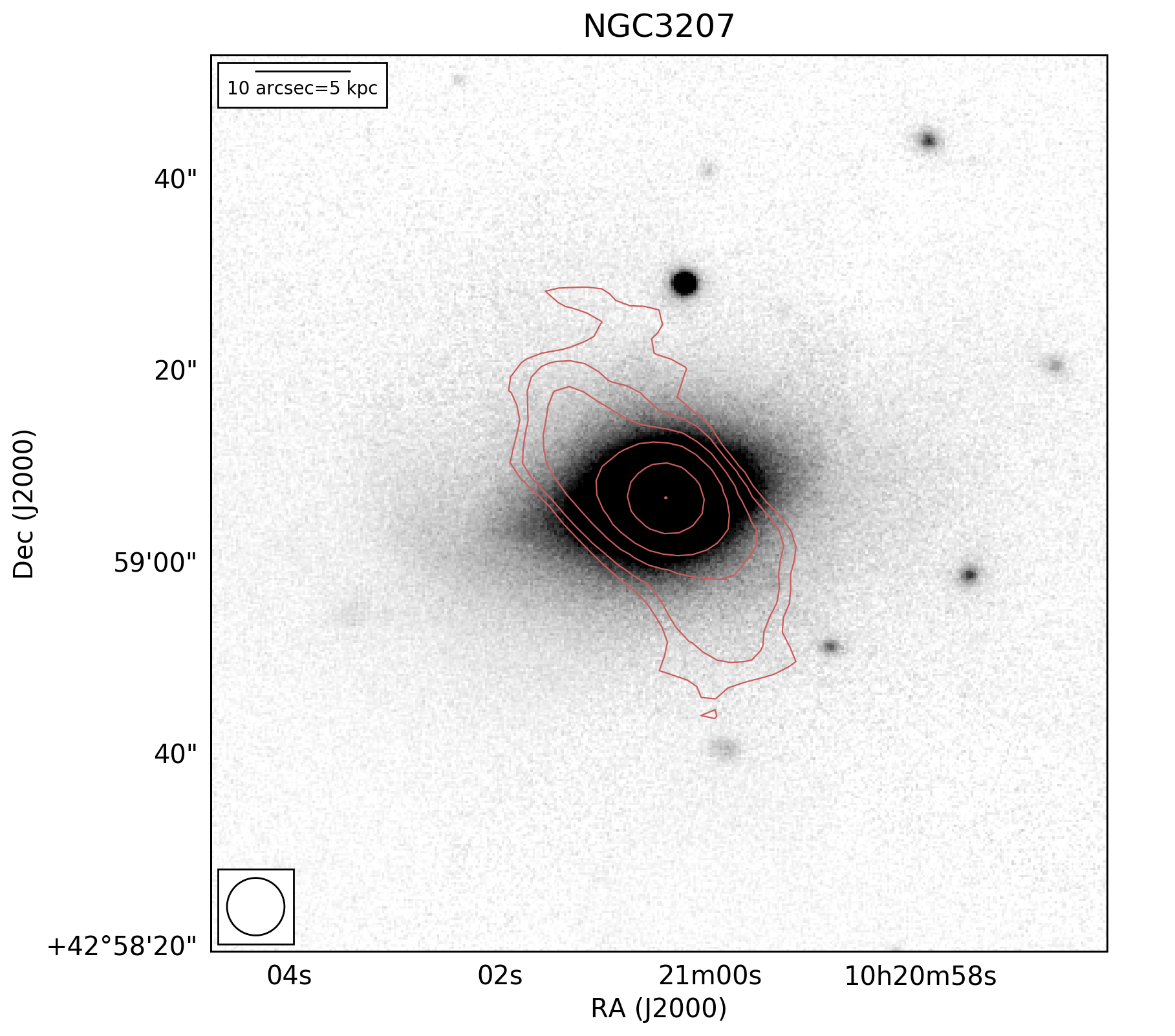}
\includegraphics[scale=0.27]{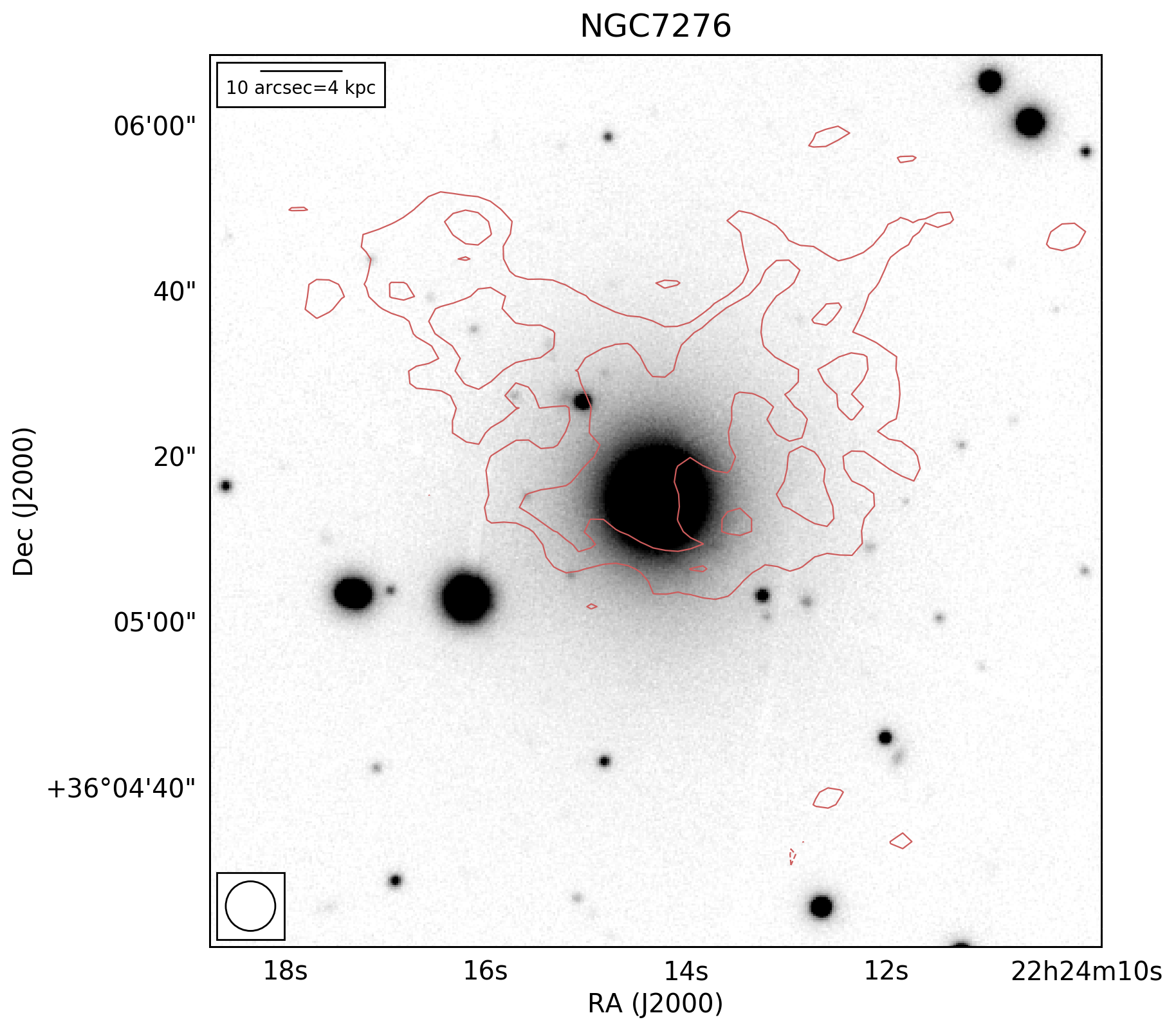}
\caption{Three examples of mETGs with extended radio emission,
  representative of the main morphological classes: (left) cospatial
  radio and optical emission, (center) jetted source, (right) diffuse
  radio source. The LOFAR images at 150 MHz are superposed to the
  optical images from Pan-STARRS. The lowest contours are drawn at
  three times the local r.m.s. (see Table \ref{tab}). Images of all 31
  extended sources are shown in the Appendix B. }
\label{estese}
\end{figure*}

In Paper I we selected the galaxies that are included in the 2MASS
Redshift Survey \citet{huchra12} by requiring a declination DEC
$>0^\circ$, a Hubble type $T \le 0$, a total absolute magnitude $M_K
<-25$, and a recession velocity (corrected for the effects of the
Virgo cluster, the Great Attractor, and the Shapley supercluster;
\citealt{mould00}) $v< 7500$ \kms.  We here extend the analysis to
less massive ETGs by requiring a total absolute magnitude in the
$-23.5 < M_K <-25$ range. The magnitude limit of the 2MASS survey,
$K=11.75$, translates into a threshold in the absolute of $M_K =
-23.45$ at $v= 7500$ \kms: The selected sources form a complete
volume-limited sample. By using the relation between the dynamical
mass and K-band absolute magnitude (log $M_* =
10.58-0.44\times(M_{Ks}+ 23)$; \citealt{cappellari13}), the mETG
masses lie in the range of $\sim 6 \times10^{10} - 3 \times10^{11}
{\rm M}_\odot$, while we considered in Paper I all the available LoTSS
pointings (including DR1 data and proprietary data).  Considering the
higher space density of lower-mass ETGs, we can now limit ourselves to
the 432 galaxies that are located in the area covered by the second
LoTSS data release (DR2; \citealt{shimwell22}) as they form a
sufficiently large sample for this statistical study. Furthermore, the
DR2 images are deeper than those obtained from individual LoTSS
pointings. Two additional sources are superposed to the lobes of
bright nearby sources, and they were dropped from the sample at this
stage.

In the same DR2 area there are 158 lmETGs.  The magnitude limit of the
2MASS survey, $K=11.75$, translates into a threshold in the absolute
of $M_K = -23.45$ at $v= 7500$ \kms: Unlike the gETGs and mETGs, the
lmETGs sample is not complete (see Fig. \ref{kz} where we present the
distribution of the lmETGs in the $cz$ versus absolute magnitude
plane).

Figure \ref{kzhist} shows the distributions in redshift and absolute
K-band magnitude of all the ETGs considered in this study. The
combined sample of mETGs and gETGs covers a range of a factor $\sim 100$
in near-infrared luminosity.

The LoTSS DR2 includes 27\% of the
northern sky with a median rms sensitivity of 83 $\mu$Jy/beam. DR2
provides fully calibrated mosaics at a resolution with a full width at
half maximum (FWHM) $\sim 6\arcsec$, catalogs, and pipeline
products. The flux density scale has an accuracy of approximately 10\%.

\section{The radio properties of mETGs.}
\label{results}

We estimated the rms of each image of the mETGs in various regions,
usually centered 45$\arcmin$ away from the source of interest. The
median rms is 88$\mu$Jy/beam, similar to the value measured over the
whole DR2. The radio flux density of the sources is available from the
released catalog, but several galaxies have large-scale and complex
radio structures that are not always fully included in the catalog
measurement. For these objects, we measured the flux densities directly
from the images and included the whole source emission within the
3$\sigma$ isophote. We obtained a detection at $>5\sigma$ significance
for 208 (48\%) of the sources of the sample.

The measurements at 1.4 GHz were obtained from the Faint Images of the
Radio Sky at Twenty centimeters survey (FIRST;
\citealt{becker95,helfand15}) or from the National Radio Astronomy
Observatory Very Large Array Sky Survey (NVSS;
\citealt{condon98}). The FIRST area contains 219 galaxies,
45 of which are detected by this survey above the 5$\sigma$ level. For
the undetected sources, we estimated upper limits at five times the
local noise, typically $\sim 0.14$ mJy beam$^{-1}$. For the 102
galaxies outside the FIRST area, we collected the NVSS measurements; 37
of these sources have an NVSS detection. For the remaining 65 sources, we
adopted a limit of 2 mJy. For the extended galaxies (defined as
described in Section 3) we measured the NVSS flux densities in the
same region as was used for the LOFAR images.

The LOFAR observations did not detect 224 mETGs at a 5$\sigma$
significance (52\%). The corresponding limits to the radio
luminosity range from $\sim$3$\times10^{19}$ to $\sim 3\times10^{21}
\WHz$, with a median of 3$\times10^{20} \WHz$. The luminosity distribution at 150 MHz of the mETGs covers a wide range in power from
$\sim 4 \times 10^{19} \WHz$ to $\sim 2 \times 10^{24} \WHz$ and is
compared to that obtained for the gETGs in Figure \ref{lum}. The mETGs
have a lower median power than the gETGs, which extend
to a higher radio power.

We defined the 31 sources in which the 3$\sigma$
level isophote extends to a radius of at least 15\arcsec (about twice
the beam FWHM) as extended objects. Their images are presented in Appendix B. In
Figure \ref{estese} we show three examples that are representative of the main
morphological classes. The sizes of the extended sources range from 2.2
to 180 kpc, with a median of 12 kpc.

In 5 galaxies with extended radio structures, optical and radio
emission are closely cospatial (see Appendix B for a morphological
description of the individual sources), and the radio emission is
elongated along the same axis as the host galaxy in another 6
cases. This suggests that the radio emission is not produced by an AGN
in these 11 galaxies, but is due to processes related to the host,
such as free-free emission in star-forming regions and cosmic rays
accelerated by supernova explosions (see, e.g.,
\citealt{tabatabaei17}). Nine sources have an elongated morphology
that suggests the presence of jets. The radio emission of four mETGs
is instead diffuse and lacks a central component, while one has a
ring-like structure. The radio structure of the remaining six sources
is unclear because they are barely resolved.

The remaining 177 mETGs are unresolved or marginally resolved. The
limit to their sizes is $\lesssim4$ kpc. These compact sources
represent 85\% of the detected sources and 41\% of the whole sample.
The fraction of detected sources (and of those with extended emission)
among the mETGs decreases with decreasing near-infrared host luminosity
(see Table \ref{tab2}).

\begin{figure}
\includegraphics[scale=0.55]{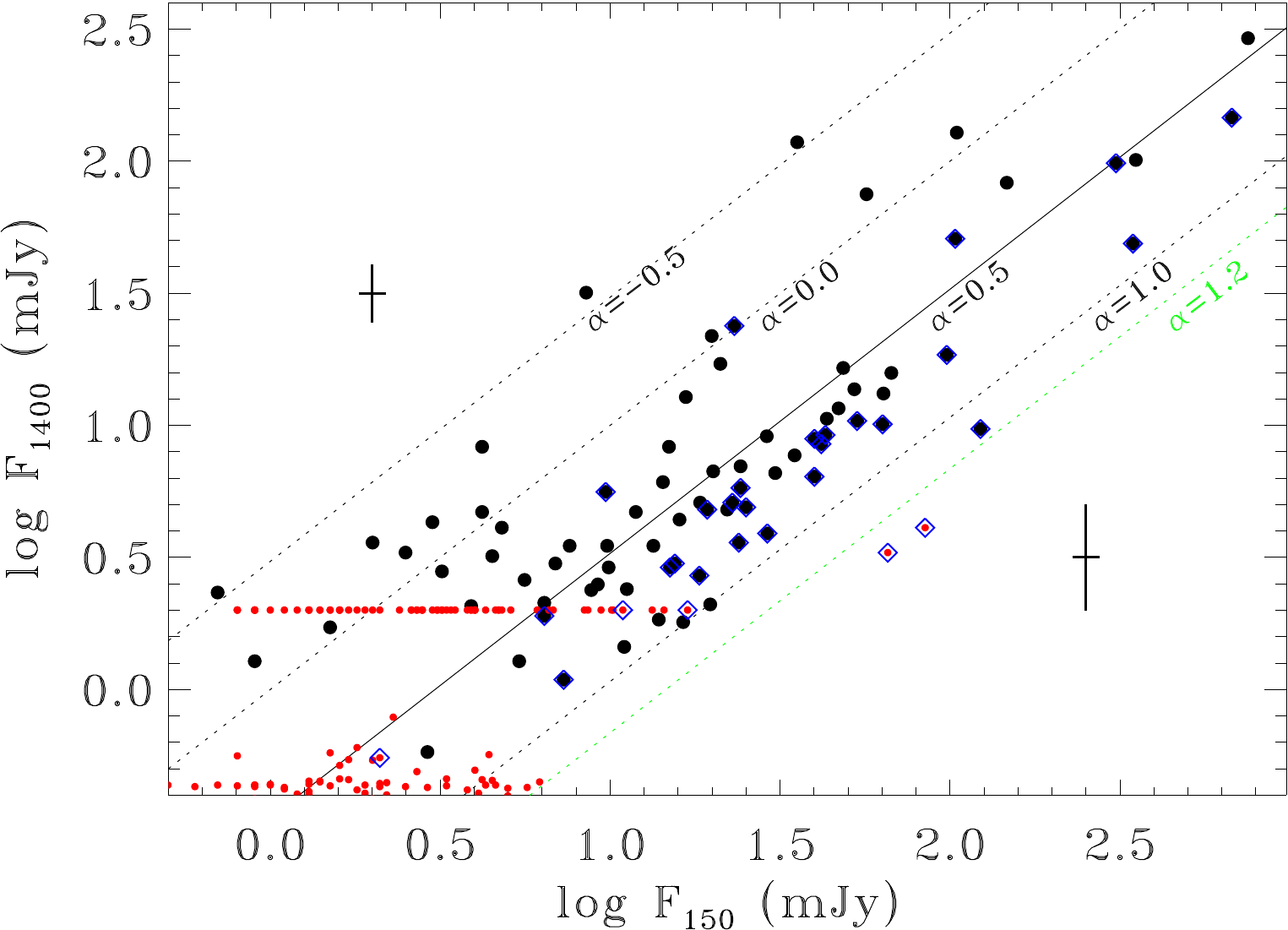}
\caption{Comparison of the flux densities at 150 and 1400 MHz for the
  mETGs. The small red dots mark the upper limits for the sources that are not
  detected at 1.4 GHz. The two clusters of upper limits correspond to
  the sources that are undetected by either the FIRST or the NVSS. The two
  black crosses represent the typical errors on the flux density
  measurements, which are $\sim 10$\% on $F_{150}$ and $\sim 30$\% and
  $\sim 5$\% for high and low values of $F_{1400}$, respectively. The
  blue diamonds mark the extended sources at 150 MHz. The lines
  represent the loci of constant spectral index.}
\label{spix}
\end{figure}

In Figure \ref{spix} we compare the flux densities at 150 MHz and 1.4
GHz of the radio sources associated with the mETGs. The compact
objects show a large spread in their spectral slopes, $-0.5 <
\alpha_{150,1400} < 1.2$, with the spectral indices $\alpha$ 
  defined as $F_{\nu}\propto\,\nu^{-\alpha}$. but the range could be
even wider considering the lower limits. Most extended
sources have $0.5 < \alpha_{150,1400} < 1.0$, which is typical of currently
active objects. Three of them (namely, NGC~3619, NGC~4148, and
NGC~6211) have flatter spectral indices ($0.0<\alpha_{150,1400}<0.5$):
These are all core-dominated sources, and their spectrum is determined
by the (likely flat spectrum) central component.

In Paper I we defined as candidate remnant and restarted sources (RRS)
those with 1) a bright small-scale radio structure accompanied by
large-scale diffuse emission (restarted sources) or 2) a radio
spectral index $\alpha_{150,1400}>1.2$ and lack of a central component
(remnant sources). Using the same definition, we find in the sample of
mETGs one candidate restarted source, CGCG499-084. Its radio emission
is dominated by two diffuse tails extending over $\sim$180 kpc and by
a central compact bright region. The overall spectral index is
$\alpha_{150,1400}=1.14$, suggesting that the radio tails are remnants
of a previous phase of activity. The limits to the radio spectral
index for two extended sources (namely, CGCG~311-017 and NGC~7276) are
$\alpha_{150,1400} > 1.34$ and $\alpha_{150,1400}>1.35$, respectively,
and we consider them as candidate remnant sources. Three more extended
sources are not detected at 1.4 GHz either (namely, NGC~2521,
UGC~1022, and UGC~1503). The lower limits to their spectral slopes are
in the $\alpha_{150,1400} > 0.54 - 0.96$ range, which means that they
are not necessarily remnant sources. However, they are all characterized by a
diffuse radio structure, which supports this interpretation. Overall,
we set an upper limit of $\lesssim$16\% to the fraction of RRSs among
the mETGs with extended radio emission and of $\lesssim$1\% for the
whole sample.

\begin{figure}
\includegraphics[scale=0.5]{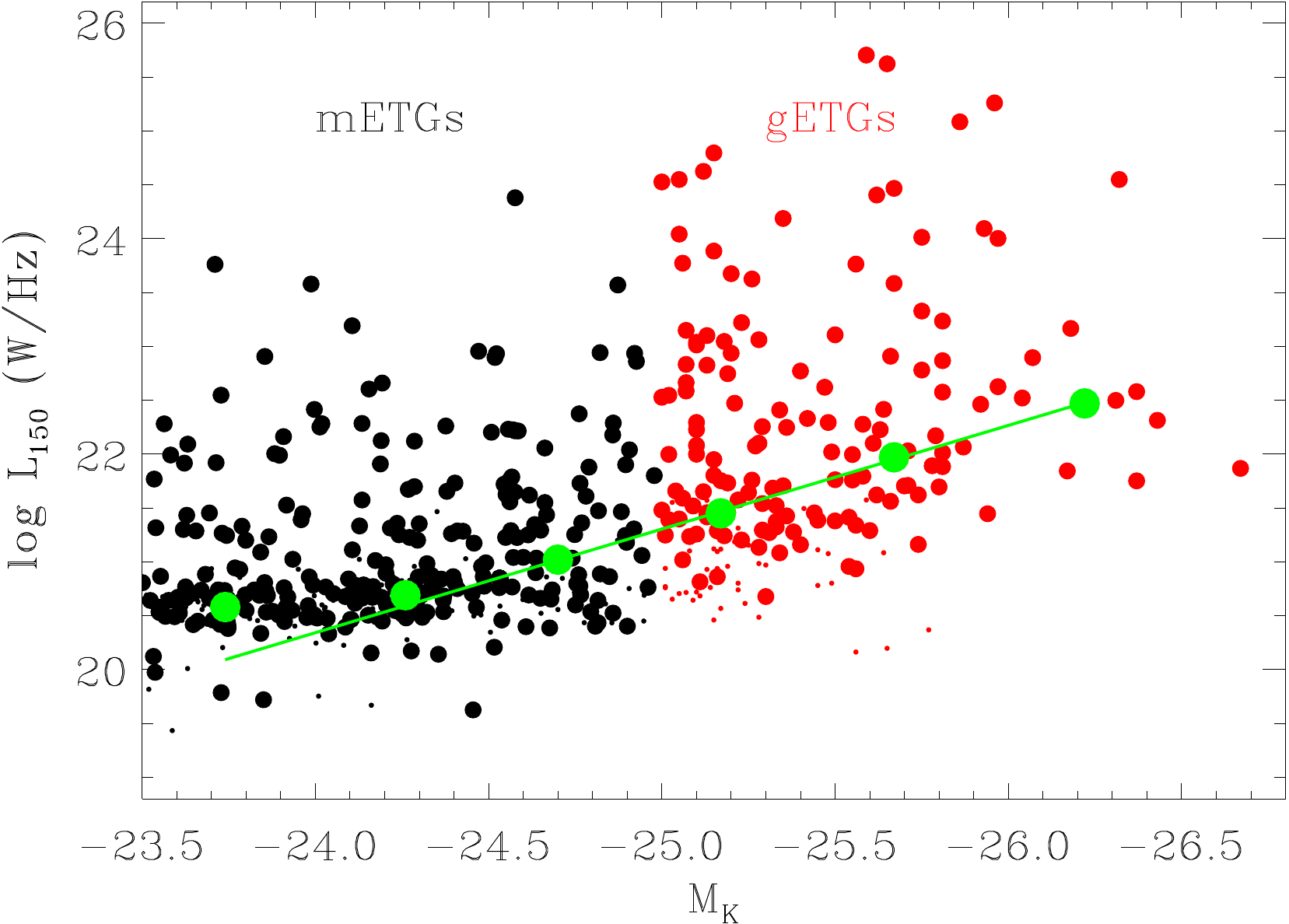}
\caption{Absolute K-band magnitude vs. the luminosity at 150
  MHz. The red symbols show the gETGs ($M_K <-25.0$), and the black symbols
  show the mETGs ($-25.0 < M_K <-23.5$). The small dots mark the upper
  limits of the undetected objects. The large green dots represent the
  median radio luminosity in six bins of absolute magnitude, and the
  green line is the best linear fit to the four brightest galaxies
  bins. \label{mlum}}
\end{figure}

\begin{figure}
\includegraphics[scale=0.5]{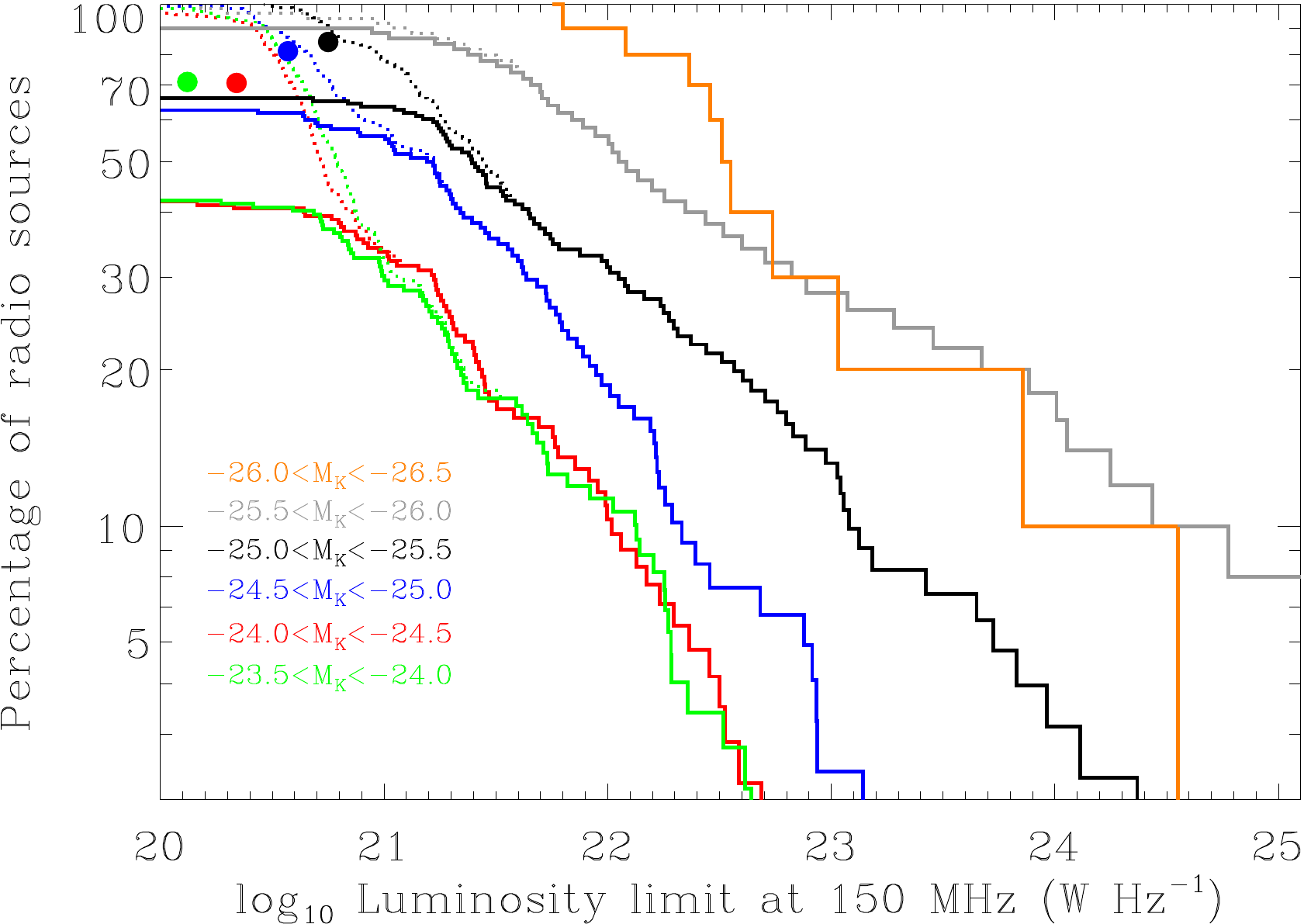}
\caption{Fraction of galaxies that host a radio source above a given
  luminosity limit. The samples of mETGs and gETGs are split into six bins with a K-band luminosity that are 0.5 magnitude wide. The galaxies in the first bin
  have $-23.5 > M_K > -24.0$ (green histogram), and the following bins
  correspond to the red, blue, black, gray, and orange
  histograms. The solid histograms correspond to the detected sources,
  and the dashed histograms include the upper limits and represent the
  lower and upper envelope of the distribution, respectively. The
  circles represent the median value of the undetected sources
  obtained with the stacking technique.}
\label{flum}
\end{figure}

\begin{figure}
\includegraphics[scale=0.52]{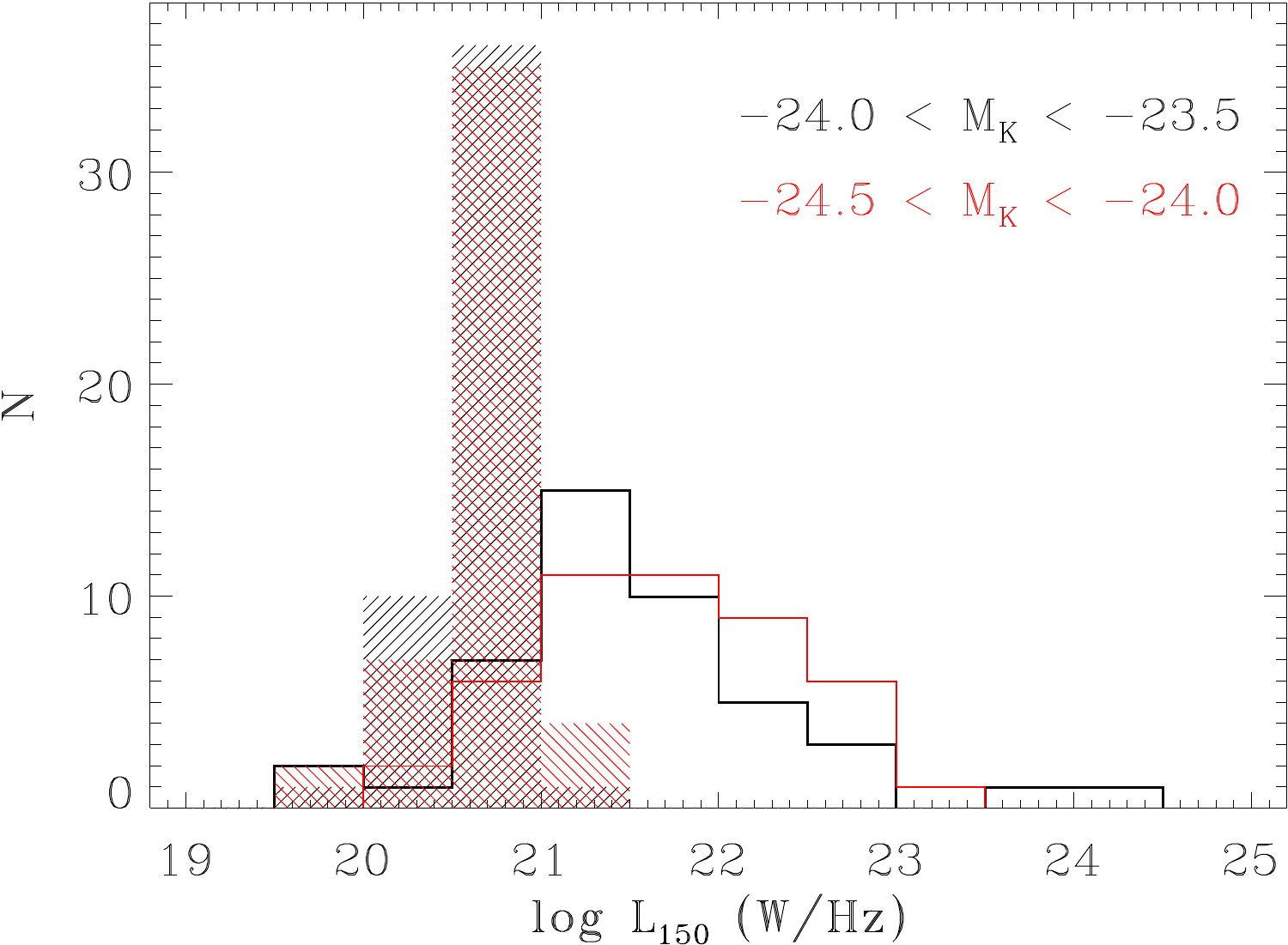}
\caption{Comparison of the distributions of the 150 MHz luminosity of
  the mETGs in the two least massive bins. The dashed histograms
  represent the upper limits for the undetected sources.}
\label{lum2}
\end{figure}

In Figure \ref{mlum} we compare the near-infrared and radio luminosity
of the mETGs and gETGs. The radio power shows a spread of about four
orders of magnitude at a given host luminosity. We estimated the median
luminosity, $\overline{L}_{150}$, of the mETGs by splitting the sample
into three bins of absolute magnitude, each 0.5 mag wide, and by using
the stacking technique (e.g., \citealt{white07}). By performing a
filtering of the images of the three subsamples in the luminosity
domain, we measured the values of $\overline{L}_{150}$ reported in
Tab. \ref{tab2}. The median $L_{150}$ increases with stellar
luminosity. A linear fit{, estimated using the Python procedure {\sl
    polyfit}, to the median $L_{150}$ in the four brightest bins of
  $M_K$ (i.e., also including the gETGs) returns log
  $L_{150}=(-0.96\pm0.02)\times(M_K+25.5)+(21.79\pm0.01)$.  The
  faintest galaxies show an excess with respect to the general
  trend. In particular, the measured median luminosity for the
  galaxies with $-23.5<M_K<-24.0$ is log$L_{150}=20.58\pm0.04$. This
  value is higher by a factor of about three  than the extrapolation of the
  correlation, which predicts log$L_{150}=20.10 \pm 0.03 $.

\begin{figure*}
\includegraphics[scale=0.52]{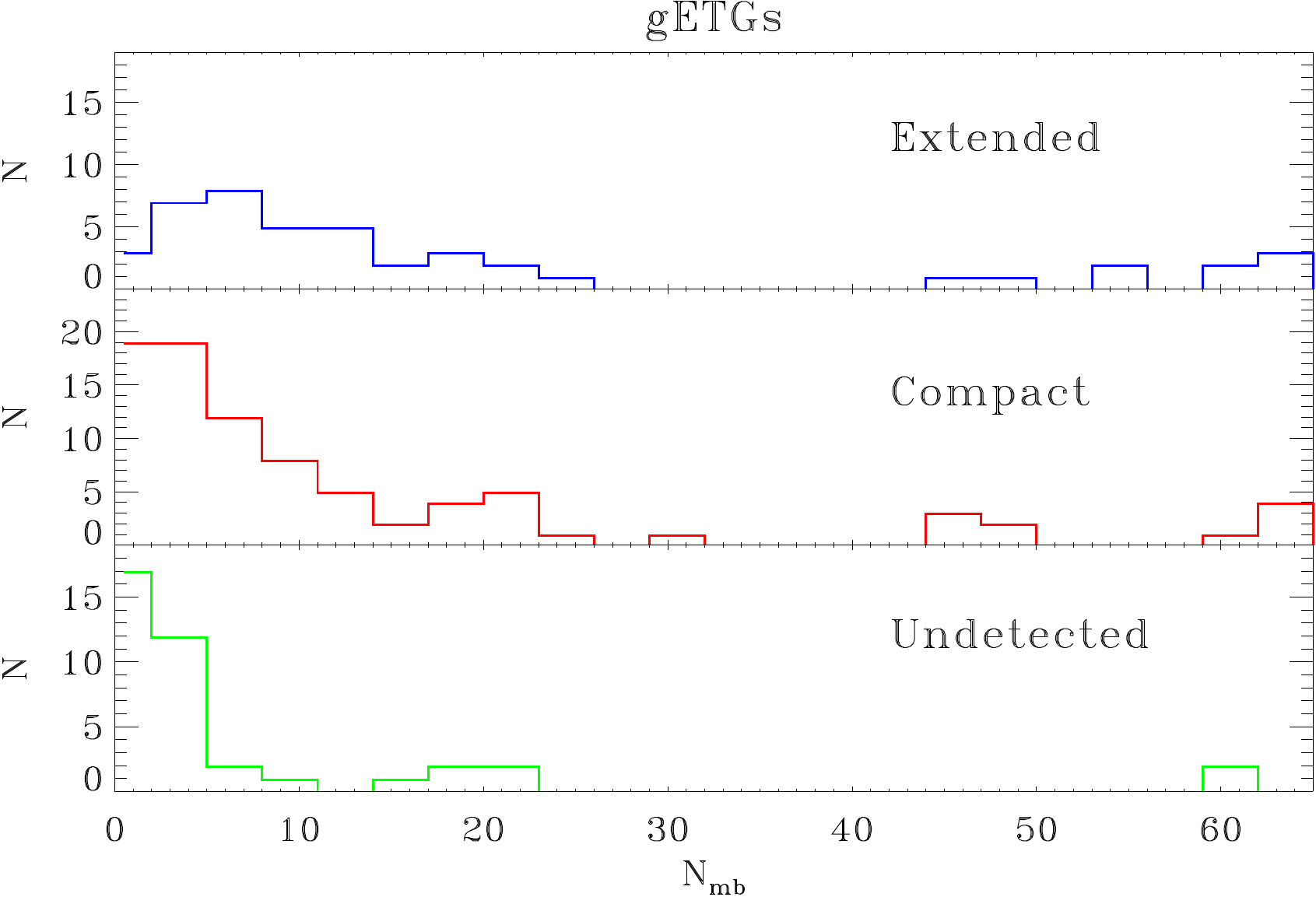}
\includegraphics[scale=0.52]{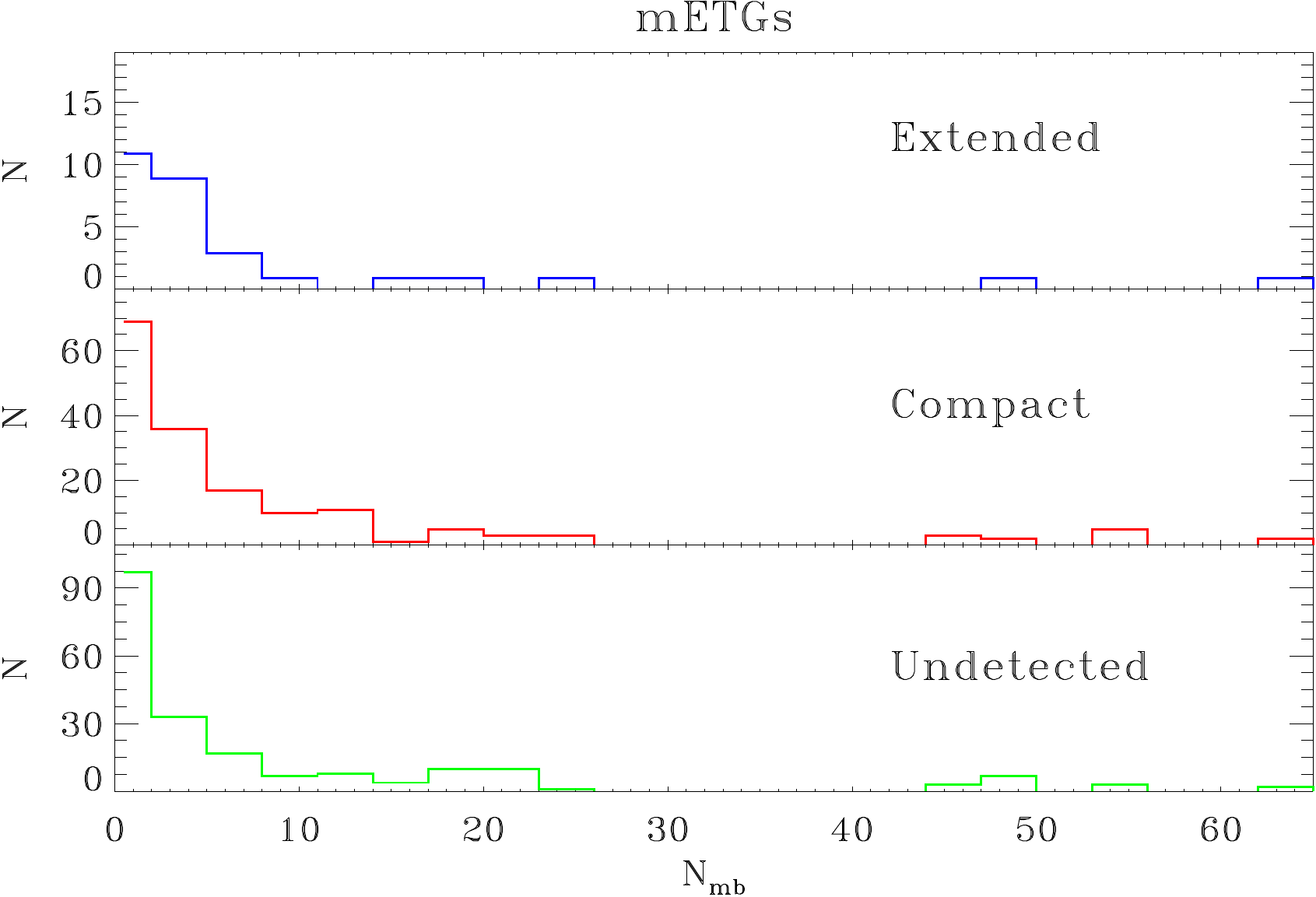}
\caption{Comparison of the number of group members, $N_{\rm mb}$, for
  the gETGs (left) and mETGs (right). In each subsample, we
  separately consider the sources based on their radio properties. The 34 mETGs
  with $N_{\rm mb}> 65$ (22 undetected, 10 compact, and 2 extended),
  all belonging to the Coma cluster, are not shown for more
  clarity. Similarly, there are 9 gETGs with $N_{\rm mb}> 65$ (3
  undetected, 5 compact, and one extended). }
\label{friends}
\end{figure*}

Figure \ref{flum} reports the fraction of ETGs hosting a radio source
with a power higher than a given radio power in various bins of
near-infrared absolute magnitude. These functions are fully determined
down to the luminosity at which we encounter the first upper
limit. Below this value, the presence of undetected sources produces
two branches that represent the lower and upper envelope of the true
distribution.  This analysis can be extended to lower luminosities
with the stacking technique, which enables us to estimate the median
power of the undetected sources. We find a power of log
L$_{150}$=20.57 at the 81th$^{\rm }$ percentile, log L$_{150}$=20.34
at the 71th$^{\rm }$ percentile, and log L$_{150}$=20.12 at the
71th$^{\rm }$ percentile for the three luminosity bins of the
mETGs. Overall, this analysis confirms that the more massive galaxies
have a higher probability to host a radio luminosity above a given
threshold. However, this trend is not seen at the lowest near-infrared
luminosities because the mETGs in the two lower-luminosity bins show
very similar distributions. This is clearly seen in Fig. \ref{lum2},
where we compare the distributions of the 150 MHz luminosity of the
least massive mETGs. In addition, the slopes of the distributions
generally increase with decreasing $L_K$. This effect is likely due to
the cutoff in the power law that describes these distributions, which
occur at a radio power that moves to lower $L_{150}$ for lower $L_K$.

\section{Large-scale environment and radio activity in mETGs and gETGs.}

We explored the connection between the large-scale environment and the
radio properties of both ETG samples. We used the galaxy group catalog
built from the 2MASS Redshift Survey by \citet{tully15} to identify
those associated with each ETG of our sample. We report the number of
group members from \citet{tully15}, $N_{\rm mb}$, for each
galaxy in Tables \ref{tab} and C.1. Figure \ref{friends} shows the
distributions of $N_{\rm mb}$ for the gETGs and the mETGs, considering extended, compact, and undetected sources separately.

\begin{table*}
  \caption{Properties of mETGs versus gETGs.}
  \begin{center}
\begin{tabular}{c | c | r r | r c c r c }
\hline
      &        &  $\overline{M}_K$ & log $\overline{M_\odot}$& Detected & \multicolumn{2}{c}{Compact} & Extended  &log $\overline{L}_{150}$ \\
      &        &                   & &          & Detected                    &  Total &  &                  \\
\hline                               
      &-26.0$<M_K<$-26.5  & -26.22 & 12.0 & 100\% & 55\% & 55\% &  45\% & 22.47 \\
gETGs &-25.5$<M_K<$-26.0  & -25.67 & 11.7 &  91\% & 63\% & 57\% &  34\% & 21.97 \\
      &-25.0$<M_K<$-25.5  & -25.17 & 11.5 &  70\% & 74\% & 52\% &  18\% & 21.45 \\
\hline                                       
      &-24.5$<M_K<$-25.0  & -24.70 & 11.3 &  64\% & 84\% & 54\% &  10\% & 21.02 \\
mETGs &-24.0$<M_K<$-24.5  & -24.26 & 11.1 &  42\% & 86\% & 36\% &   6\% & 20.69 \\
      &-23.5$<M_K<$-24.0  & -23.74 & 10.9 &  43\% & 86\% & 37\% &   6\% & 20.58 \\
\hline
\end{tabular}
\end{center}
\label{tab2}
\smallskip
\small{Columns 1 and 2: Range and median value of
  infrared absolute magnitude. Column 3: Fraction of sources detected at
  150 MHz. Columns 4 and 5: Fraction of compact sources in the detected sources and
  in the whole bin. Column 6: Fraction of extended sources. Column 7: Median
  radio luminosity.}
\end{table*}

The radio morphology and $N_{\rm mb}$ in gETGs are strongly
connected. The median $N_{\rm mb}$ value for the extended sources is
$\overline{N}_{\rm mb}$=13, for the compact sources, it is
$\overline{N}_{\rm mb}$=7, and for the undetected sources, it is
$\overline{N}_{\rm mb}$=3. The Kolmogorov-Smirnov test indicates that
the $N_{\rm mb}$ distributions for the three subsamples are all
significantly different. The Kolmogorov-Smirnov test returns a
probability of P=0.042 when the extended and the compact sources are
compared, P=5$\times10^{-5}$ for the comparison between extended and
undetected sources, and P=0.013 for the compact and the unresolved
sources.  For the mETGs, the median value for all classes is
$\overline{N}_{\rm mb}$=4, and their distributions are not
statistically different at a 5\% confidence level.

\citet{capetti20b} compared the large-scale environment of the compact
FR~0 radio galaxies with those with a FR~I morphology. They found that
FR~0s are located in a poorer environment, 63\% of them living in
groups formed by less than 15 galaxies, compared to only 17\% of the
FR~Is. Their analysis is based on the number of cosmological
neighbors, $N_{\rm cn}^{1000}$, i.e., galaxies lying within a region
of projected radius of 1 Mpc and having a spectroscopic redshift
differing by less than 0.005 from the radio galaxy, based on SDSS
data.

The gETGs sample is only partially covered by the SDSS survey and this
method cannot be applied. To put the present result into context we
estimated $N_{\rm cn}^{1000}$ and $N_{\rm mb}$ for the 78 gETGs with
available SDSS spectra. The number of companions estimated by the two
methods shows a strong correlation, with $N_{\rm cn}^{1000} \sim
2\times N_{\rm mb}$, the larger number of cosmological neighbors being
due to the fact that the SDSS data reach lower magnitudes with respect
to 2MASS.

After applying a factor two correction to the galaxy members of each
system, following the aforementioned analysis, we find that the
fractions of gETGs located in 2MASS groups with less than eight
members are 28\% and 53\% for the extended and compact sources,
respectively, in substantial agreement with the previous analysis.

The link between the radio morphology of the gETGs and large-scale
galaxy densities translates into a similar connection between radio
luminosity and environment, likely due to the fact that the extended
sources are generally those of higher radio power. The median
luminosities at 150 MHz are $4.2\times 10^{20}, 12.6\times 10^{20}$, and
$19.5\times 10^{20} \WHz$, for galaxies with $N_{\rm mb} \leq 10 $,
$10 < N_{\rm mb} \leq 30 $, and $N_{\rm mb} > 30 $, respectively.

Furthermore, the galaxy densities around compact sources is
higher than in undetected ones. For the compact gETGs we find a median
$\overline{L}_{150}=5.0\times 10^{21} \WHz$ while the stacking
analysis indicates that this value for the undetected sources is $5.1
\times 10^{20} \WHz$. This effect does not stem from the dependence on
the host luminosity, because these two classes have very similar
median values ($\overline{M}_K=-25.33$ and $\overline{M}_K=-25.19$ for
the compact and undetected sources, respectively). These results
indicate that the large scale environment plays an important role in
setting the morphology and the luminosity of the radio sources
associated with ETGs.

As already mentioned, these results apply only to the gETGs. No link
between environment and radio properties is seen for the mETGs.

\section{Optical spectroscopic properties of mETGs.}

\begin{figure*}
\includegraphics[scale=1.0]{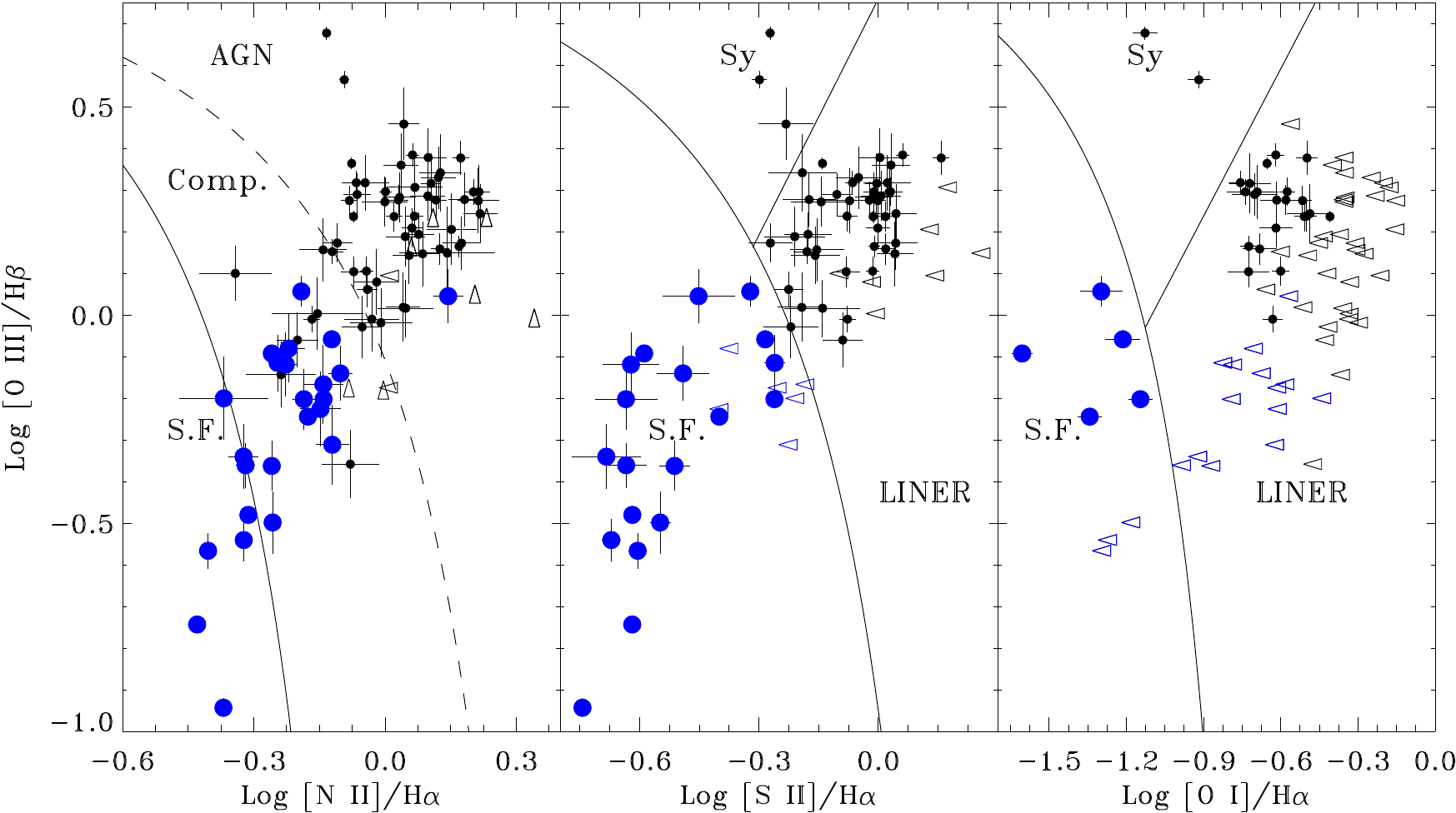}
\caption{Optical spectroscopic diagnostic diagrams for the mETGs with
  available SDSS spectra. We adopted a conservative threshold of S/N$>$5
  for the lines detection. The solid lines are from \citet{kewley06}
  and separate S.F. galaxies, LINER, and Seyfert; in
  the first panel, the region between the two curves is populated by
  the composite galaxies. Triangles are used to mark upper or lower
    limits. The blue symbols represent mETGs located within the region
    populated by S.F. galaxies in the [S~II]/\Ha\ vs.
    [O~III]/\Hb\ diagram.}
\label{diag}
\end{figure*}

Optical spectra can be used to identify the gas ionization mechanism
by measuring the ratios of selected emission lines
(e.g. \citealt{heckman80,baldwin81,veilleux87,kewley06}). We used the
fluxes of the diagnostic optical emission lines from the MPA-JHU added
value catalog of SDSS spectral measurements.\footnote{available at
  {\sl http://www.mpa-garching.mpg.de/SDSS/DR7/}} SDSS spectra are
available for 158 of the ETGs of our sample, but only in 73 of these
spectra it is possible to measure a sufficient number of emission
lines (at a significance $> 5\sigma$) to locate them in the diagnostic
diagrams, see Figure~\ref{diag}. Their spectral classification is
reported in Table \ref{tab}. Three ETGs in the
[O~III]/\Hb\ versus [S~II]/\Ha\ diagram fall into the region of Seyferts,
and 45 of them are among the LINERs. The remaining 25 (34 \%) are
compatible with a classification as star-forming (S.F.)

The 25 S.F. candidates span the whole range of near-infrared luminosity
with a median of $M_K = -24.0$, which is slightly lower than the median for the
whole sample. Most S.F. candidates (20 out of 25) are detected by LOFAR
with luminosities in the range $L_{150} = 1.7 \times 10^{21} - 6
\times 10^{22}$ W Hz$^{-1}$ (median $L_{150} = 5 \times 10^{21}$
Hz$^{-1}$). All but two (MCG+10-12-141 and UGC~09473) are compact
sources.

By adopting the relation from \citet{gurkan18}, we translated the radio luminosities
at 150 MHz of the S.F. candidates detected by LOFAR
into a star formation rate (SFR). We obtained a range between 0.2 and
4.3 M$_\odot$yr$^{-1}$ (median 0.4 M$_\odot$yr$^{-1}$).

The masses of the 25 S.F. candidates are in the range of $\sim6 \times10^{10} - 2
\times10^{11} {\rm M}_\odot$ . The values for the specific SFR
(sSFR) are $0.6 - 30 \times 10^{-12} {\rm yr}^{-1}$ . The median
value (including the sources not detected at 150 MHz) is $2.5 \times
10^{-12} {\rm yr}^{-1}$.

From the point of view of the radio spectra, only eight of the S.F.
candidates are detected in the surveys at 1.4 GHz, and the resulting
spectral slopes are in the range of $\alpha_{150}^{1400} = 0.31 - 0.90$ ,
with a median of $\alpha_{150}^{1400} = 0.70$ that is consistent with the
low-frequency radio slope of S.F. galaxies ($\alpha=0.59$,
\citealt{klein18}).

\section{Extending the study to lower-mass ETGs.}

The results of the previous sections show several trends linking the radio
properties to the host luminosity. In particular, we found an increase
in median radio luminosity with the host mass. While this trend in the
brightest galaxies is well described by $L_{150} \propto
L_K^{2.6}$, the $L_{150}$ versus $L_K$ flattens in the faintest
galaxies. However, this effect is not well defined. It is only visible in
the $-24.5 < M_K < -23.5$ luminosity range. The lmETGs sample can be
used to place this result on a stronger basis.

The luminosity distribution at 150 MHz of the 55 detected lmETGs
covers a wide range in power, from $\sim 3 \times 10^{19} \WHz$ to
$\sim 10^{23} \WHz$ (see Figure \ref{lum}). One hundred and three lmETGs are not detected
by the LOFAR observations, and the median limit to their radio
luminosities is 2.5$\times10^{20} \WHz$. The fraction of detected
lmETGs is smaller than that measured in the samples of more massive ETGs. It is 35\%, 48\%, and 78\% for the lmETGs, mETG, and gETGs,
respectively.
We list the main properties of the lmETG sample in Table \ref{tablm} .

In Figure \ref{mlums} we compare the near-infrared and radio
luminosity of the three samples. As we found for the more massive
ETGs, the radio power of lmETGs is spread widely, about four orders of
magnitude) at a given host luminosity. By including the lmETGs, we
confirm that the median radio luminosity decreases at lower stellar
luminosity. We estimated the median luminosity, $\overline{L}_{150}$,
of the lmETGs by splitting the sample into four bins of absolute
magnitude, each 0.5 mag wide, and using the stacking technique. The
stacking analysis only returns a robust measurement for the two bins
including the brightest galaxies. We obtained log $\overline{L}_{150}
= 20.28$ and 20.12 for $-22.5 > M_K > -23.0$ and $-23.0 > M_K >
-23.5$, respectively. Similarly to the lower-mass mETGs, the lmETGs
show an excess with respect to the trend measured at the highest
masses. The overall connection between $L_{150}$ and near-infrared
luminosity $L_K$ is well described by log $L_{150} \sim $log$
L_K^{2.6}$ for $M_K \lesssim -24.5,$ while it flattens to a slope of
1.03 for the four bins including the less luminous ETGs. From a linear
fit, we obtain log $L_{150} =
(-0.41\pm0.04)\times(M_K+23.5)+(20.41\pm0.04)$.

\begin{figure}
\includegraphics[scale=0.5]{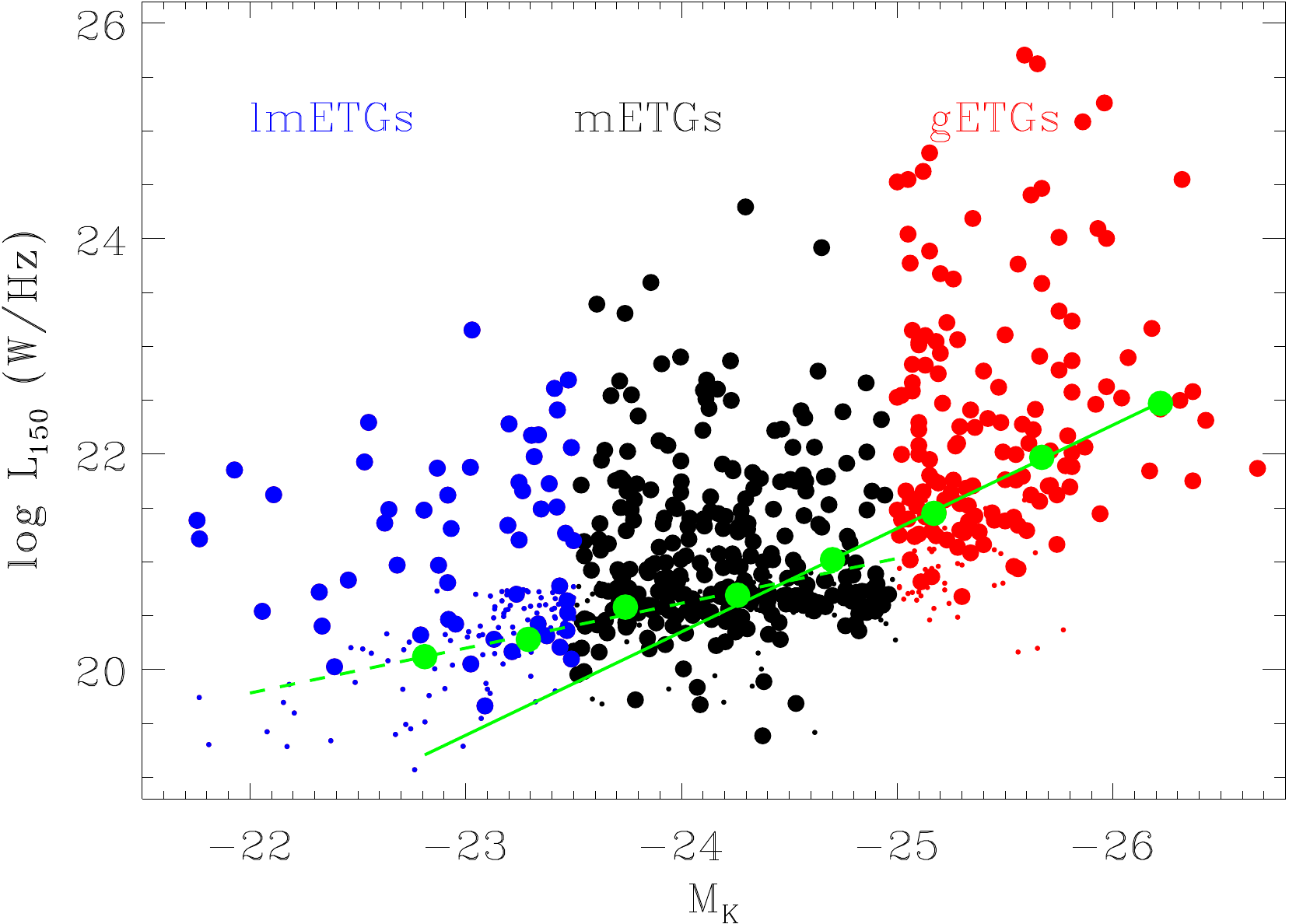}
\caption{Absolute K-band magnitude vs. luminosity at 150
  MHz. The red symbols are the gETGs ($M_K <-25.0$), the black symbols
  are the mETGs ($-25.0 < M_K <-23.5$), and the blue symbols are the
  lmETGs ($-23.5 < M_K <-21.5$). The small dots mark the upper limits
  of the undetected objects. The large green dots represent the median radio
  luminosity in eight bins of absolute magnitude, and the solid (dashed)
  green line represents the best linear fit to the four brightest
  (faintest) galaxies bins.}
\label{mlums}
\end{figure}

Another effect shown by the ETGs is the dependence of the morphology
of the radio sources on the host mass. In this respect, the
morphologies of the radio sources associated with the lmETGs are
significantly different from those seen in the brighter ETGs. In the
gETGs, we found 42 (out of 188) extended sources, at least 80\% of
which having a class I \citet{fanaroff74} morphology.
The fraction of sources with a jetted morphology in mETGs is strongly reduced.
Only nine out of 432 galaxies ($\sim 2\%$) are jetted, while the morphology of none of the lmETGs
is indicative of collimated jets. Only
4 (out of 158) lmETGs satisfy the definition of an extended radio
source. In all four cases, the radio emission is diffuse and cospatial
with the host galaxy.

While most radio sources in the gETGs and mETGs are compact or barely
resolved, the most common morphology in lmETGs is a
diffuse emission. To draw a more quantitative comparison, we show in
Fig. \ref{fwhm}  the distribution of the radio source
extension (obtained by fitting a two-dimensional Gaussian profile as
reported in the DR2 catalog) versus the flux density at 150 MHz for
the lmETGs and the mETGs split into two bins of absolute
magnitude. Overall, there is a clear trend of increasing radio source
size for lower-mass ETGs. More quantitatively, 51\% of the lmETGs have
measured angular sizes between $\sim 10 \arcsec$ and $\sim 25
\arcsec$, corresponding to deconvolved linear sizes of $\sim 4 - 12$
kpc. Conversely, for the two luminosity bins of mETGs, the fractions
are 36\% and 14\%.

\begin{figure}
\includegraphics[scale=0.55]{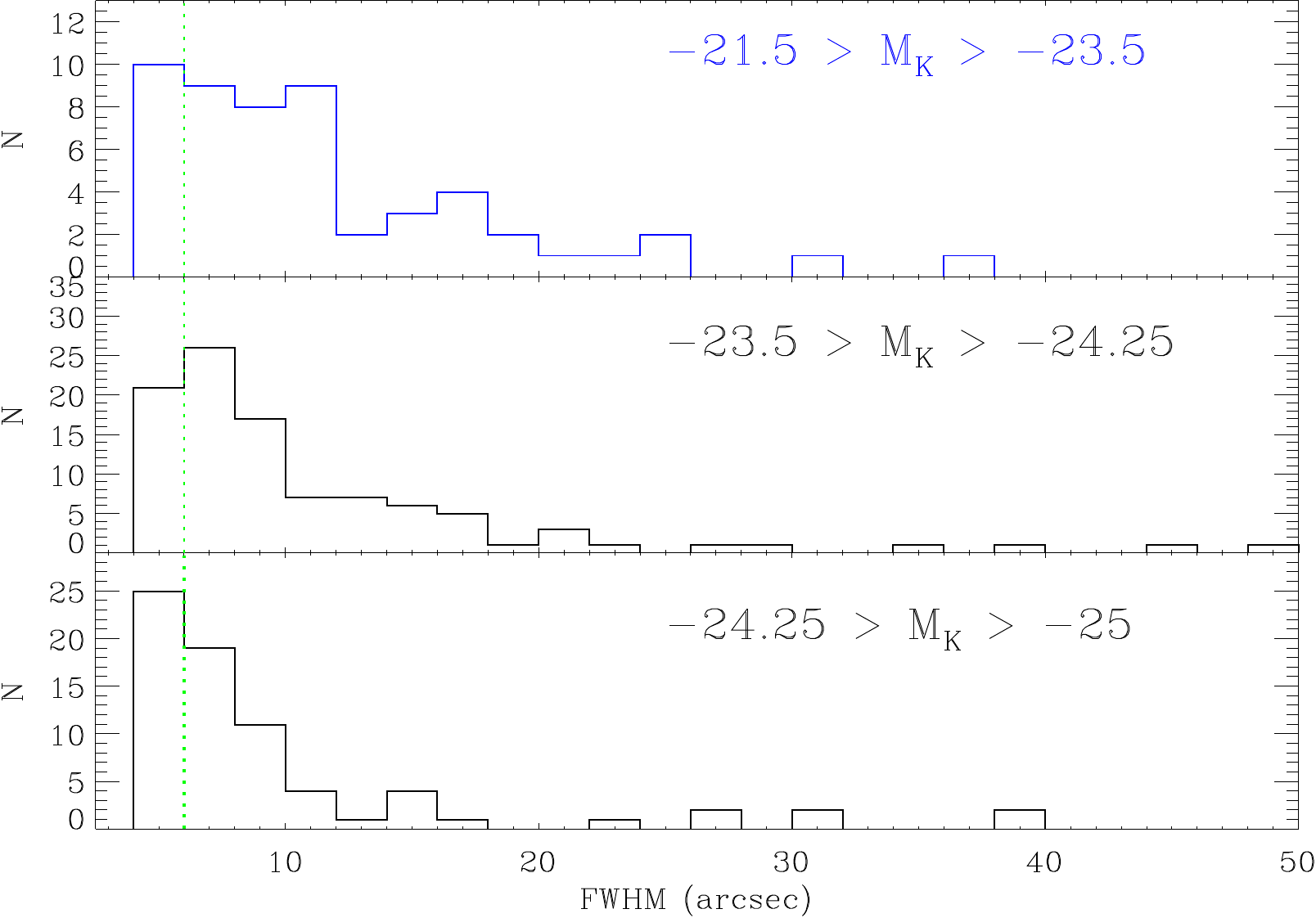}
\caption{Distribution of the radio sources extension vs. flux
  density at 150 MHz for the lmETGs (top) and the mETGs, split into
  two bins of near-infrared luminosity (middle and bottom). The dashed
  green lines mark the 6\arcsec\ resolution of the LoTSS images.}
\label{fwhm}
\end{figure}

\section{Discussion.}

The study of a volume-limited sample of ETGs covering a factor $\sim
100$ in host luminosity indicates that several physical parameters
show a positive correlation with the stellar luminosity.

\begin{itemize}
\item 1) The fraction of galaxies detected by LOFAR increases with
increasing $L_K$: While the more luminous gETGs (with
$-26.0<M_K<-26.5$) are all detected at 150 MHz, this fraction
decreases to $\sim$40\% for the least luminous mETGs and to 35\%
  for lmETGs.

  \item 2) The fraction of extended radio sources decreases from 45\% in the
more massive gETGs to 6\% in the less massive mETGs. While the extended sources in gETGs have a median size of 57 kpc, this value is
12 kpc in the mETGs. At least 80\% of the extended gETGs have a FR~I
class morphology, which clearly indicates jets. Only 9
mETGs ($\sim$ 2\%) are likely jetted sources, six of which lie in the
brightest host luminosity bin. The radio emission is
cospatial (or aligned) with the optical emission in 11 mETGs, suggesting a
prevalence of processes at the origin of the radio emission related to
the host galaxies. None of the lmETGs has a morphology indicative of the
presence of collimated jets. Only 4 (out of 158) lmETGs satisfy the
definition of an extended radio source, and in all cases, the radio
emission is diffuse and cospatial with the host galaxy.

\item 3) The median radio luminosity shows an increasing trend with
  the host luminosity. When only the brightest galaxies are
  considered, this trend is well described by $L_{150} \propto
  L_K^{2.6}$. The faintest galaxies instead show a much shallower,
  almost linear, dependence on $L_K$. This is interpreted as due to
  the transition from an AGN- to host-dominated radio emission.

\item 4) The fraction of candidate RRSs among the gETGs is $\sim 7\%$. This is $\sim
28\%$ of those with extended emission. Conversely, we
find only one candidate restarted source among the mETGs and two sources with a
spectral index higher than 1.2, which are therefore candidate remnants. Three
additional extended sources that are not detected at 1.4 GHz might be remnant
sources. This interpretation is supported by their diffuse
morphology. However, the limits to their spectral slopes are all well
below the 1.2 threshold. Overall, we set an upper limit of $\lesssim
1\%$ to the fraction of RRSs in the mETG sample, which is  $\lesssim$16\%
of those with extended emission. This is lower than in gETGs. However, when we
only consider the mETGs with a jetted morphology, this fraction
increases significantly.

\item 5) The source radio morphology and the
large-scale environment of gETGs are connected. The extended (and more powerful
radio sources) are found in regions of higher galaxy density than the compact and undetected sources. This effect is absent for
the mETGs.
\end{itemize}

To summarize, the median radio power, the fraction of the detected
sources, and the fraction (and size) of the jetted sources all depend
on the host luminosity. The fraction of RRS is also related to the
host mass because it decreases from $\sim 7\%$ in the gETGs to $<1\%$
in the mETGs. These same parameters are also related to the
large-scale environment: We find a connection between the large-scale
galaxy density and the presence of extended radio emission and its
luminosity. This effect is seen only in the gETGs and in the most
massive mETGs. However, these results only describe statistical
trends. Even when we consider sources of similar stellar mass and
environment, for example, a wide spread in radio luminosity is
found. This implies that additional physical quantities such as the
accretion rate and the source age are likely to be involved in setting
the level of radio activity. In addition, as we already argued in
Paper I, compact and extended radio sources differ in their jet
properties. The compact sources might be produced by slower jets
because of a lower black holes spin.

These trends break at the lowest near-infrared luminosity.  While at
high mass, the $L_{150}$ versus $L_K$ relation is well described
by a power law with an index of $\sim 2.6$, at lower masses, the index
decreases to $\sim 1$. In addition, while the more massive ETGs are
mostly associated with point-like or jetted radio sources, in the
least luminous galaxies, these are often diffuse and extended on a
scale of $\sim 4 - 12$ kpc. These results suggest that in the less
massive galaxies probed by our study, the dominant process at the
origin of the radio emission is not associated with a radio-loud AGN,
but with the host galaxy. This interpretation is supported by the
significant fraction of mETGs and lmETGs in which the radio emission
is aligned (or cospatial) with the host galaxy. In these sources, the
radio emission is likely to be dominated by processes related to the
host galaxies such as young stars and supernovae. The fraction
of mETGs in which the optical spectra indicate that the nuclear gas is
ionized by young stars is $\sim 16\%$, similar to what we found for
the gETGs. The star formation rate, which is derived by ascribing the radio
emission to young stars, is similar for two luminosity classes of
ETGs, but the specific SFR is about three times higher in mETGs than in gETGs because their mass is lower. However, the SDSS spectra
cover a nuclear region of less than 1 kpc in radius, and they are not
representative of the whole galaxy. The only two sources where
the extended emission is cospatial (or aligned) with the host and with
an available SDSS spectra are one Seyfert and one LINER. This suggests
that the radio emission in low-mass ETGs might also be powered by a
radio-quiet AGN.

The transition between radio sources that are dominated by a radio-loud AGN to
host galaxy processes (or radio-quiet AGN) occurs at a mass that is similar to the mass at
which ETGs change their stellar brightness profile from a pure
S\'ersic law to a core-S\'ersic law. As already mentioned in the
Introduction, the different brightness profiles correspond to the
dichotomy between radio-loud and radio-quiet AGN: Radio-loud AGN are
only hosted by core galaxies, while radio-quiet AGN are found only in
S\'ersic galaxies. 

\section{Summary and conclusions}

We explored the radio properties of ETGs in the local Universe
(recession velocity $<7,500$ \kms) as seen by LOFAR observations at
150 MHz. In Paper I we considered 188 giant ETGs with $M_K < -25$,
while here we analyzed the mETGs, which have a near-infrared luminosity in
the $-23.5 > M_K > -25$ range, that is, with a mass $6 \times10^{10}
\lesssim M_* \lesssim 3 \times10^{11} {\rm M}_\odot$. We also considered
even less luminous galaxies, the lmETGs, down to $M_K = -21.5$. The
combined volume-limited sample covers an interval of a factor $\sim
100$ in stellar mass.

The fraction of sources detected above a typical luminosity of $\sim 3
\times 10^{20} \WHz$ decreases from 78\% of the gETGs to $\sim 48\%$
of the mETGs and $\sim 35\%$ of the lmETGs. Of the 31 mETGs with
a radio emission that extends over at least 15\arcsec, only a minority
(9, representing $\sim$ 2\% of the whole sample) have a morphology
suggesting the presence of collimated jets. This fraction is $\sim$
20\% for the gETGs. No jetted source is found among the
lmETGs. Although we observed mETGs with extended radio emission (with
a median size of $\sim 12$ kpc), the majority of them ($\sim 85\%$ of
the detected sources) is unresolved, with a characteristic limit to
their size of $\sim$ 4 kpc. This strengthen previous results that
compact FR~0 radio sources represent the dominant population of radio
galaxies in the nearby Universe.

Our study confirms the strong connection between the radio properties
and the host mass: 1) the positive correlation between the stellar luminosity
and the median radio power, 2) the fraction of compact sources that
increases with decreasing host mass, 3) the high incidence of jetted
sources among the most massive ETGs (about 20\%), and their scarcity
($\sim 2\%$) among the least massive ones.

However, the trends above break for the least luminous ETGs. The
extension of the study to even less massive ETGs, the lmETGs, which reach
masses as low as $\sim 0.8\times10^{10} {\rm M}_\odot$, shows that
while at high mass, the log $L_{150} \sim $log$ L_K$ relation is well
described by a power law with an index of $\sim 2.6$, at lower masses,
the index decreases to $\sim 1$. Combined with the radio
morphology of the less massive ETGs, which is often aligned with the optical
axis, this result suggests that we are observing the transition from the
population of the most massive ETGs, $M_* \gtrsim 2 \times10^{11} {\rm
  M}_\odot$, in which the dominant process of radio emission can be
associated with a radio loud AGN, to host-dominated
sources. At approximately this mass, the stellar surface
brightness profile also transitions from S\'ersic galaxies to those
with a depleted stellar core. Previous studies have shown that only
core galaxies host radio-loud AGN. This is in line with the the present
analysis.

The selection of a complete sample based only on stellar luminosity is
optimally suited to exploring the duty cycle of the radio activity in
ETGs. We are now in the position to further constrain the energy
output of radio-loud AGN. The mETGs showing extended radio
emission include a candidate restarted source, with a bright central
region associated with two low-brightness tails, and two galaxies
are characterized by a steep radio spectral index ($\alpha_{150}^{1400}
\sim 1.35$) that are likely remnant sources. Three additional extended
sources have a diffuse morphology: They are not detected in the
surveys at 1.4 GHz and should be considered candidate remnant
sources. Overall, by considering the mETGs and gETGs samples, based on a combination of morphology an spectral index, we isolated 19
restarted and remnant (or candidate remnant) sources. With
multiband radio observations, we will be able to estimate the
timescale of the various activity phases.

\begin{acknowledgements}
  MB acknowledges financial support from the agreement ASI-INAF
  n. 2017-14-H.O, from the PRIN MIUR 2017PH3WAT “Blackout” and from
  the ERC-Stg ``MAGCOW", no. 714196.

  LOFAR, the Low Frequency Array designed and constructed by ASTRON, has
facilities in several countries, which are owned by various parties
(each with their own funding sources), and are collectively operated
by the International LOFAR Telescope (ILT) foundation under a joint
scientific policy. The ILT resources have benefited from the following
recent major funding sources: CNRS-INSU, Observatoire de Paris and
Universit\'e d'Orl\'eans, France; BMBF, MIWF-NRW, MPG, Germany;
Science Foundation Ireland (SFI), Department of Business, Enterprise
and Innovation (DBEI), Ireland; NWO, The Netherlands; the Science and
Technology Facilities Council, UK; Ministry of Science and Higher
Education, Poland; The Istituto Nazionale di Astrofisica (INAF),
Italy.

Part of this work was carried out on the Dutch national
e-infrastructure with the support of the SURF Cooperative through
grant e-infra 160022 \& 160152. The LOFAR software and dedicated
reduction packages on {\sl https://github.com/apmechev/GRID\_LRT} were
deployed on the e-infrastructure by the LOFAR e-infragrop, consisting
of J.\ B.\ R.\  (ASTRON \& Leiden Observatory), A.\ P.\ Mechev
(Leiden Observatory) and T. Shimwell (ASTRON) with support from
N.\ Danezi (SURFsara) and C.\ Schrijvers (SURFsara). The J\"ulich
LOFAR Long Term Archive and the German LOFAR network are both
coordinated and operated by the J\"ulich Supercomputing Centre (JSC),
and computing resources on the supercomputer JUWELS at JSC were
provided by the Gauss Centre for supercomputing e.V. (grant CHTB00)
through the John von Neumann Institute for Computing (NIC).

This research made use of the University of Hertfordshire
high-performance computing facility and the LOFAR-UK computing
facility located at the University of Hertfordshire and supported by
STFC (ST/P000096/1), and of the Italian LOFAR IT computing
infrastructure supported and operated by INAF, and by the Physics
Department of Turin University (under an agreement with Consorzio
Interuniversitario per la Fisica Spaziale) at the C3S Supercomputing
Centre, Italy.

The Pan-STARRS1 Surveys (PS1) and the PS1 public science archive have
been made possible through contributions by the Institute for
Astronomy, the University of Hawaii, the Pan-STARRS Project Office,
the Max-Planck Society and its participating institutes, the Max
Planck Institute for Astronomy, Heidelberg and the Max Planck
Institute for Extraterrestrial Physics, Garching, The Johns Hopkins
University, Durham University, the University of Edinburgh, the
Queen's University Belfast, the Harvard-Smithsonian Center for
Astrophysics, the Las Cumbres Observatory Global Telescope Network
Incorporated, the National Central University of Taiwan, the Space
Telescope Science Institute, the National Aeronautics and Space
Administration under Grant No. NNX08AR22G issued through the Planetary
Science Division of the NASA Science Mission Directorate, the National
Science Foundation Grant No. AST-1238877, the University of Maryland,
Eotvos Lorand University (ELTE), the Los Alamos National Laboratory,
and the Gordon and Betty Moore Foundation.
\end{acknowledgements}

\bibliographystyle{./aa}

\clearpage
\onecolumn

\begin{appendix}
\section{Properties of the mETG sample.}

\begin{landscape}
\begin{longtable}{l | r r c c | c r r l | r | r | r}

\caption{Radio properties of the sample}
\label{tab} \\                                   

\hline 
Name & R.A. & DEC. & v & M$_K$ & rms  & F$_{150}$ & L$_{150}$ & Morph. & F$_{1.4}$ & SDSS & $N_{\rm mb}$ \\ 
\hline  
\endfirsthead

\multicolumn{3}{c}{{\tablename} \thetable{} -- Continued} \\                                   [0.5ex]
\hline 
Name & R.A. & DEC. & v & M$_K$ & rms  & F$_{150}$ & L$_{150}$ & Morph. & F$_{1.4}$ & SDSS &  $N_{\rm mb}$\\ 
\hline
\endhead

\hline
  \multicolumn{10}{c}{{Continued on next page}} \\                                   
\endfoot

  \\                                   [-1.8ex] 
\endlastfoot
                  NGC7805 &     0.3616 &  31.4337 & 4766 & -24.10 & 0.10 &   $<$ 0.5 & $<$ 20.46 &   &         --- &       &     4\\ 
                  NGC0005 &     1.9537 &  35.3623 & 5062 & -23.74 & 0.17 &   $<$ 0.8 & $<$ 20.73 &   &         --- &       &     1\\ 
                  NGC7836 &     2.0068 &  33.0707 & 4854 & -24.00 & 0.09 &     43.0  &     22.41 & E &     9.2     &       &     6\\ 
                  NGC0016 &     2.2679 &  27.7294 & 3097 & -24.52 & 0.13 &   $<$ 0.7 & $<$ 20.21 &   &         --- &       &     1\\ 
                  NGC0006 &     2.3862 &  33.3087 & 4919 & -24.88 & 0.08 &      4.7  &     21.46 & C &         --- &       &     6\\ 
                  NGC0042 &     3.2348 &  22.1003 & 5876 & -24.58 & 0.10 &      1.9  &     21.23 & C &         --- &       &     2\\ 
                  NGC0043 &     3.2531 &  30.9153 & 4728 & -24.78 & 0.10 &   $<$ 0.5 & $<$ 20.44 &   &         --- &       &     3\\ 
                 UGC00130 &     3.4872 &  30.8830 & 4724 & -23.62 & 0.10 &   $<$ 0.5 & $<$ 20.46 &   &         --- &       &     3\\
\hline
\end{longtable}

\smallskip
\small{Column description: (1) name, (2 and 3) right ascension and declination,
(4) recession velocity (\kms), (5) K-band absolute magnitude, (6)
r.m.s. noise of the LOFAR image (mJy beam$^{-1}$), (7 and 8) flux
density (mJy) and luminosity (W Hz$^{-1}$) at 150 MHz from LOFAR, (9)
radio morphology (C=compact, E=extended), (10) 1.4 GHz flux density
(mJy) from FIRST or from NVSS when the FIRST measurement is not
available, or not reliable, the long dash represents a source undetected in
the NVSS at a typical limit of 2 mJy), (11) SDSS spectrum available
and optical spectral classification: Sy = Seyfert, LI = LINER, SF =
star forming, --- = unclassified}, (12) number of 2MASS group
members.
\end{landscape}

\clearpage
\section{Images and notes on the radio morphology of the extended sources.}

\begin{figure*}
\includegraphics[scale=0.27]{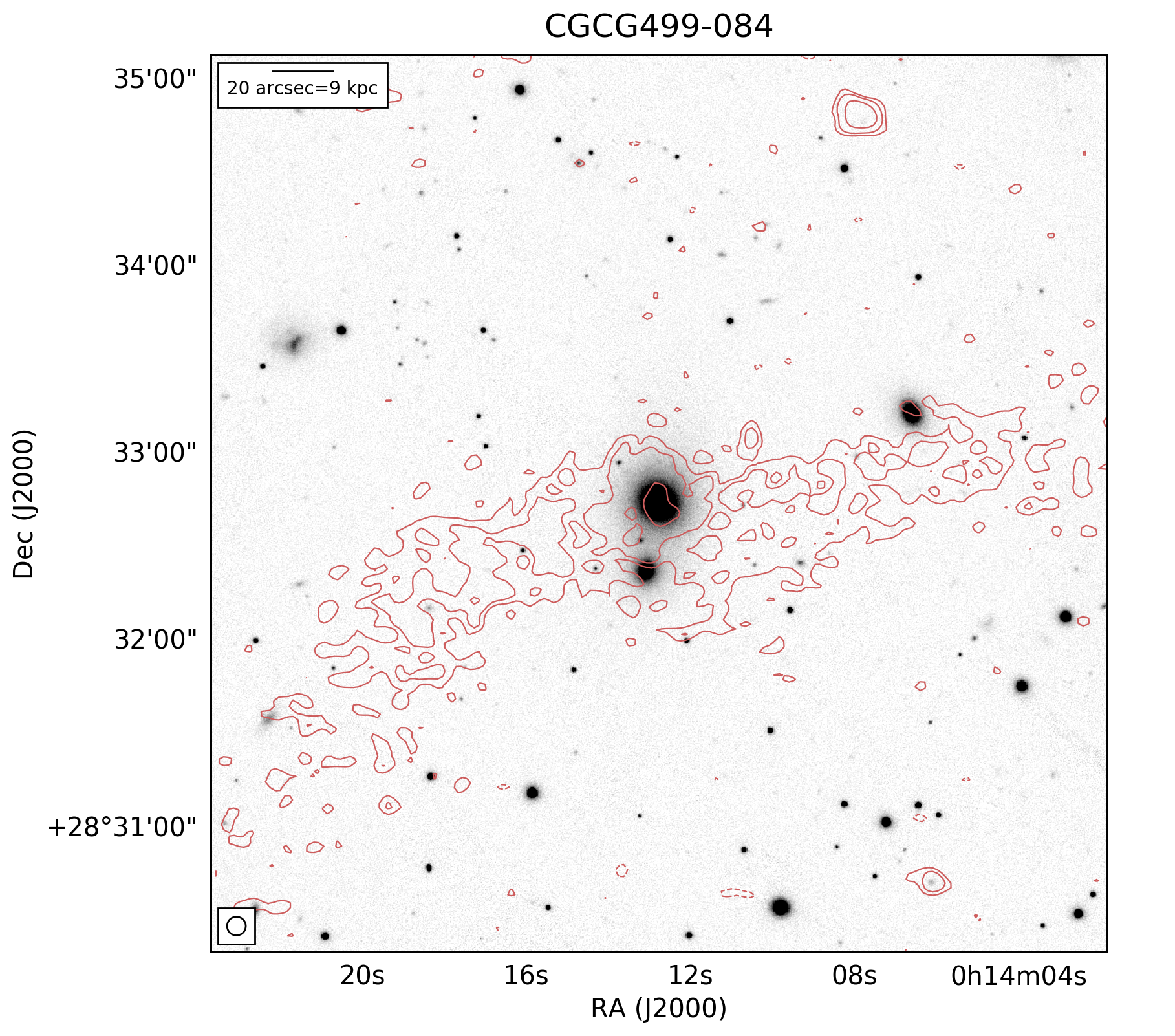}
\includegraphics[scale=0.27]{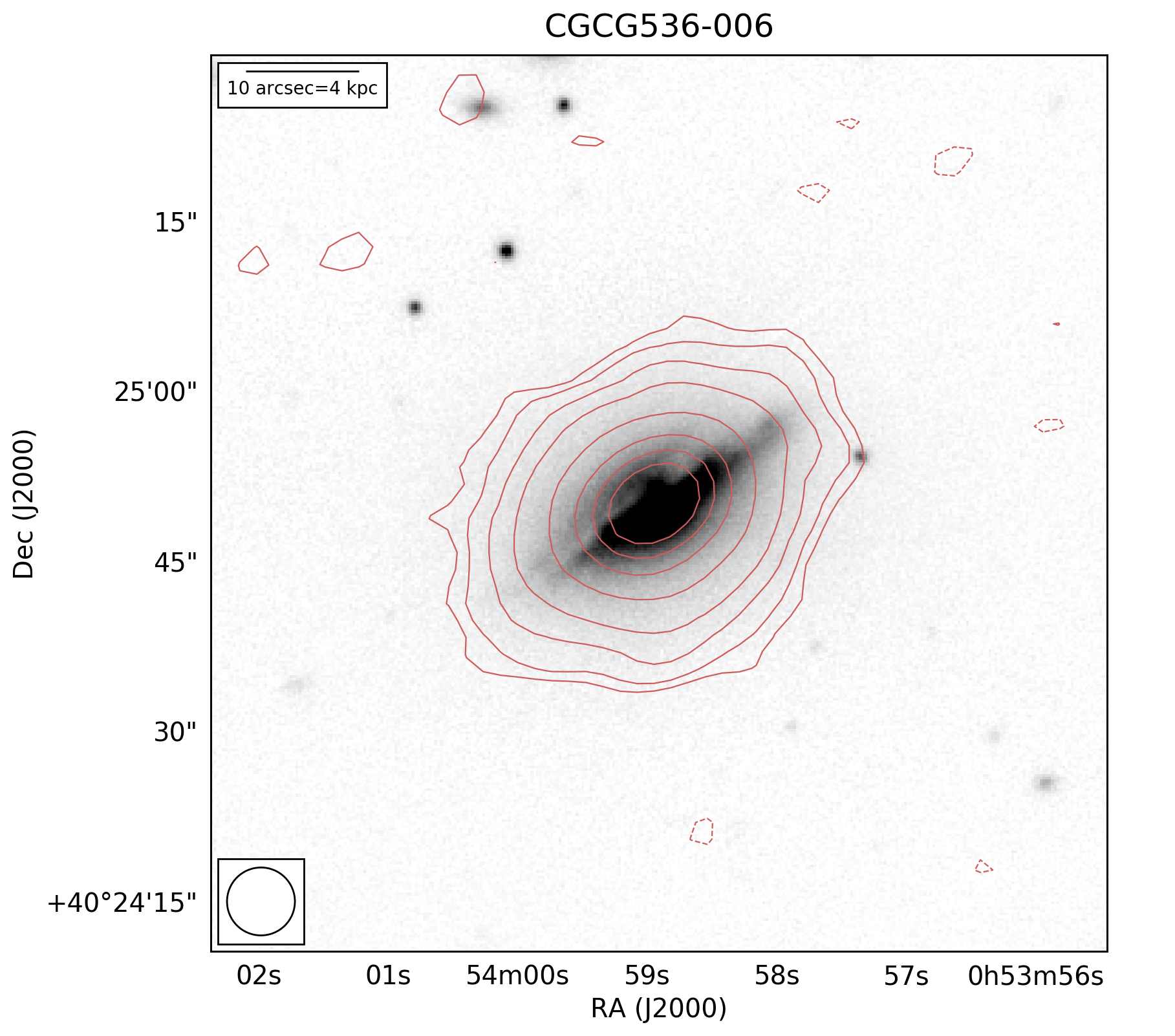}
\includegraphics[scale=0.27]{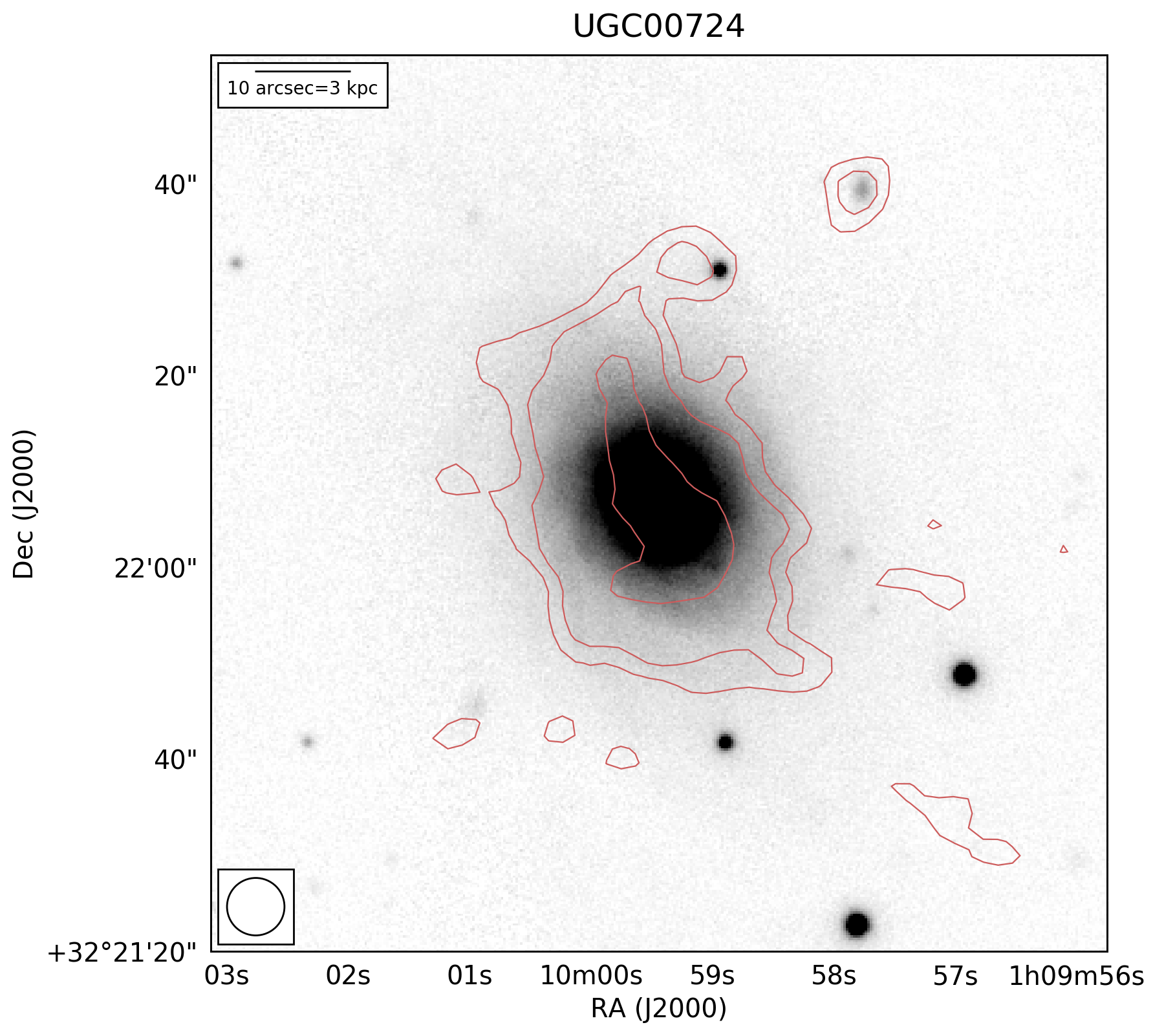}
\includegraphics[scale=0.27]{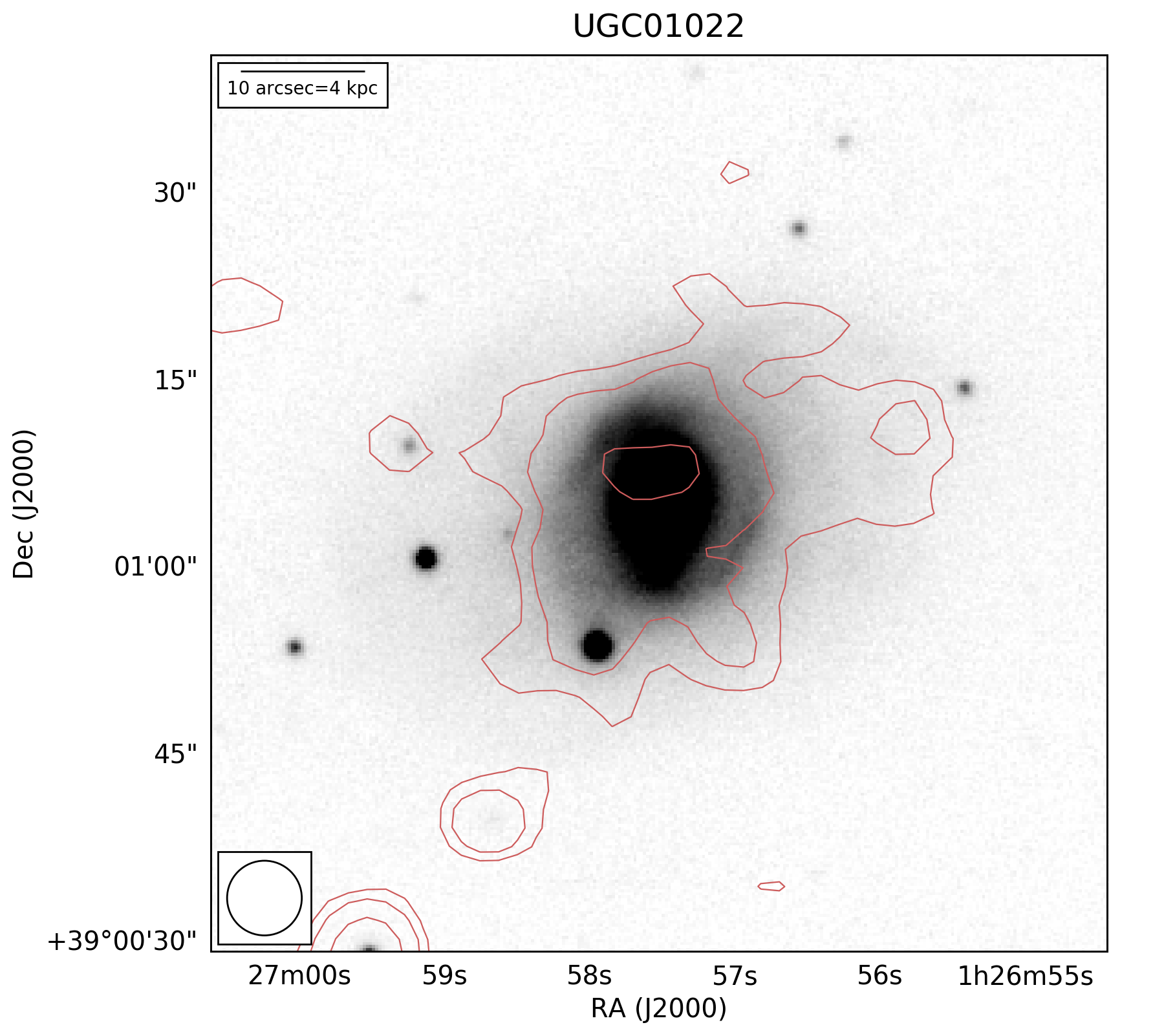}
\includegraphics[scale=0.27]{NGC0670-red.png}
\includegraphics[scale=0.27]{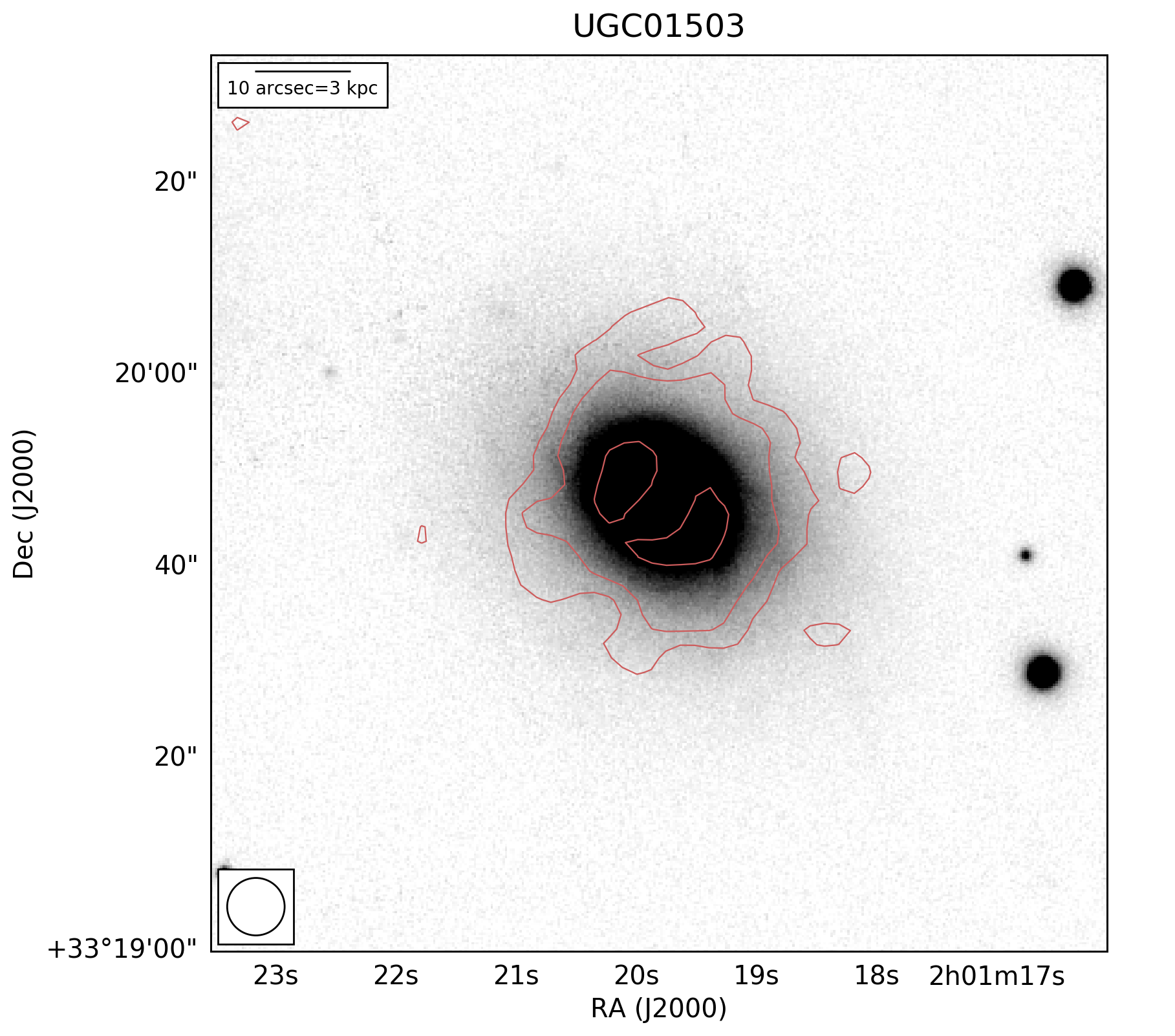}
\includegraphics[scale=0.27]{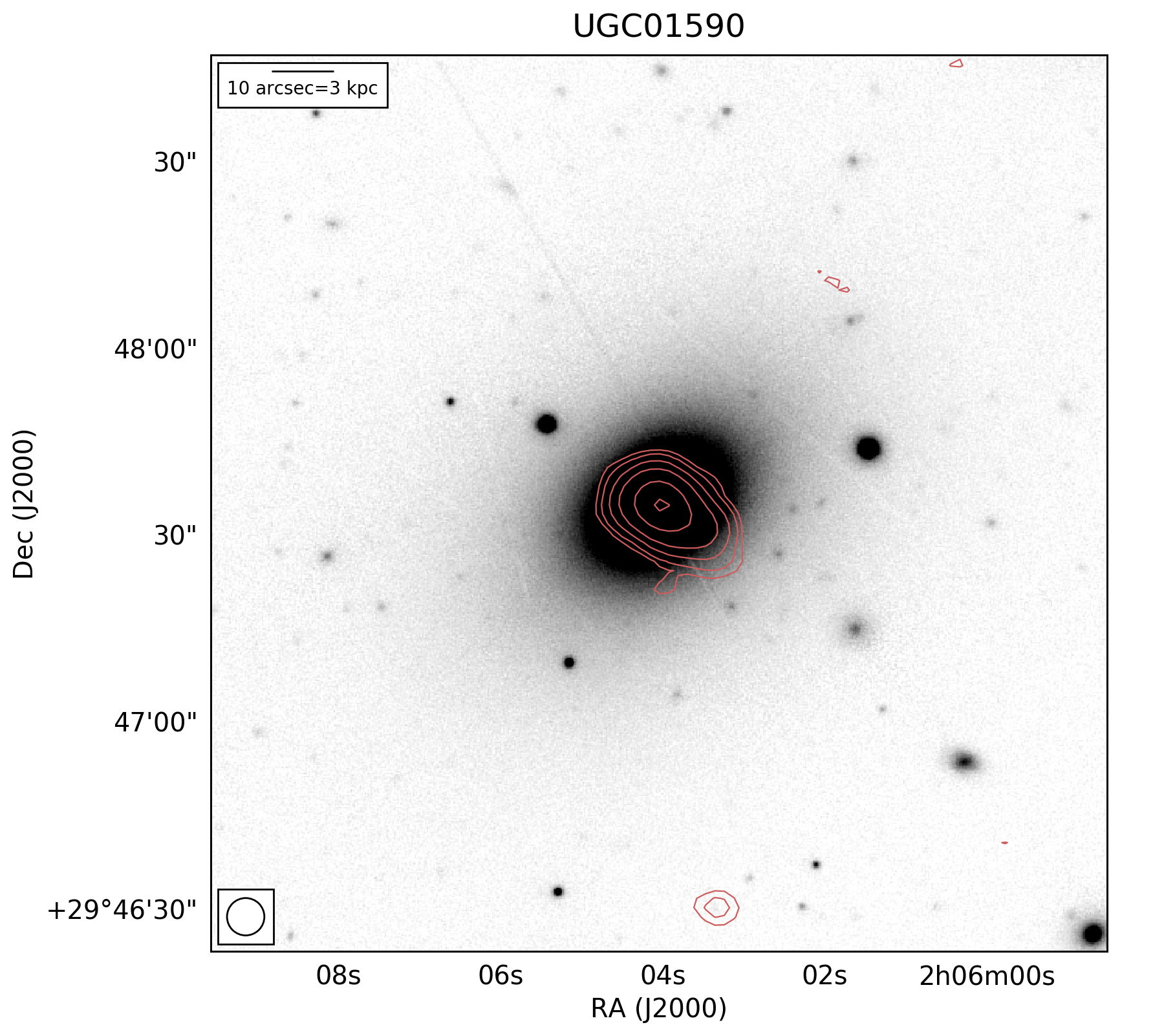}
\includegraphics[scale=0.27]{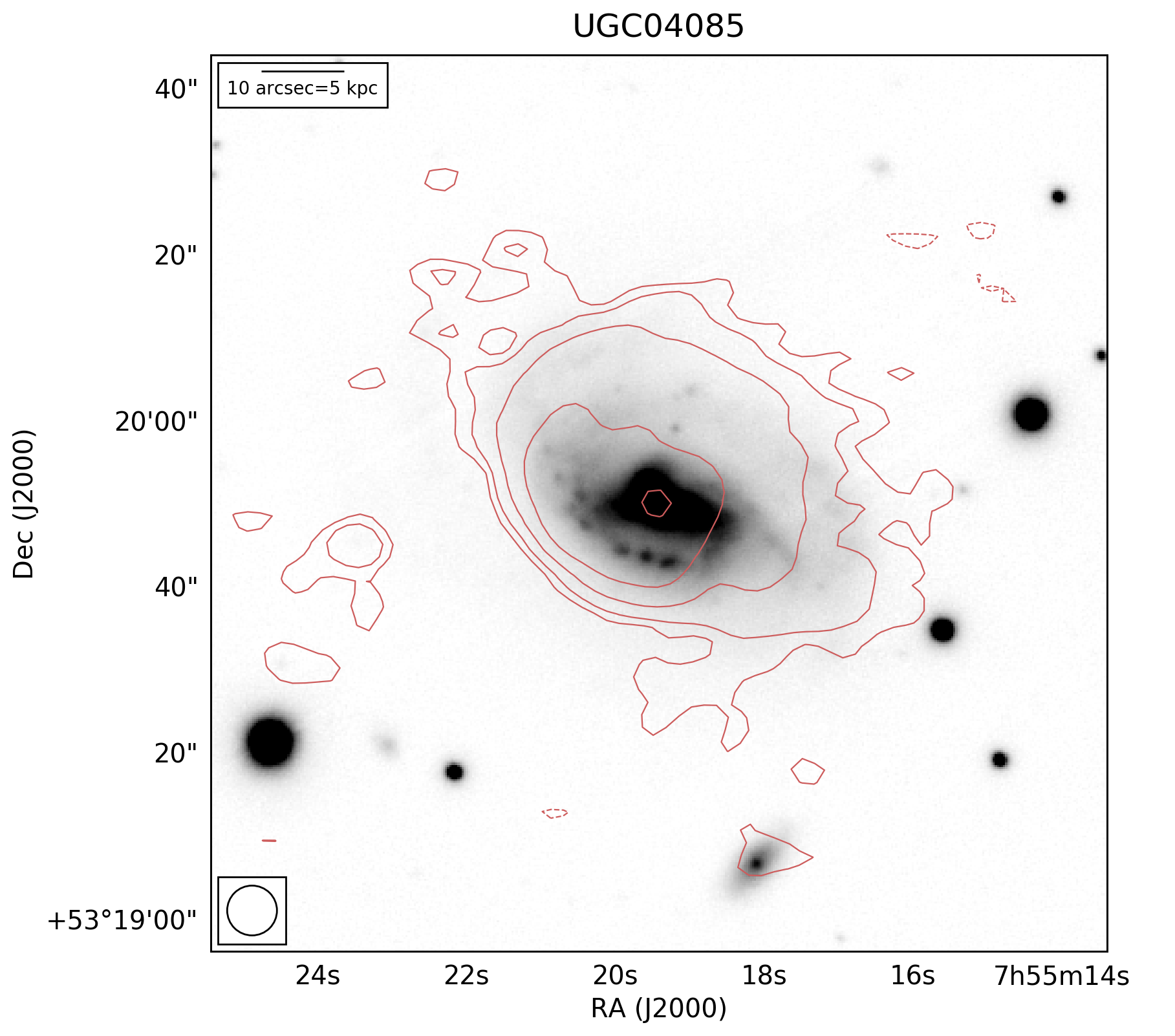}
\includegraphics[scale=0.27]{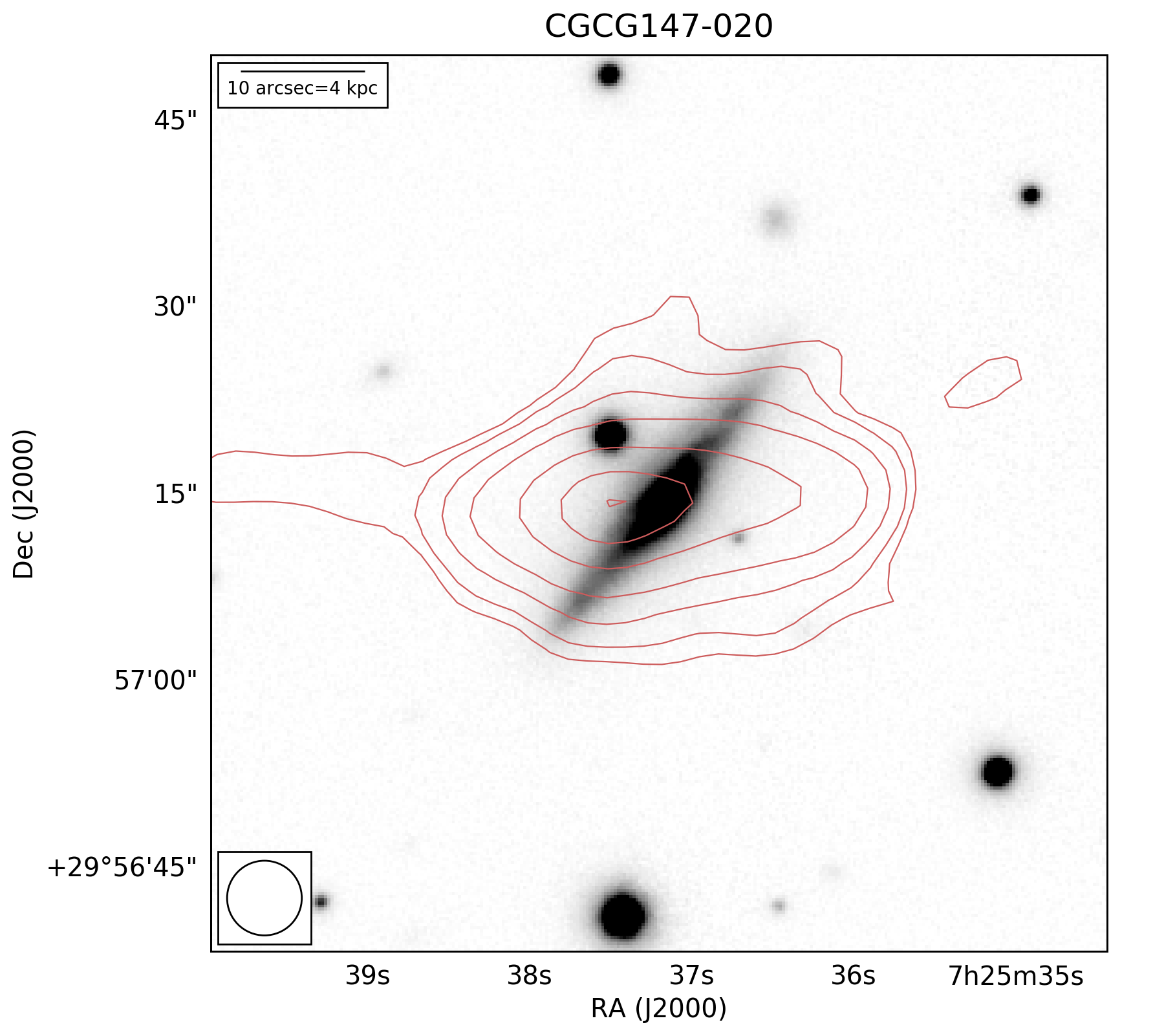}
\includegraphics[scale=0.27]{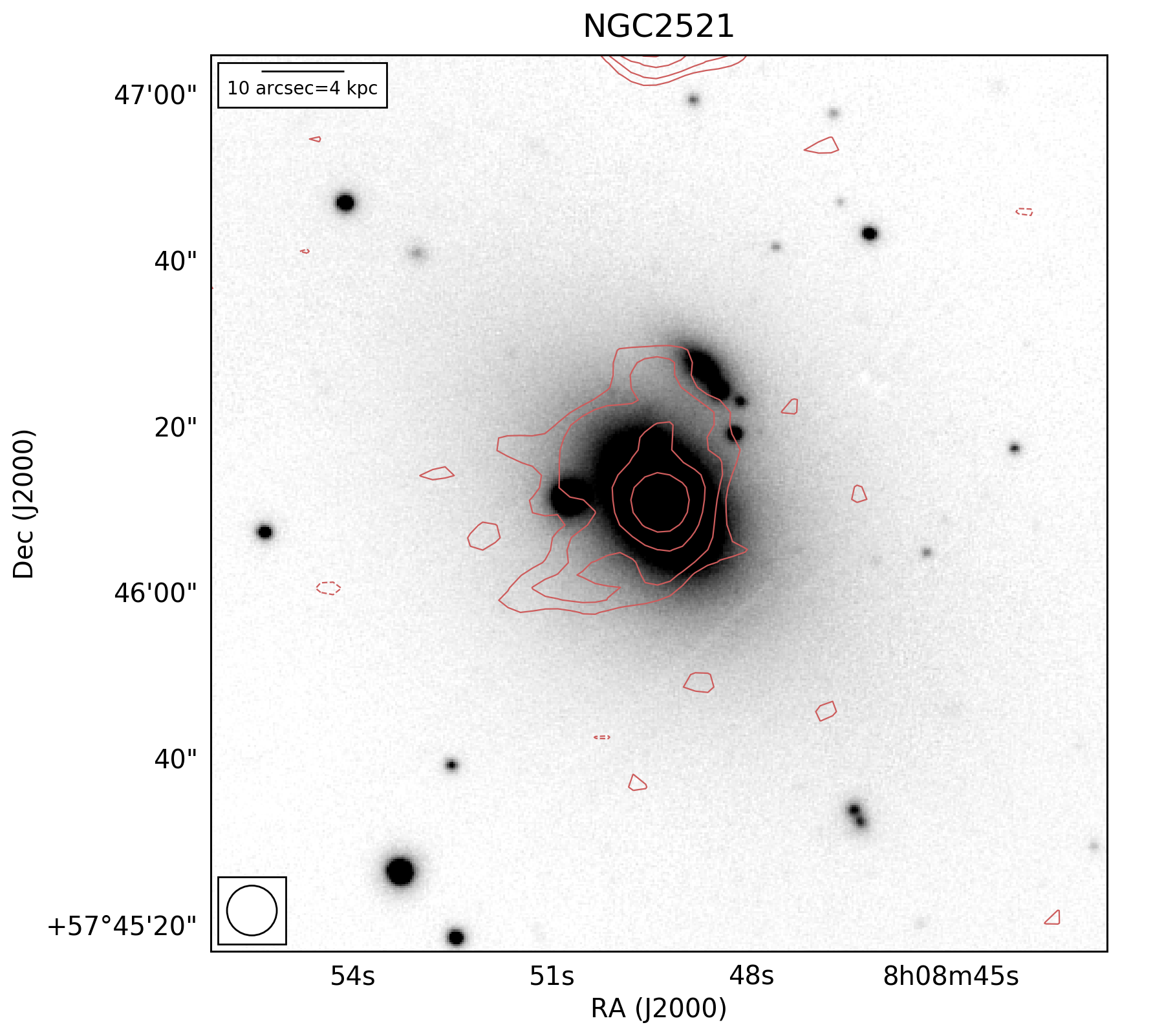}
\includegraphics[scale=0.27]{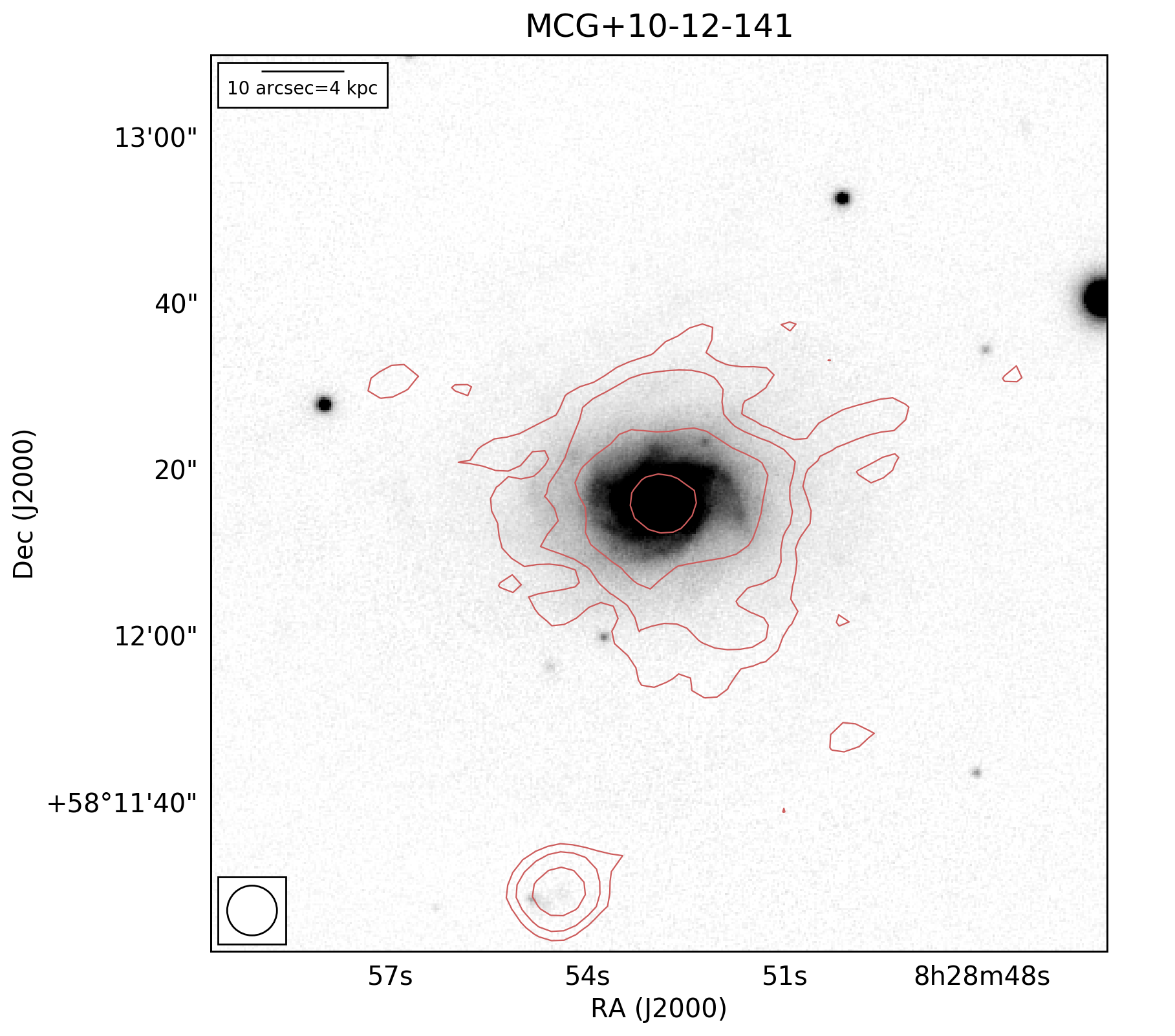}
\includegraphics[scale=0.27]{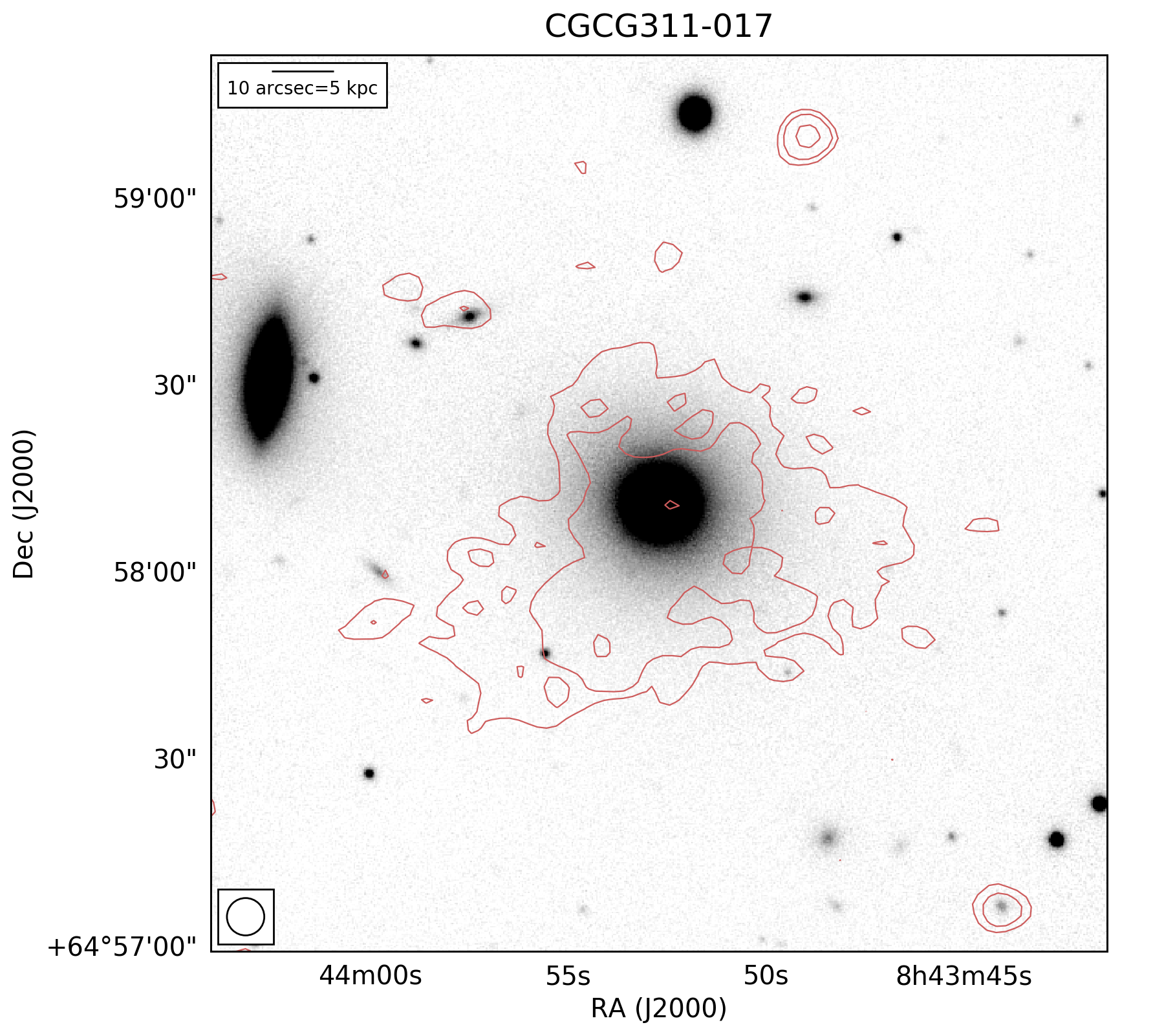}
\caption{LOFAR images at 150 MHz of the 31 galaxies showing extended
  emission. The lowest contour is drawn at three times the local
  r.m.s., as reported in Table \ref{tab}.}
\label{estese2}
\end{figure*}

\addtocounter{figure}{-1}
\begin{figure*}
\includegraphics[scale=0.27]{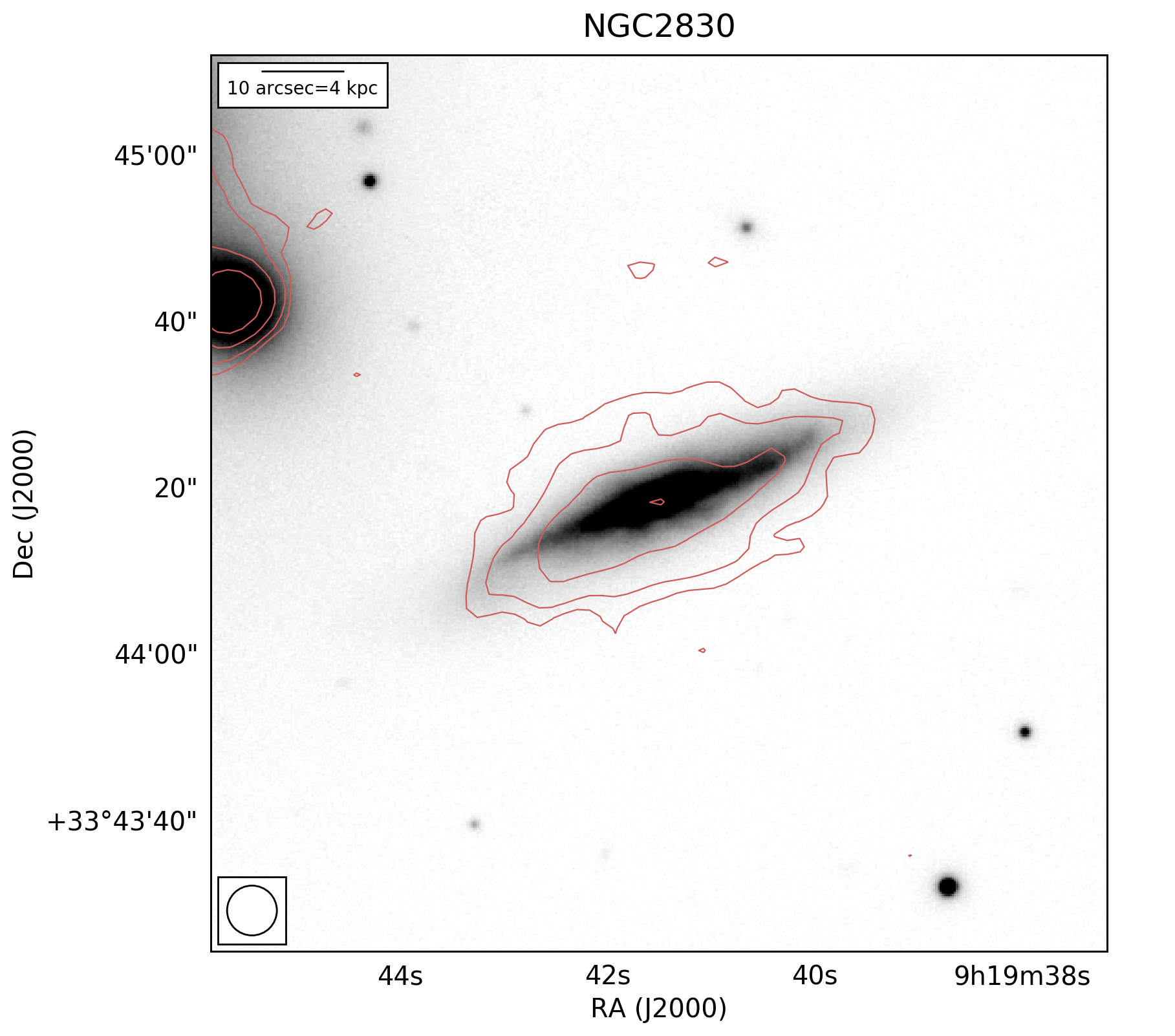}
\includegraphics[scale=0.27]{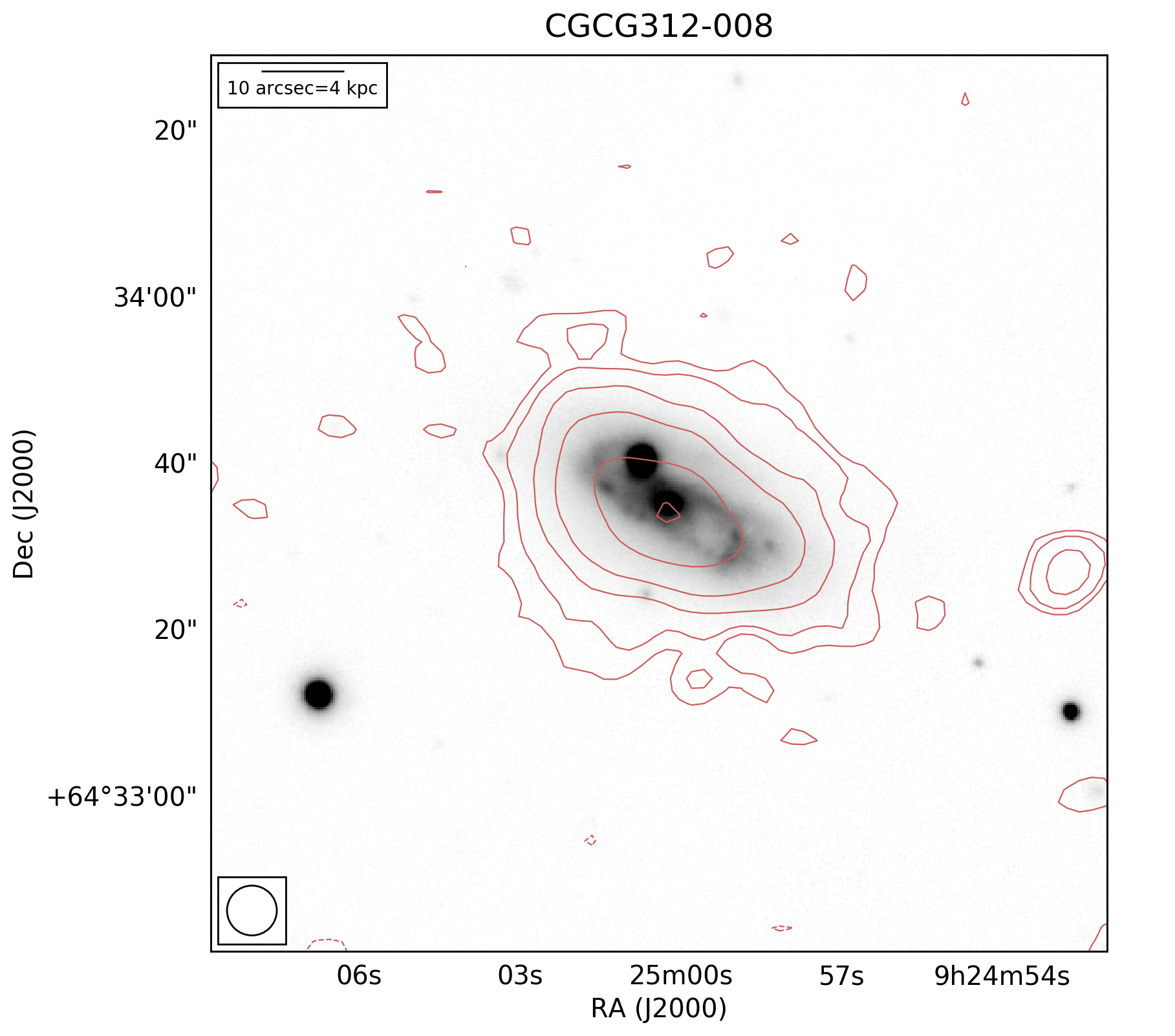}
\includegraphics[scale=0.27]{NGC3207-red.png}
\includegraphics[scale=0.27]{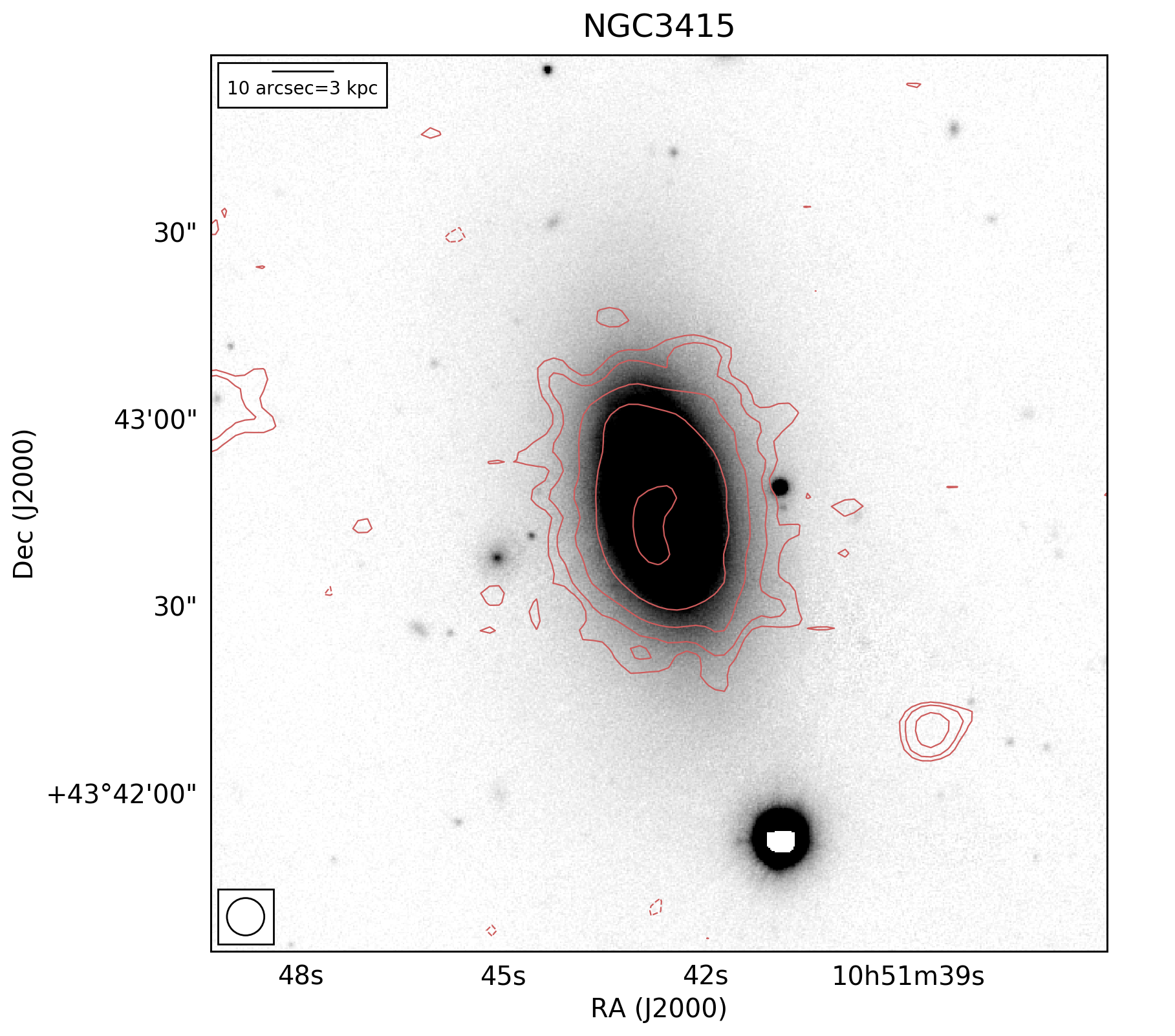}
\includegraphics[scale=0.27]{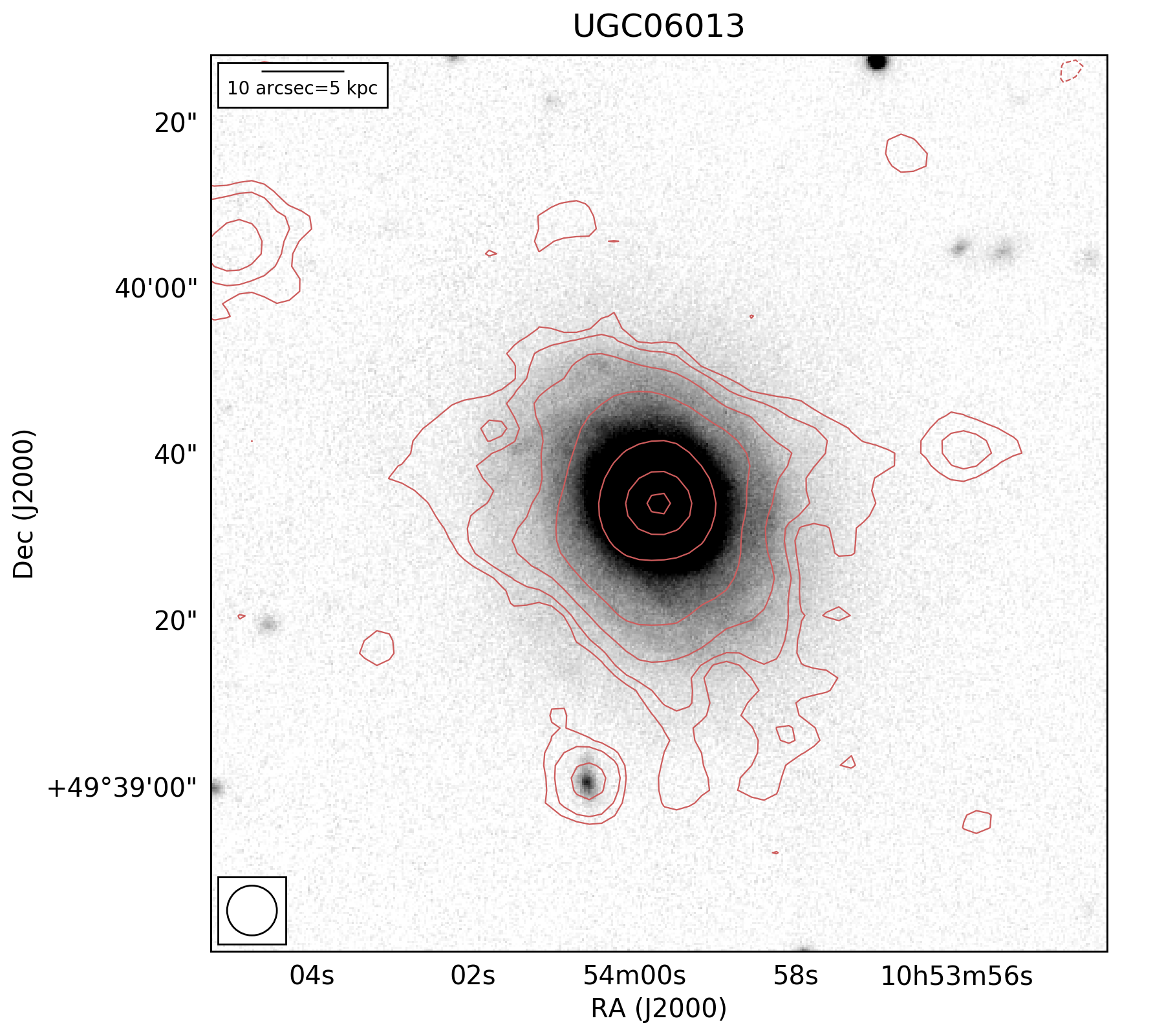}
\includegraphics[scale=0.27]{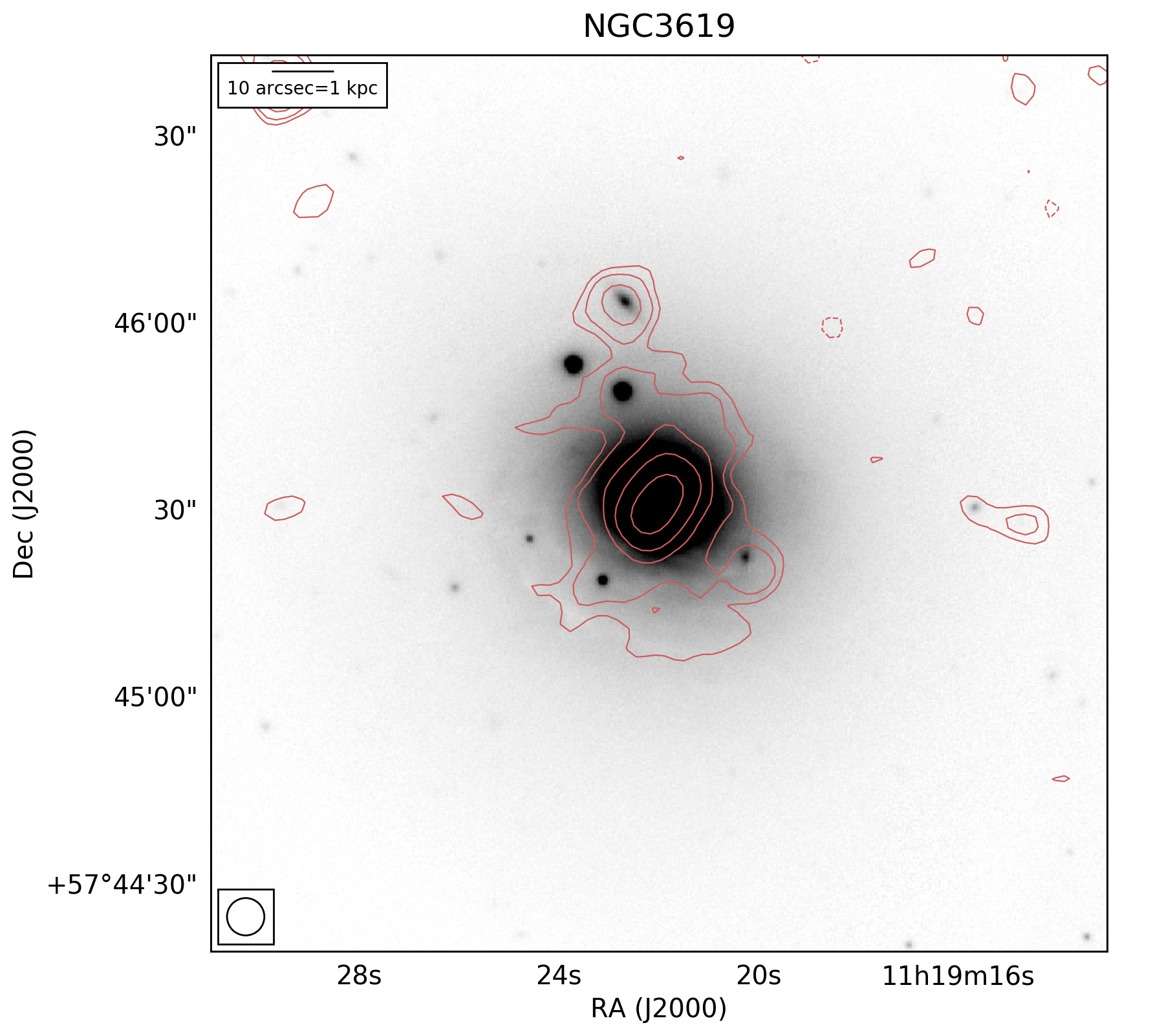}
\includegraphics[scale=0.27]{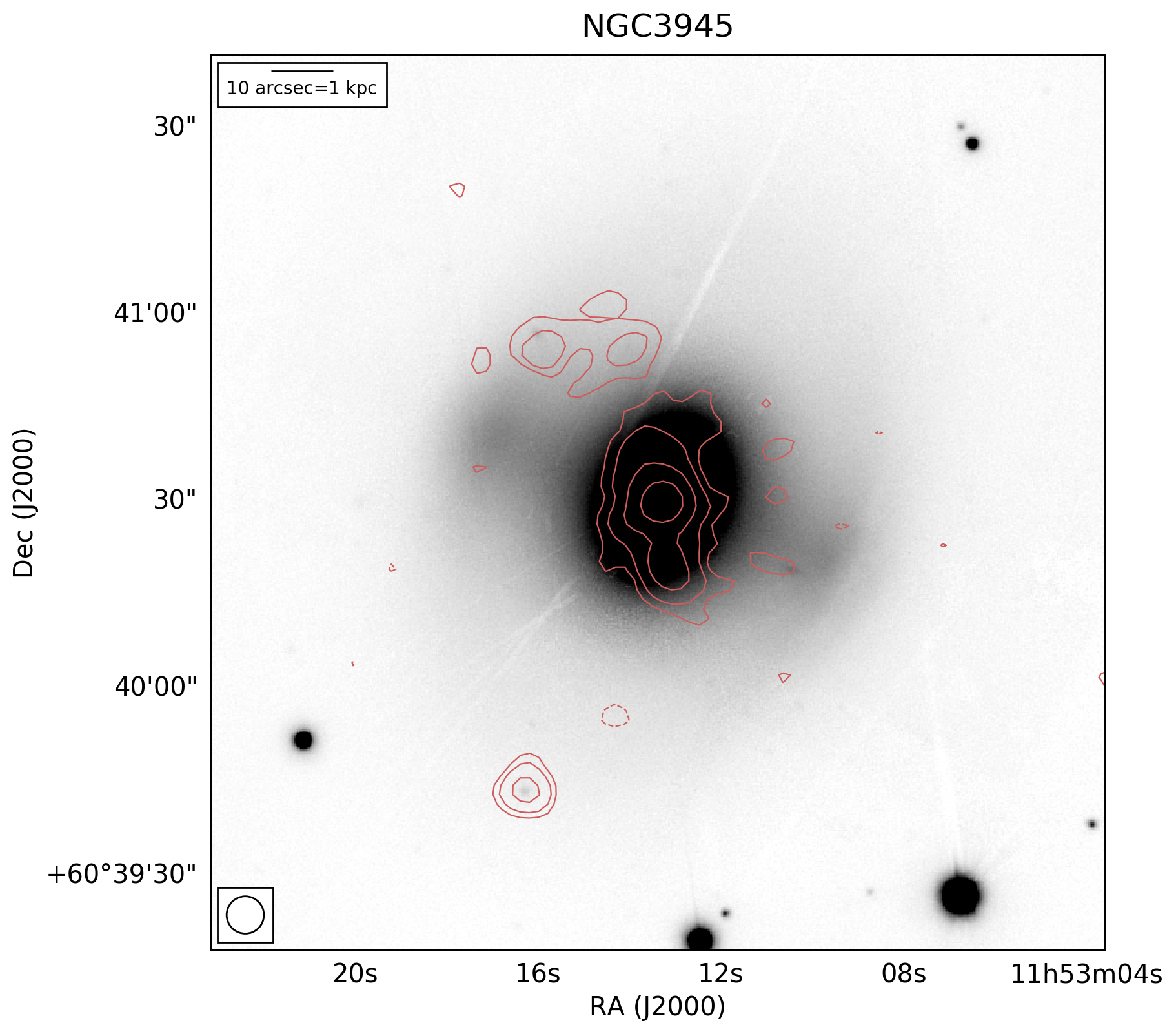}
\includegraphics[scale=0.27]{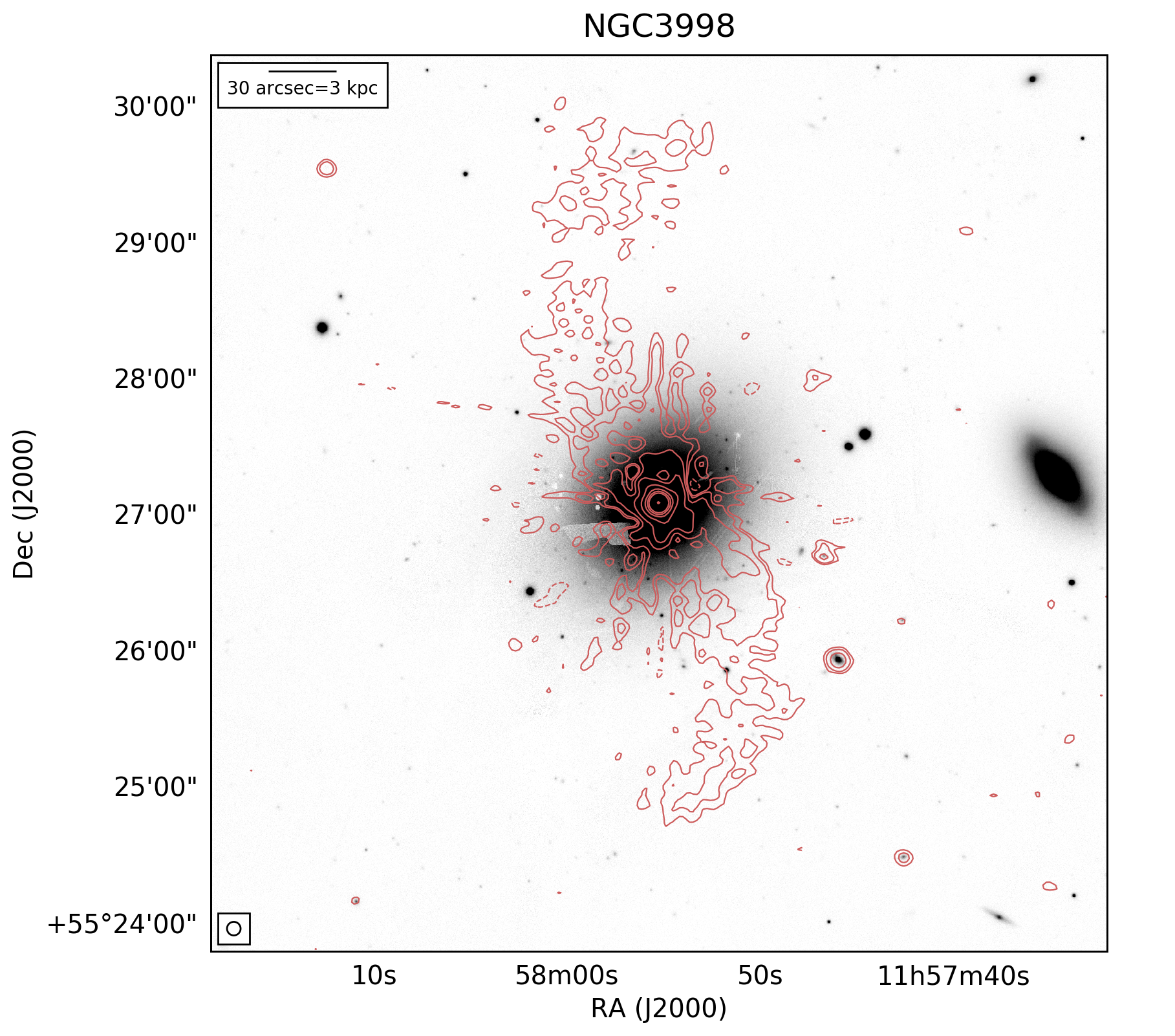}
\includegraphics[scale=0.27]{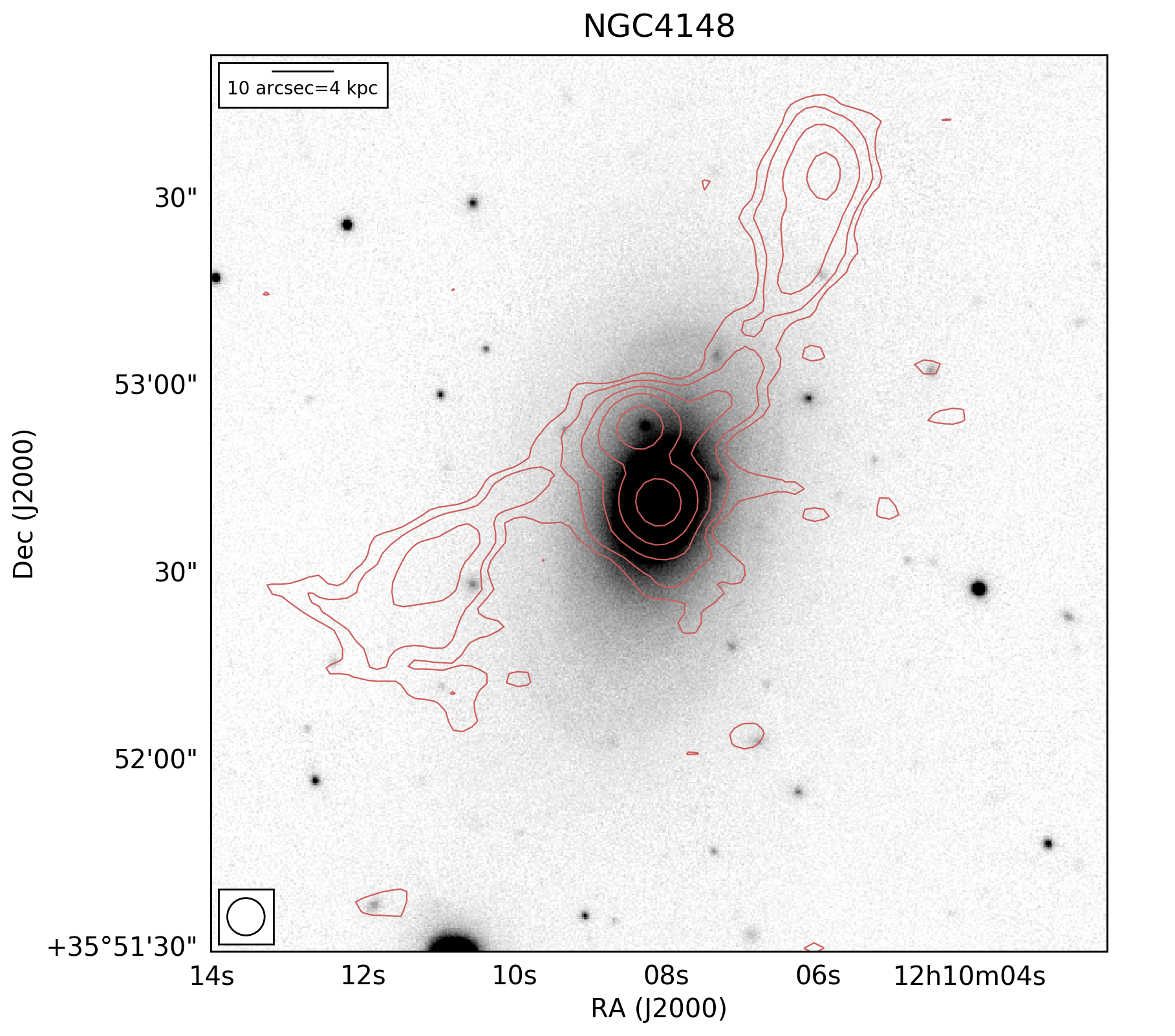}
\includegraphics[scale=0.27]{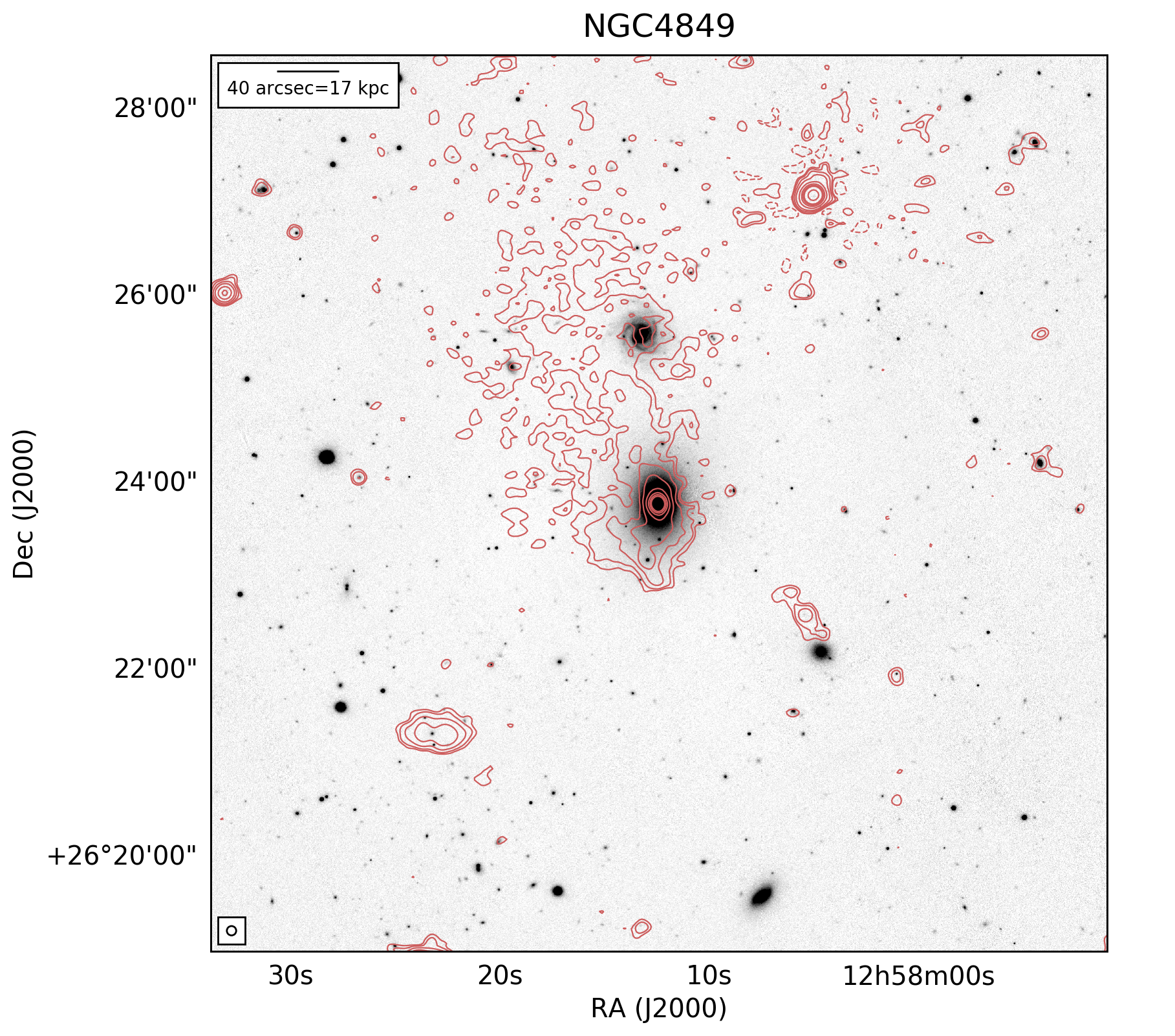}
\includegraphics[scale=0.27]{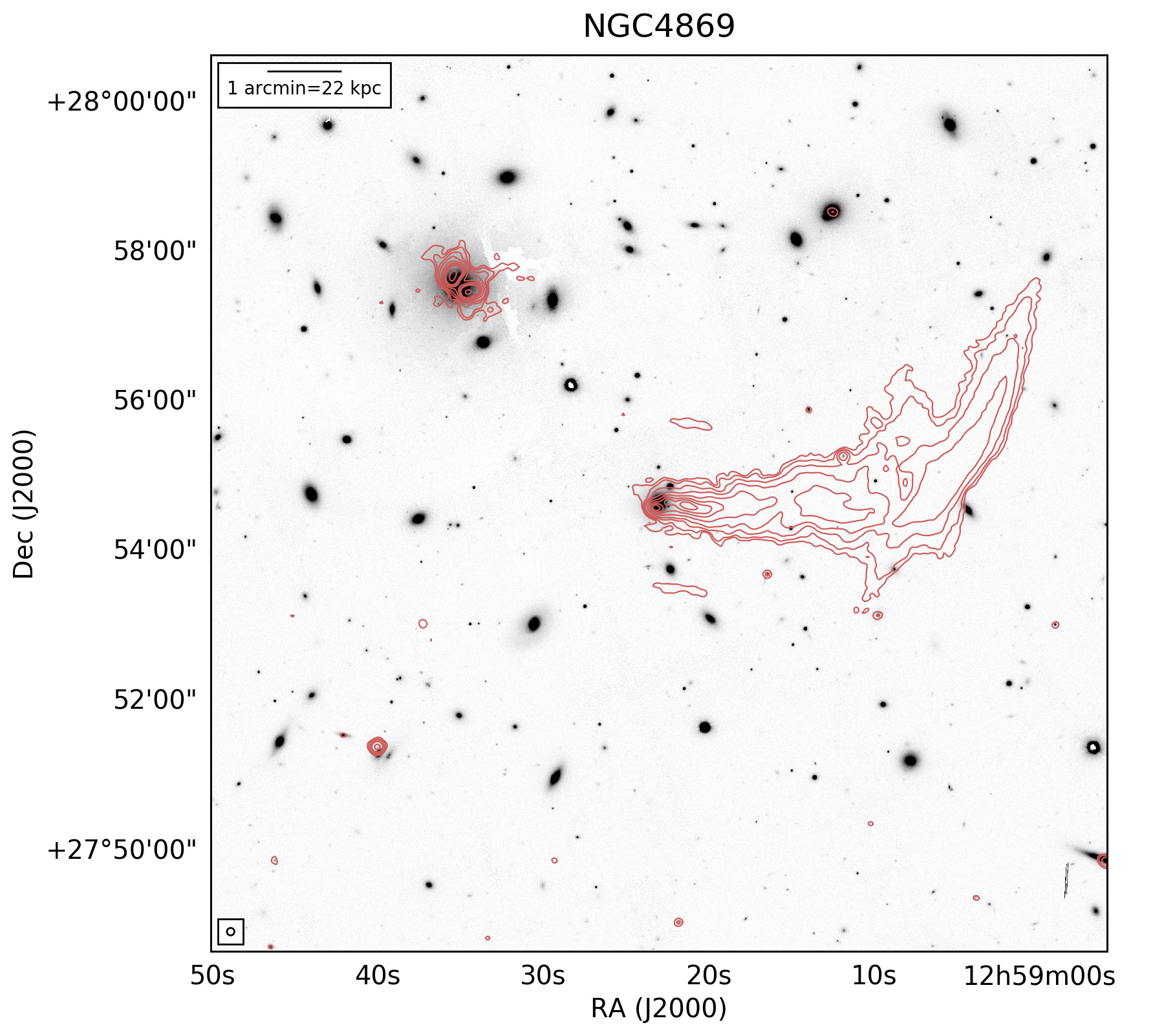}
\includegraphics[scale=0.27]{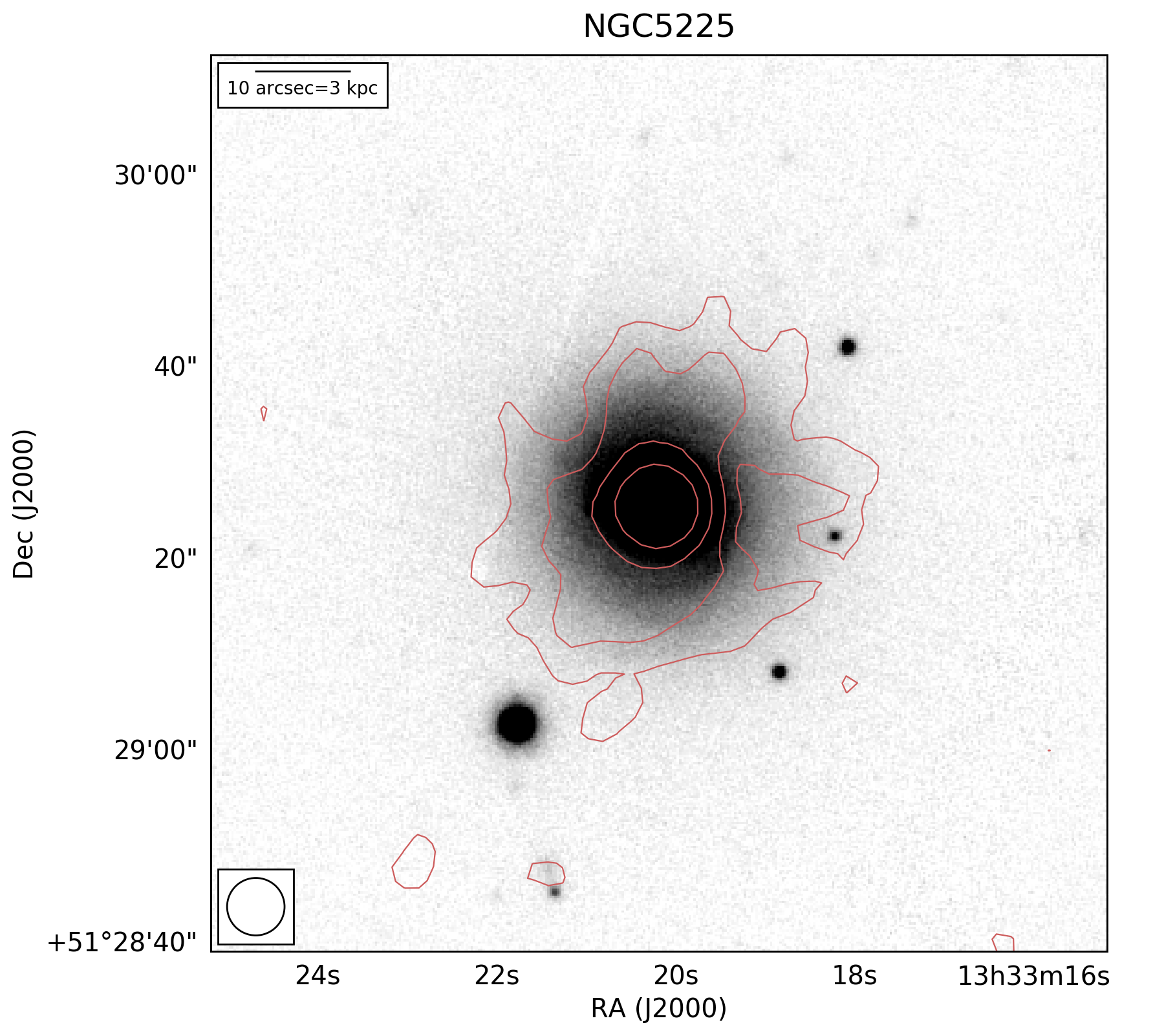}
\caption{(continued)}
\end{figure*}

\addtocounter{figure}{-1}
\begin{figure*}
\includegraphics[scale=0.27]{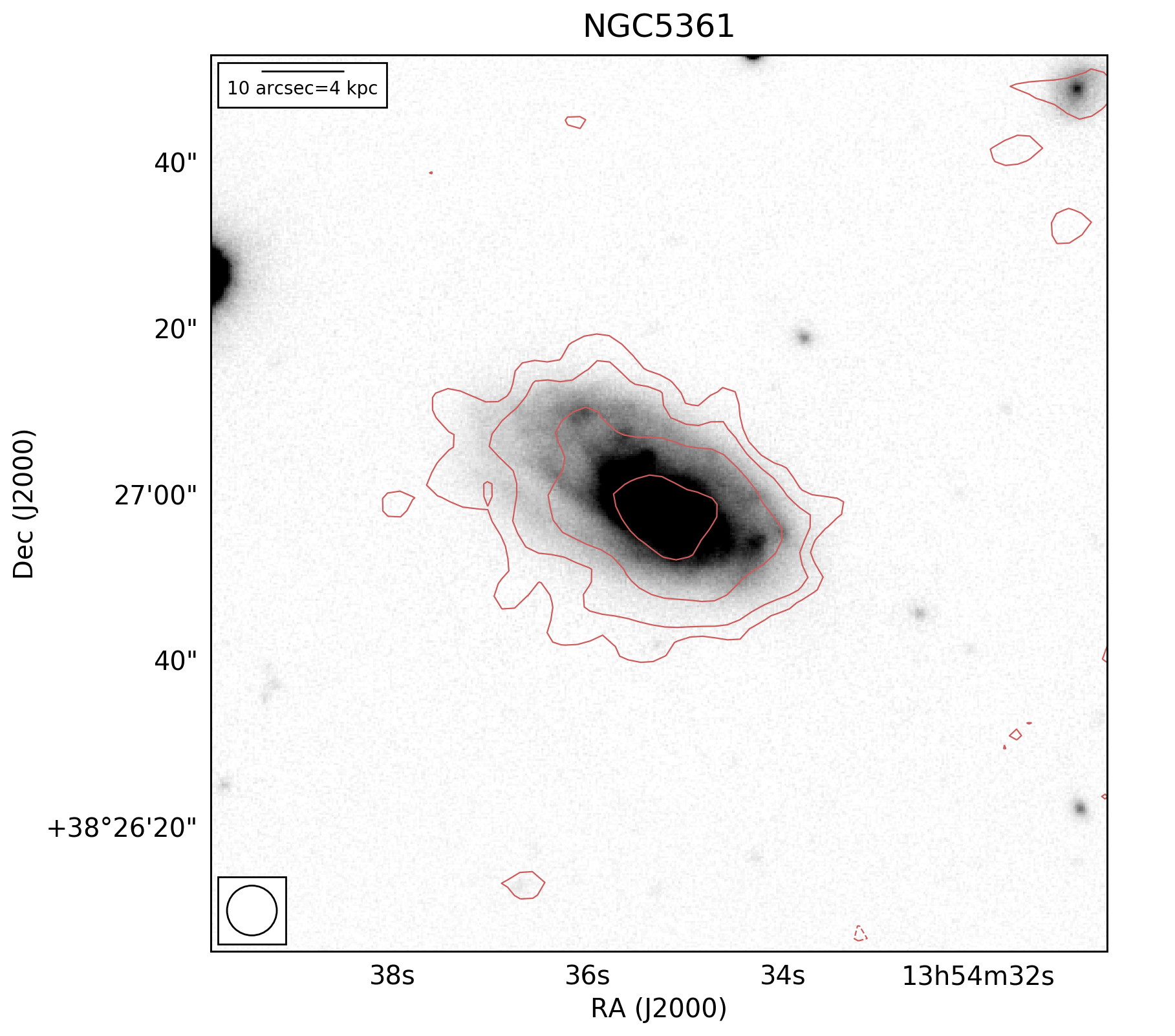}
\includegraphics[scale=0.27]{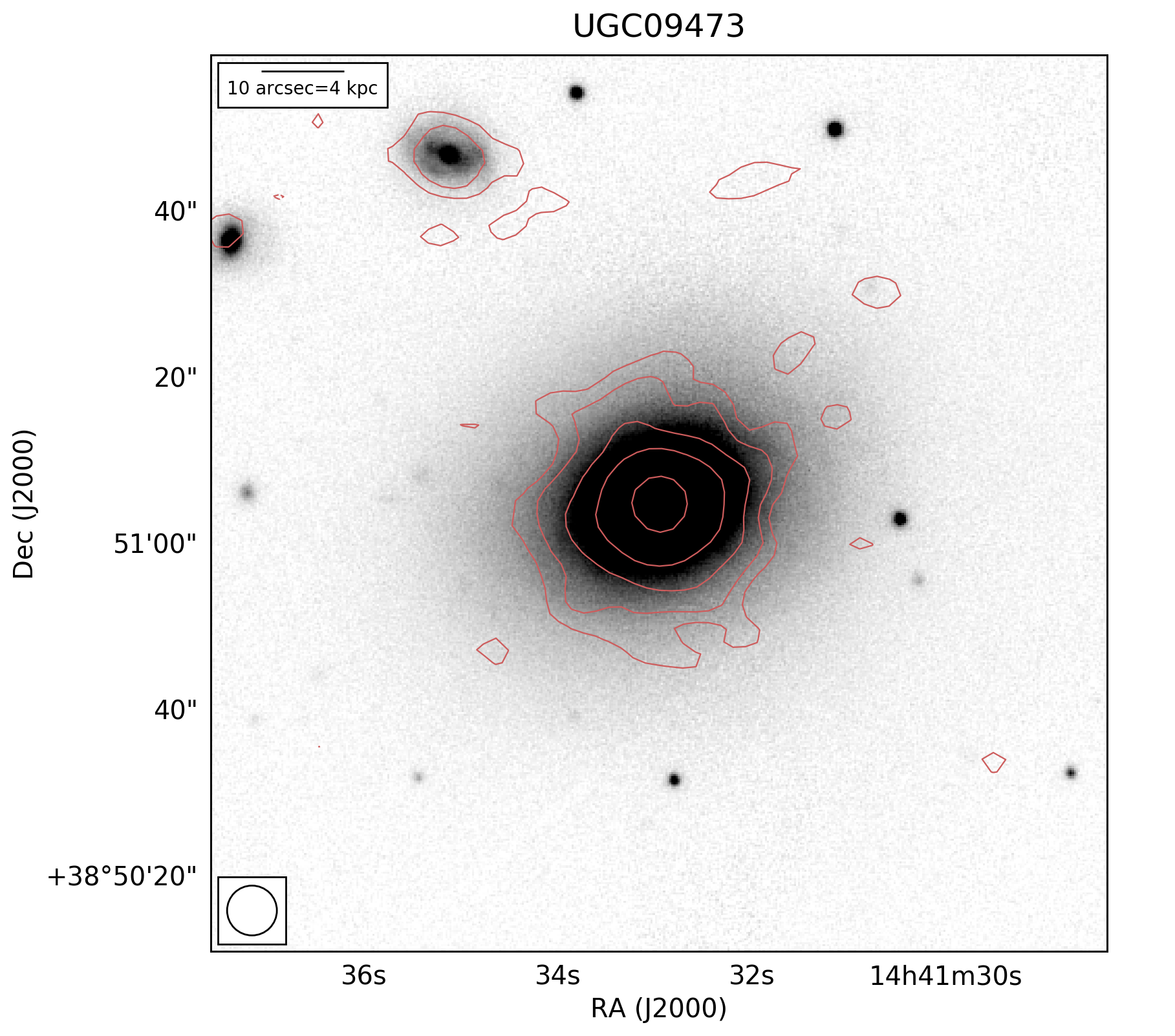}
\includegraphics[scale=0.27]{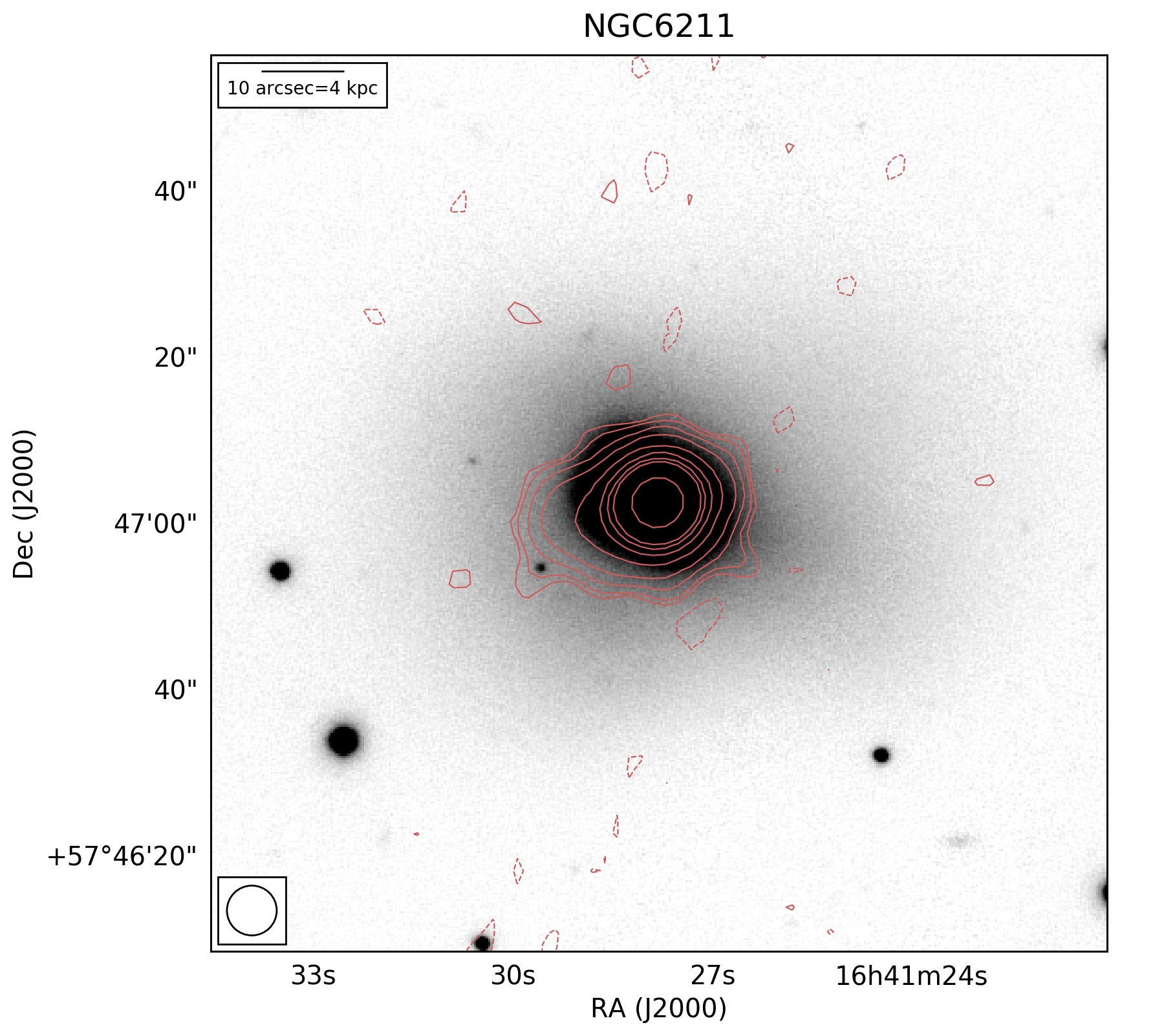}
\includegraphics[scale=0.27]{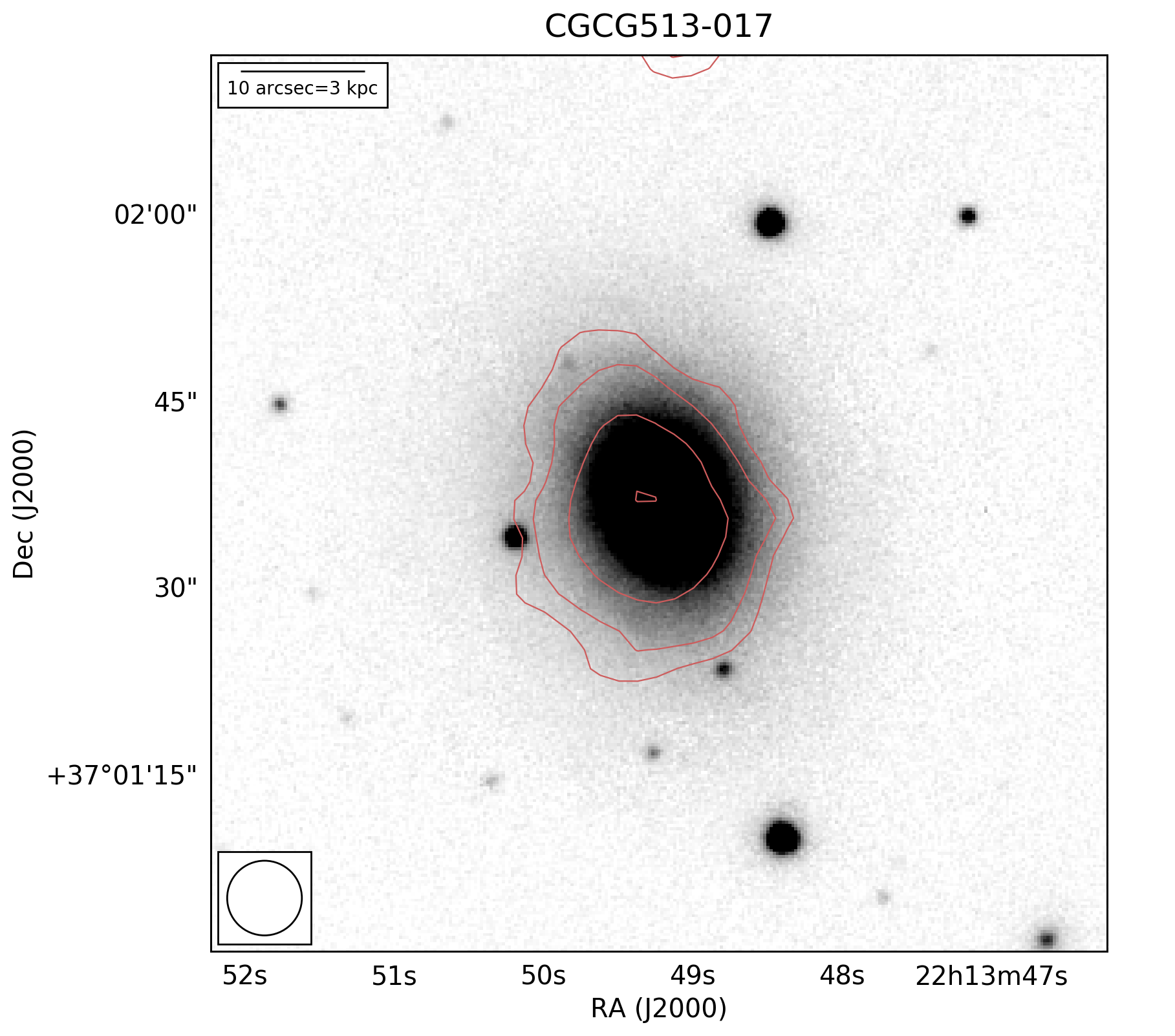}
\includegraphics[scale=0.27]{NGC7276-red.png}
\includegraphics[scale=0.27]{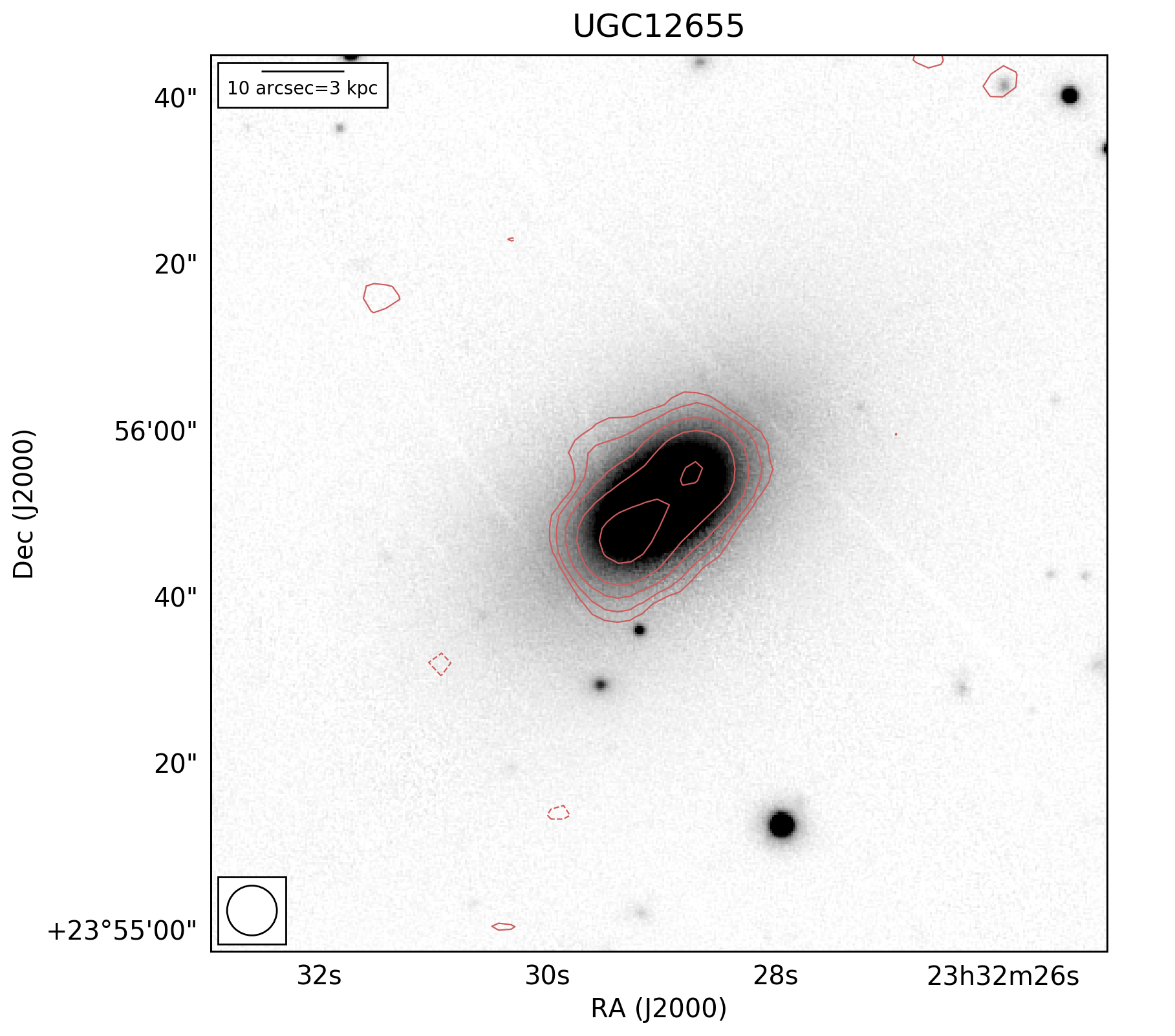}
\includegraphics[scale=0.27]{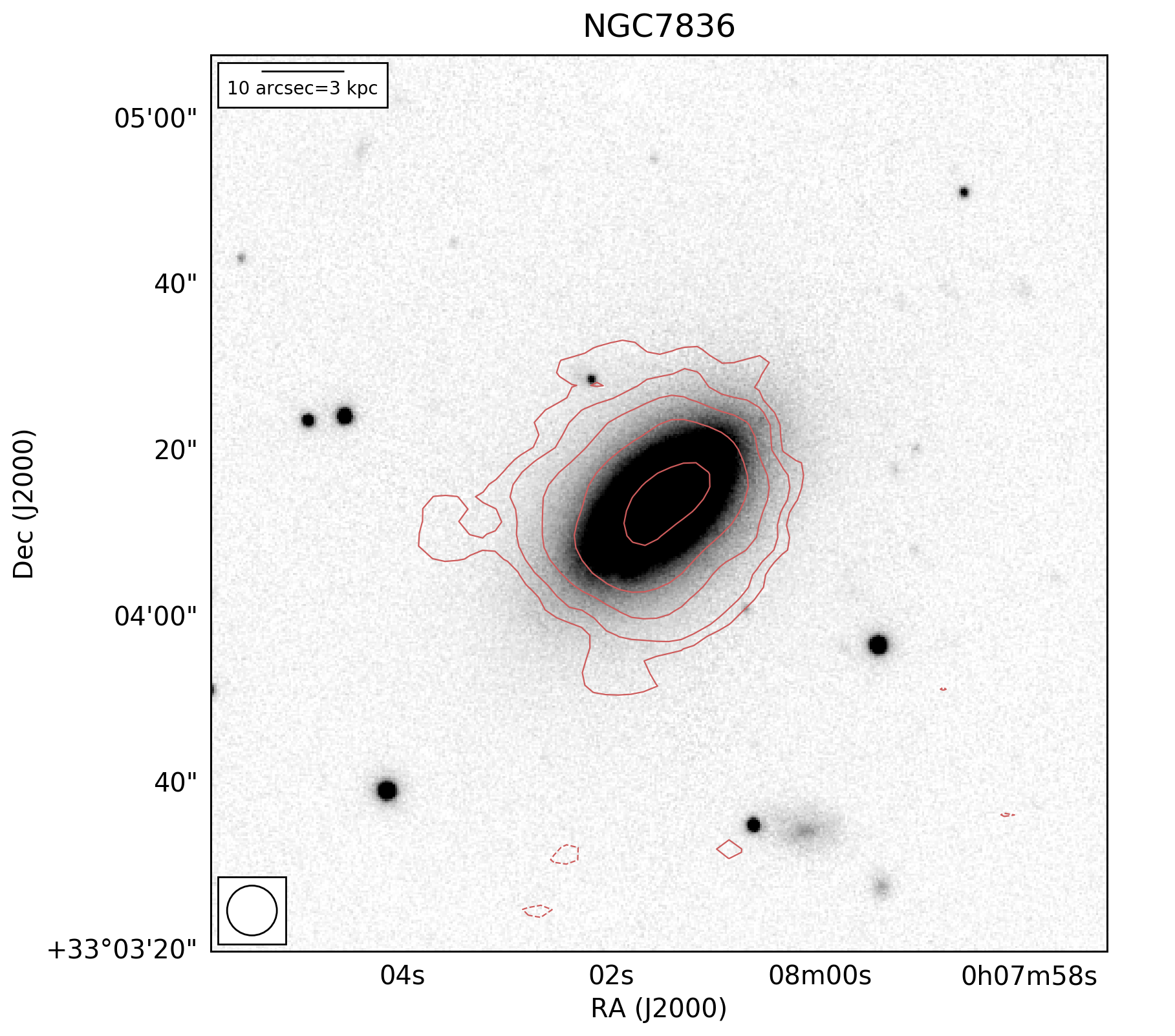}
\caption{(continued)}
\end{figure*}

We present here the images and a description of the properties of the
31 mETGs with extended radio emission, that is, those in which the
3$\sigma$ level isophote extends to a radius of at least 15\arcsec
(about twice the beam FWHM).

\noindent {\bf CGCG~499-084.} Two diffuse tails of emission, extending
for $\sim$ 180 kpc, are suggestive of the presence of jets. This
source has a rather steep spectral slope ($\alpha_{150}^{1400}$=1.14),
and we consider it candidate restarted source.

\noindent {\bf CGCG~536-006.} Radio emission aligned with the
optical axis.

\noindent {\bf UGC~00724.} Diffuse and elongated emission.

\noindent {\bf UGC~01022.} Diffuse source. Possible remnant.

\noindent {\bf NGC~0670.} Cospatial radio and optical emission.

\noindent {\bf UGC~01503.} Ring-like structure. Possible remnant
source.

\noindent {\bf UGC~01590.} Possible one-sided jet. 

\noindent {\bf UGC~04085.} Radio emission aligned with the optical
axis.  

\noindent {\bf CGCG~147-020.} Possible two-sided jetted source.

\noindent {\bf NGC~2521.} Diffuse radio structure. Possible remnant source.

\noindent {\bf MCG~+10-12-141.} Unclear radio structure.

\noindent {\bf CGCG~311-017.} Diffuse radio source with a steep spectral
index ($\alpha_{150,1400} > 1.34$): We consider it candidate
remnant source.

\noindent {\bf NGC~2830.} Cospatial radio and optical emission.

\noindent {\bf CGCG~312-008.} Radio emission aligned with the optical
axis.

\noindent {\bf NGC~3207.} Two-sided jetted source. 

\noindent {\bf NGC~3415.} Cospatial radio and optical emission.

\noindent {\bf UGC~06013.} Cospatial radio and optical emission.
 
\noindent {\bf NGC~3619.} Four nearby galaxies cause the radio
emission to appear extended.

\noindent {\bf NGC~3945.} Possible two-sided jetted source. 

\noindent {\bf NGC~3998.} Two-sided jetted source. 

\noindent {\bf NGC~4148.} The central emission has a double structure,
with the secondary compact source cospatial with a fainter galaxy
located $\sim 12$\arcsec \ north. The extended lobes appear to originate
from this secondary radio galaxy.

\noindent {\bf NGC~4849.} Possible two-sided jetted source. 

\noindent {\bf NGC~4869.} Narrow-angle tail jetted source.

\noindent {\bf NGC~5225.} Unclear radio structure.

\noindent {\bf NGC~5361.}  Cospatial radio and optical emission.

\noindent {\bf UGC~09473.} Unclear radio structure.

\noindent {\bf NGC~6211.} Possible one-sided jetted source.

\noindent {\bf CGCG~513-017.} Radio emission aligned with the optical
axis.

\noindent {\bf NGC~7276.} Diffuse radio source with a steep spectral
index ($\alpha_{150,1400} > 1.35$): We consider it candidate remnant source.

\noindent {\bf UGC~12655.}  Radio emission aligned with the optical
axis.

\noindent {\bf NGC~7836.} Radio emission aligned with the optical
axis.
          
\clearpage
\newpage

\section{Environment of the gETG sample.}

\begin{table*}
\label{tabnmb}
  \caption{Number of group members for the gETGs.}
\begin{tabular}{l r | l r | l r | l r | l r | l r}
  \hline
         Name     & $N_{\rm mb}$ & Name     & $N_{\rm mb}$ & Name     & $N_{\rm mb}$ & Name     & $N_{\rm mb}$  & Name & $N_{\rm mb}$ & Name     & $N_{\rm mb}$ \\ 
         \hline
         NGC0057&   6 &           UGC01841&  54 &       CGCG286-070 &  7   &         NGC3816&  61&              NGC4978&   1&                 IC5180&  14 \\
         NGC0068&  21 &            NGC0910&  54 &          UGC04051 &  9   &         NGC3837&  61&               IC0858&   3&               UGC11950&  14 \\
         NGC0071&  21 &           UGC01930&   2 &          UGC04052 &  9   &         NGC3842&  61&               IC0885&   2&                NGC7242&  17 \\
         NGC0076&  21 &            NGC0940&   5 &       CGCG310-022 &  4   &         NGC3862&  61&              NGC5127&   3&                NGC7248&  14 \\
         NGC0080&  18 &           NGC0978A&  11 &           NGC2475 &  1   &         NGC3886&  61&              NGC5141&   5&                NGC7265&  26 \\
         NGC0083&  18 &            NGC1004&  16 &           NGC2492 &  3   &         NGC3894&   7&              NGC5322&   8&                NGC7274&  26 \\
         NGC0128&   5 &            NGC1016&  16 &           NGC2493 &  2   &         NGC3919&  18&              NGC5332&   1&                NGC7315&  22 \\
\hline
\end{tabular}
\end{table*}

\clearpage
\section{Properties of the lmETG sample}

 \begin{longtable}{l | r r c c | c r r l | r | r | r}

\caption{Radio properties of the lmETG sample}
\label{tablm} \\                                   

\hline 
Name & R.A. & DEC. & v & M$_K$ & rms  & F$_{150}$ & L$_{150}$ \\ 
\hline  
\endfirsthead

\multicolumn{3}{c}{{\tablename} \thetable{} -- Continued} \\                                   [0.5ex]
\hline 
Name & R.A. & DEC. & v & M$_K$ & rms  & F$_{150}$ & L$_{150}$ \\ 
\hline
\endhead

\hline
  \multicolumn{10}{c}{{Continued on next page}} \\                                   
\endfoot

  \\                                   [-1.8ex] 
\endlastfoot
                  NGC7805 &     0.3616 &    31.4337 & 4766 & -24.10 & 0.10 &   $<$ 0.5              & $<$ 20.46     &   \\
                  NGC0005 &     1.9537 &    35.3623 & 5062 & -23.74 & 0.17 &   $<$ 0.8              & $<$ 20.73     &   \\
                  NGC7836 &     2.0068 &    33.0707 & 4854 & -24.00 & 0.09 &     43.0 $\pm$     0.3 &     22.41     & E \\
                  NGC0016 &     2.2679 &    27.7294 & 3097 & -24.52 & 0.13 &   $<$ 0.7              & $<$ 20.21     &   \\
                  NGC0006 &     2.3862 &    33.3087 & 4919 & -24.88 & 0.08 &      4.7 $\pm$     0.5 &     21.46     & P \\
                  NGC0042 &     3.2348 &    22.1003 & 5876 & -24.58 & 0.10 &      1.9 $\pm$     0.0 &     21.23     & P \\
                  NGC0043 &     3.2531 &    30.9153 & 4728 & -24.78 & 0.10 &   $<$ 0.5              & $<$ 20.44     &   \\
                 UGC00130 &     3.4872 &    30.8830 & 4724 & -23.62 & 0.10 &   $<$ 0.5              & $<$ 20.46     &   \\
\hline
\end{longtable}

\smallskip
\small{Column description: (1) name, (2 and 3) right ascension and declination,
(4) recession velocity (\kms), (5) K-band absolute magnitude, (6)
r.m.s. noise of the LOFAR image (mJy beam$^{-1}$), (7 and 8) flux
density (mJy) and luminosity (W Hz$^{-1}$) at 150 MHz from LOFAR.}

\end{appendix}

\end{document}